\documentclass{aa} 

\usepackage{graphicx}
\usepackage{txfonts}
\usepackage[dvipsnames]{xcolor}

\usepackage{float}
\usepackage{array}
\usepackage{hyperref}
\usepackage{physics}
\usepackage{amsfonts}
\usepackage{amsmath}
\usepackage{amssymb}
\usepackage{xcolor}
\usepackage{placeins}
\usepackage{multirow}
\usepackage{soul}

\def\jobname{aa57106-25}

\begin{document} 

   \title{The radial acceleration relation at the EDGE of galaxy formation:}
   \subtitle{testing its universality in low-mass dwarf galaxies}

   \author{Mariana P. Júlio \inst{1, 2} \and Justin I. Read \inst{3} \and Marcel S. Pawlowski \inst{1} \and Pengfei Li \inst{4} \and Daniel Vaz\inst{5, 6} \and Jarle Brinchmann \inst{5, 6, 7} \and Martin P. Rey \inst{8} \and Oscar Agertz \inst{9} \and Tom Holmes \inst{3}}

   \institute{Leibniz-Institut für Astrophysik Potsdam (AIP), An der Sternwarte 16, D-14482 Potsdam, Germany \and Institut für Physik und Astronomie, Universität Potsdam, Karl-Liebknecht-Straße 24/25, D-14476 Potsdam, Germany \and University of Surrey, Physics Department, Guildford, GU2 7XH, UK \and School of Astronomy and Space Science, Nanjing University, Nanjing, Jiangsu 210023, China \and  Instituto de Astrofísica e Ciências do Espaço, Universidade do Porto, CAUP, Rua das Estrelas, 4150-762 Porto, Portugal \and Departamento de Física e Astronomia, Faculdade de Ciências, Universidade do Porto, Rua do Campo Alegre 687, 4169-007 Porto, Portugal \and Leiden Observatory, Leiden University, P.O. Box 9513, 2300 RA Leiden, The Netherlands \and University of Bath, Department of Physics, Claverton Down, Bath, BA2 7AY, UK \and Lund Observatory, Division of Astrophysics, Department of Physics, Lund University, Box 43, SE-221 00 Lund, Sweden}
             
   \date{Received 04 September 2025 / Accepted 03 October 2025}

   \abstract{A tight correlation between the baryonic and observed acceleration of galaxies has been reported over a wide range of mass ($10^8 < M_{\rm bar}/{\rm M}_\odot < 10^{11}$) -- the Radial Acceleration Relation (RAR). This has been interpreted as evidence that dark matter is actually a manifestation of some modified weak-field gravity theory. In this paper, we study the radially resolved RAR of 12 nearby dwarf galaxies, with baryonic masses in the range $10^4 < M_{\rm bar}/{\rm M}_\odot < 10^{7.5}$, using a combination of literature data and data from the MUSE-Faint survey. We use stellar line-of-sight velocities and the Jeans modelling code \textsc{GravSphere} to infer the mass distributions of these galaxies, allowing us to compute the RAR. We compare the results with the EDGE simulations of isolated dwarf galaxies with similar stellar masses in a $\Lambda$CDM cosmology. We find that most of the observed dwarf galaxies lie systematically above the low-mass extrapolation of the RAR. Each galaxy traces a locus in the RAR space that can have a multi-valued observed acceleration for a given baryonic acceleration, while there is significant scatter from galaxy to galaxy. Our results indicate that the RAR does not apply to low-mass dwarf galaxies, and that the inferred baryonic acceleration of these dwarfs does not contain enough information, on its own, to derive the observed acceleration. The simulated EDGE dwarfs behave similarly to the real data, with a higher observed acceleration at a fixed baryonic acceleration than the extrapolated RAR. We show that, in the context of modified weak-field gravity theories, these results cannot be explained by differential tidal forces from the Milky Way, nor by the galaxies being far from dynamical equilibrium, since none of the galaxies in our sample seems to experience strong tides. As such, our results provide further evidence for the need for invisible dark matter in the smallest dwarf galaxies.}

    \keywords{dark matter – galaxies: dwarf - galaxies: kinematics and dynamics – stars: kinematics and dynamics – techniques: imaging spectroscopy}
   
    \titlerunning{The RAR in the dwarf regime}
	\authorrunning{M. P. Júlio et al.}
	
	\maketitle

\section{Introduction}\label{introduction}

A key puzzle of galaxy formation in the Standard $\Lambda$-Cold Dark Matter ($\Lambda$CDM) cosmology is the apparently tight correlation between the distribution of baryonic matter in galaxies and their observed dynamics, even in systems where the dark matter dominates (see e.g. \citealt{bullock} for a review). These correlations have been known about since the 1970s and include the Tully-Fisher relation \citep{tully_new_1977} that links the luminosity to the peak rotation velocity for rotationally-supported galaxies, and the Faber-Jackson relation \citep{faber_velocity_1976} that links stellar mass and stellar velocity dispersion for pressure-supported systems. More recently, the mass-discrepancy acceleration relation, which relates the distribution of baryonic matter with the observed acceleration of a galaxy \citep{mcgaugh_mass_2004}, and the radial acceleration relation (RAR, \citealt{mcgaugh_radial_2016} and \citealt{lelli_one_2017}), which relates the observed dynamical radial acceleration with the acceleration expected from the visible baryons in galaxies, were discovered. 

To establish the RAR, \cite{mcgaugh_radial_2016} used the Spitzer Photometry and Accurate Rotation Curves (SPARC) database \citep{lelli_sparc_2016} to determine the observed and the baryonic acceleration of over 150 late-type galaxies. These covered a wide range of morphological types (S0 to Irr), surface brightnesses ($\sim 4$ dex) and luminosities ($\sim 5$ dex), corresponding to baryonic masses over the range $10^8 < M_{\rm bar}/{\rm M}_\odot < 10^{11}$\,M$_\odot$. This relation is surprisingly tight, with an observed scatter of just 0.13\,dex which is mostly driven by observational errors. In \cite{lelli_one_2017}, it was confirmed that early-type galaxies and classical dwarf spheroidals followed the same relation as the one in \cite{mcgaugh_radial_2016}, within the uncertainties at that time. Furthermore, the residuals are well described by a Gaussian of width 0.11\,dex and did not show any correlation with galaxy properties, such as baryonic mass, gas fraction or radius. \cite{freundlich_probing_2022} showed that ultra-diffuse galaxies in the Coma cluster also follow the RAR. \cite{mistele_radial_2024} used kinematic and gravitational lensing data to derive the RAR over a large dynamic range. They found that the RAR inferred from weak-lensing data smoothly continues the trend inferred from kinematic data towards lower accelerations by about 2.5 dex. They also found that early- and late- type galaxies lie on the same joint RAR when a sufficiently strict isolation criterion is adopted. Recently, \cite{varasteanu_mightee-hi_2025} used resolved HI kinematics from the MIGHTEE-HI survey \citep{jarvis_meerkat_2016} to extend the RAR down to accelerations of $\log g_{\mathrm{bar}}\sim-12$m\,s$^{-2}$ and found that gas-rich galaxies continue to follow the same relation over several orders of magnitude ($M_\star\sim10^8 -10^{10}$M$_\odot$) up to $z=0.08$. Taken together, these works suggested that the RAR could be a fundamental relation obeyed by all galaxies.

However, for very low and high mass systems, and when considering distinct radii within galaxies, cracks begin to show in the idea that the RAR is universal. At high mass, both the brightest galaxies in galaxy clusters, and galaxy clusters themselves, appear to obey an RAR, but one distinct from the galactic RAR \citep{tian_radial_2020,tian_distinct_2024, li_measuring_2023}. At low mass, \citet{lelli_one_2017} showed preliminary evidence that dwarf spheroidal galaxies tend to scatter above the low-mass extrapolation of the RAR, with large galaxy-to-galaxy scatter (though within the quoted observational uncertainties at that time). \citet{read_dark_2019} showed that two of the Milky Way's satellites, Draco and Carina, have similar baryonic mass distributions and similar orbits around the Milky Way, but distinct dynamical mass distributions, inconsistent with a universal RAR. Finally, individual galaxies can deviate from the RAR deep in the centre. For example, some galaxies at lower accelerations trace a curved locus in the RAR space, with two different observed accelerations for the same baryonic acceleration \citep{Eriksen21}, but this is less frequent in observations than in simulations \citep{Mercardo24}. As such, it is yet unclear if the RAR is truly a one-to-one map from baryonic to observed accelerations.

Several theoretical studies have attempted to explain the RAR in a $\Lambda$CDM cosmology (for example, \citealt{ludlow_mass-discrepancy_2017}  \citealt{tenneti_radial_2018}, and \citealt{garaldi_radial_2018}). These studies find a correlation between the total and the baryonic accelerations can arise in $\Lambda$CDM as well, which is perhaps not surprising -- more massive galaxies contain more gas and form more stars \citep[e.g.][]{wang15}. However, matching the RAR in detail has proven more challenging. For instance, the relation found in \cite{ludlow_mass-discrepancy_2017} using galaxies with stellar masses ranging from $10^5$M$_\odot$ to $10^{12}$M$_\odot$, has a significantly higher characteristic acceleration scale as compared to the measured RAR. Similarly, \cite{tenneti_radial_2018} find, for galaxies with stellar masses $>10^{12}$M$_\odot$, a linear correlation, without any trace of an acceleration scale. In fact, as pointed out by \cite{li_effect_2022}, comparing the intrinsic scatter of the observed RAR with that from simulated galaxies is not straightforward (see also \citealt{keller_cdm_2017}, \citealt{garaldi_radial_2018}, and \citealt{dutton_nihao_2019}), because it requires modelling observational errors, rotation curve sampling, and the covariance between the total and the baryonic acceleration. As such, it remains unclear whether the emergence of the RAR can be understood in the Standard Cosmology, or not.

If galaxy formation simulations in $\Lambda$CDM continue to struggle to reproduce the RAR in detail, then it could be that the RAR indicates a need for new physics. One such set of models that could explain the RAR is Modified Newtonian Dynamics (MOND, \citealt{milgrom_modification_1983}). This is a modified weak-field gravitational force law hypothesised as an alternative to dark matter. In MOND, the classical laws of Newtonian dynamics are modified at low accelerations instead of adding dark matter, directly coupling the observed baryonic mass distribution to the observed galaxy dynamics \citep[e.g.][]{li_fitting_2018}. As such, MOND naturally reproduces the RAR with formally zero scatter \citep[e.g.][]{lelli_one_2017}. However, it predicts deviations from the RAR for satellite galaxies, which is opposite to what is observed \citep{lelli_one_2017}. MOND does not follow the equivalence principle, so the internal dynamics of satellite galaxies orbiting in the tidal field of their host galaxy experience an `external field effect' (EFE; e.g. \citealt{milgrom_modification_1983}, \citealt{bekenstein_does_1984}, \citealt{famaey_modified_2012}). This effectively lowers the MOND acceleration scale, causing the internal dynamics of the satellite to behave more like the Newtonian gravity case. This, in turn, causes satellites orbiting in strong tidal fields to scatter systematically {\it below} the RAR (\citealt{mcgaugh_andromeda_2013}, \citealt{pawlowski_perseus_2014}, \citealt{pawlowski_new_2015}, \citealt{mcgaugh_radial_2016}, \citealt{muller_predicted_2019}, \citealt{chae_testing_2020}, \citealt{chae_testing_2021}, \citealt{freundlich_probing_2022}), which is opposite to what is observed for dwarf spheroidal satellites in the Local Group \citep{lelli_one_2017}. More modern relativistic alternative gravity theories show promise \citep[e.g.][]{skordis21} and have weak-field behaviour that is rich and distinct from MOND \citep[e.g.][]{durakovic24}. It remains to be seen if such more complete theories can explain the latest data.

Given that the deviations from the RAR lack a full explanation in both the Standard Cosmology and modified gravity theories, it is interesting to further explore its extremities with observational data. In this paper, we use literature stellar line-of-sight velocities for the classical dwarf spheroidal satellite galaxies of the Milky Way, combined with MUSE-Faint velocities of ultra-faint dwarfs, in order to study the radially resolved RAR at much lower accelerations than has been previously possible. Moving to a regime 1-2 orders of magnitude lower in baryonic mass than the SPARC sample, we probe the internal acceleration profiles of each dwarf galaxy, and the scatter between them. We discuss how the EFE impacts our results, given that many galaxies in our sample orbit within the Milky Way, and we compare our findings to the RAR calculated for simulated dwarf galaxies in $\Lambda$CDM taken from the `Engineering Dwarfs at Galaxy Formation's Edge' (EDGE) project \citep{agertz_edge_2020}. Our goal is to determine whether the RAR really is a one-to-one map from the baryonic to observed acceleration in galaxies. If yes, then the RAR could indicate the need for new physics at low accelerations. If not, then the RAR is more likely to be an emergent phenomenon -- a challenge for modern galaxy formation theories to solve. 

This paper is organised as follows. In Sect.~\ref{sec:data}, we describe the data we will use to measure the RAR (\ref{sec:obsdata}) and we describe the EDGE simulations that we compare our results to (Sect.~\ref{sec:simulations}). In Sect.~\ref{sec:methods}, we briefly describe the mass modelling method we use in this work to compute the baryonic and observed accelerations of these galaxies (\textsc{GravSphere}; Sect.~\ref{sec:gravsphere}), and we test its recovery of the RAR using mock data drawn from the EDGE simulations (Sect.~\ref{sec:recovery}). We present our results in Sect.~\ref{sec:results}, where we show the RAR for our galaxy sample (Sect.~\ref{sec:rar-observations}) and we compare this to the EDGE simulations (Sect.~\ref{sec:rar-simulations}). Finally, we discuss our results in Sect.~\ref{sec:discussion} and present our conclusions in Sect.~\ref{sec:conclusions}. For completeness, we provide supplementary material and relevant code modifications in support of the results that we discuss in the main text in Appendices~\ref{app:rotation},~\ref{app:scatter},~\ref{app:fits},~\ref{app:tracerdensity} and~\ref{app:gas}. Throughout this paper, all logarithms are base 10, i.e., $\log \equiv \log_{10}$.

\section{Data}\label{sec:data}

\subsection{Observational data}\label{sec:obsdata}

\begin{figure}[!htbp]
    \includegraphics[width=\columnwidth]{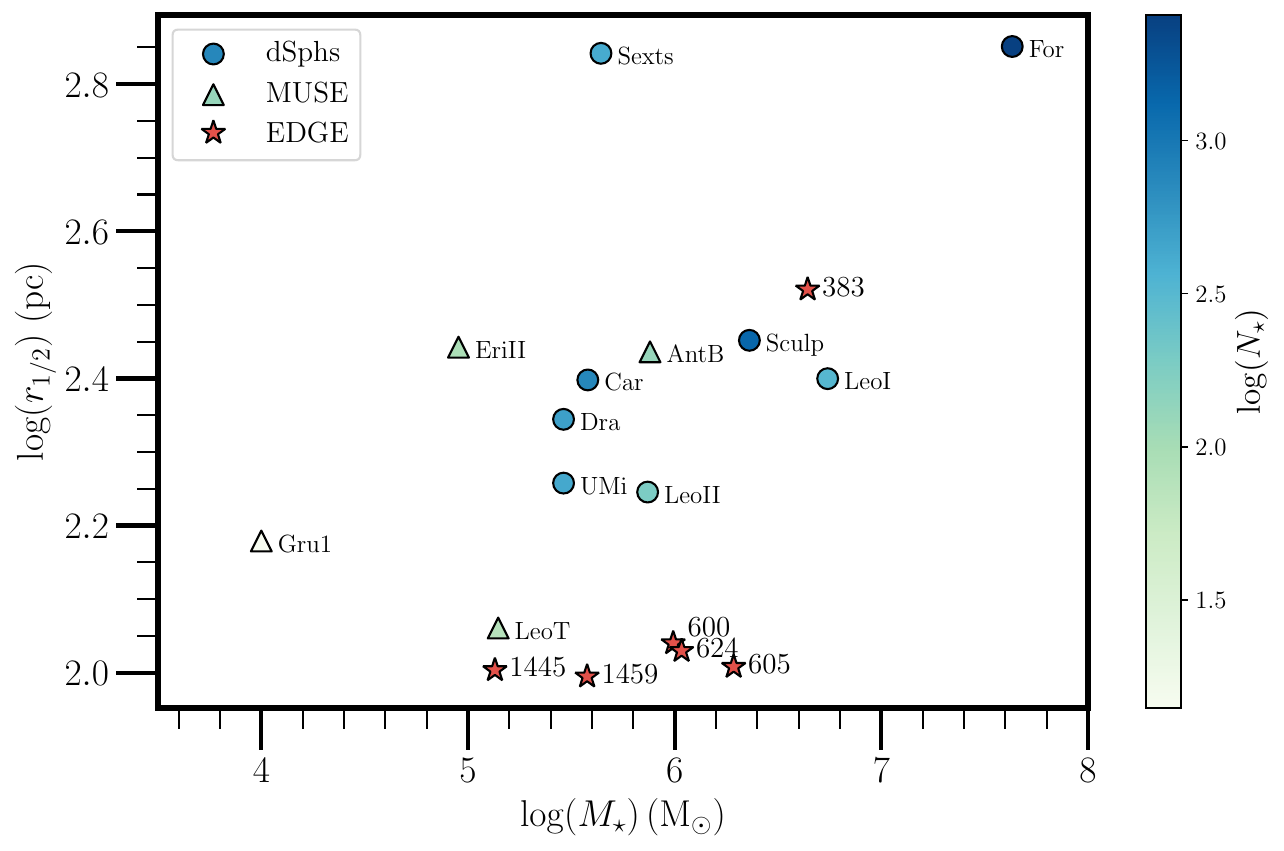}
    \caption{Stellar mass $M_\star$ against the half-light radius $r_{1/2}$ for our galaxy sample. The circles indicate the classical dSphs, and the triangles represent the dwarfs from MUSE-Faint. Objects are colour-coded by the number of stars with velocities available for each object. The red stars represent the EDGE simulated dwarfs.} \label{fig:rh-stellarmass}
\end{figure}

Our data sample comprises 12 dwarf galaxies, with eight of them being the classical dwarf spheroidals (dSphs) of the Milky Way, and the other four being faint and ultra-faint dwarfs of the Milky Way and beyond, taken from the MUSE-Faint survey \citep{zoutendijk_muse-faint_2021}. Figure~\ref{fig:rh-stellarmass} shows their structural properties, described in Table~\ref{tab:properties} and the structural properties of the EDGE simulated dwarfs for comparison, described in Table~\ref{tab:sims} of Sect.~\ref{sec:simulations}.

\subsubsection{Dwarf spheroidals}\label{sec:data-literature}
Our sample of dwarf spheroidals is comprised of Carina (Car), Draco (Dra), Fornax (For), Leo I, Leo II, Sextans (Sexts), Sculptor (Sculp) and Ursa Minor (UMi). For all of these galaxies, we have kinematics for $\gtrsim 190$ stars and their photometric light profiles are well-measured (see \citealt{read_dark_2019} for an extensive analysis). \par

For the photometric data, we use the Panoramic Survey Telescope and Rapid Response System (Pan-STARRS) DR1 catalogue \citep{flewelling_pan-starrs1_2020} for Draco, Leo I, Leo II, Sextans, and Ursa Minor, VLT/ATLAS DR1 catalogue (as re-processed by \citealt{koposov_discovery_2014}) for Fornax and Sculptor, and a catalogue derived from observations with the Dark Energy Camera by \citet{mcmonigal_sailing_2014} for Carina. \par 
For the kinematic data, we adopt spectroscopic data samples from \citet{walker_stellar_2009} for Carina, Fornax, Sculptor and Sextans, from \citet{mateo_velocity_2008} for Leo I, from \citet{spencer_multi-epoch_2017} for Leo II, from \citet{spencer_binary_2018} for Draco, and finally, from \citet{read_dark_2019} for Ursa Minor. For a comprehensive description of the selection of member stars for all dwarfs, see \citet{read_dark_2019}. \par

All of the dwarfs have tidal radii today that are significantly larger than their stellar half-light radii (\citealt{Pace22} and see Sect.~\ref{sec:efe}), though some could have experienced stronger tidal forces in the past \citep[e.g.][]{genina22}.

\subsubsection{MUSE-Faint dwarfs}\label{sec:data-muse}
Our sample of faint dwarfs comprises the ultra-faint dwarfs of the Milky Way Eridanus II (EriII), Grus 1 and Leo T, and the faint dwarf Antlia B (AntB), associated with NGC 3109. \par 
We use mock photometry catalogues with 10,000 sample photometric positions drawn from the best-fit exponential profiles found for each dwarf (see \citealt{zoutendijk_muse-faint_2021} for further details). \par  
We use spectroscopic observations from the MUSE-Faint survey, a survey of ultra-faint dwarfs conducted using the Multi Unit Spectroscopic Explorer, for these objects. The extensive data reduction and analysis for these objects can be found in \cite{zoutendijk_muse-faint_2020} for Eridanus II, in \cite{zoutendijk_muse-faint_2021} and Vaz et al. in prep for Grus 1, in \cite{julio_muse-faint_2023} for Antlia B, and in \cite{vaz_muse-faint_2023} for Leo T.  For these dwarfs, we have from 14 stars (for Grus 1) to 127 stars (for Antlia B).

As with the classical dwarfs, our faint dwarfs also have tidal radii far larger than their stellar light profiles (\citealt{Pace22} and see Sect.~\ref{sec:efe}).

\begin{table*}[ht]
\begin{center}
\caption{Parameters adopted from the literature.}

\begin{tabular}{lcccccccr}
\hline
         Galaxy  & $D$ & $r_{1/2}$ & $r_t$ & $M_V$ & $M_\star$ &$M_\mathrm{gas}$ & $N_\star$ &References \\
  & (kpc) & (pc) & (pc) & (mag) & ($10^6$M$_\odot$) & ($10^6$M$_\odot$) &
\\ \hline
Carina        & $106\pm6$   & $250\pm39$ & 2387 & $-9.1$  & 0.38 & - & 767 & 1, 2 \\
Fornax        &  $138\pm8$  & $710\pm77$ & 6083 & $-13.4$ & 43   & - & 2573 & 3, 2 \\
Draco         &  $76\pm6$   & $221\pm19$ & 2172 & $-8.8$  & 0.29 & - & 504 & 4, 5 \\
Leo I         &  $254\pm15$ & $251\pm27$ & 5299 & $-12.0$ & 5.5  & - & 328 & 6, 2\\
Leo II        &  $233\pm14$ & $176\pm42$ & 3875 & $-9.8$  & 0.74 & - & 186 & 7, 2 \\
Sculptor      &  $86\pm6$   & $283\pm45$ & 2502 & $-11.1$ & 2.3  & - & 1351 & 8, 2 \\
Sextans       &  $86\pm4$   & $695\pm44$ & 2930 & $-9.3$  & 0.44 & - &  417 & 9, 2 \\
Ursa Minor    &  $76\pm3$   & $181\pm27$ & 2279 & $-8.8$  & 0.29 & - & 430 & 10, 2 \\
%& & & & & & \\ 
\hline 
Antlia B      & $1350\pm60$ & $273\pm29$ & 7777$^*$ & $-9.7$ & $0.76$ & 0.28 & 127 & 11, 12, 13\\
Eridanus II   & $366\pm17$  & $277\pm14$ & 5154  & $-7.1$ & 0.09 & - & 92  & 14, 15 \\
Grus 1        & 125         & $151^{+21}_{-31}$  & 1076 & $-4.1$ & 0.01 & - & 14  & 16, 17, 13 \\
%Hydra II      & 151         & $68\pm11$          & $-4.8$ & 0.02 & - & 15  & 18, 13 \\
Leo T         & $409\pm28$ & $115\pm17$          & 5639 & $-7.6$ & $0.14$ & 0.41 & 75  & 18, 19, 13 \\
\hline
\label{tab:properties}
\end{tabular}
\end{center}

\small {Notes. From left to right, the columns show: name of the galaxy, galactocentric distance ($D_\odot$); half-light radius ($r_{1/2}$); tidal radius ($r_t$) estimated with the dynamical masses from \cite{pace_local_2025}; absolute $V$-band magnitude ($M_V$); stellar mass ($M_\star$); gas mass ($M_\mathrm{gas}$); and number of kinematic member stars ($N_\star$). The units for each column are marked in the second row.} \par
\vspace{2mm}
\small $^*$Obtained using the distance from Ant B to its host $D=72$ kpc \citep{sand2015} and $M_{\mathrm{NGC 3109}} = 1.1\times10^{10}$M$_\odot$ \citep{huchtmeier_distribution_1973}. \par
\vspace{2mm}
\small {References. (1) \cite{pietrzynski_araucaria_2009}, (2) \cite{irwin_structural_1995}, (3) \cite{boer_star_2012}, (4) \cite{bonanos_rr_2004}, (5) \cite{martin_comprehensive_2008}, (6) \cite{bellazzini_distance_2004}, (7) \cite{bellazzini_red_2005}, (8) \cite{pietrzynski_araucaria_2008}, (9) \cite{lee_star_2009}, (10) \cite{carrera_star_2002}, (11) \cite{sand2015}, (12) \cite{hargis2020}, (13) \cite{zoutendijk_muse-faint_2021b}}, (14) \cite{crnojevic_deep_2016}, (15) \cite{zoutendijk_muse-faint_2021}, (16) \cite{koposov_beasts_2015}, (17) \cite{chiti_magellanimacs_2022}, (18) \cite{clementini_variability_2012}, (19) \cite{wolf_accurate_2010}.  \par 

\end{table*}

\subsection{EDGE simulation data}\label{sec:simulations}
The ‘Engineering Dwarfs at Galaxy Formation's Edge’ (EDGE) project is described in detail in \cite{agertz_edge_2020}. We briefly summarise the key features of the simulations here for the reader’s convenience. The simulations we use were first published in \citet{orkney2021} and \citet{gray_edge_2025} and are summarised in Table \ref{tab:sims}.

\begin{table*}[ht]
\begin{center}
\caption{Properties of the simulated EDGE dwarf galaxies at redshift $z=0$ used in this paper.} 

\begin{tabular}{lcccccccc}
\hline
Name & Resolution & $M_{200}$ & $r_{200}$ & $M_{\star}$ & $M_{V}$ & $r_{1/2}$ & [Fe/H] \\
 & (m$_{\rm{DM}}$,m$_{\rm{gas}}$,m$_{\star}$)/M$_{\odot}$ & ($10^9$M$_{\odot}$) & (kpc) & ($10^6$M$_{\odot}$) & (mag) & (pc) & (dex) \\ 
\hline
Halo383&[945,161,300]        & 5.7  & 38.00 & 4.4   & $-10.88$      & 331.80 & $-2.00$\\
{Halo600}&[117,18,300]       & 2.65 & 31.17 & 0.984 & $-9.19$& 109.65 & $-2.48$\\
{Halo605}&[117,18,300]        & 3.20 & 31.08 & 1.930 & $-9.84$& 101.83 & $-1.96$\\
{Halo624}&[117,18,300]        & 3.23 & 29.18 & 1.080 & $-9.44$& 107.04 & $-2.12$\\
Halo1445&[117,18,300]        & 1.32 & 23.10 & 0.135 & $-6.93$& 100.79 & $-2.49$\\
Halo1459&[117,18,300]        & 1.43 & 23.75 & 0.377 & $-8.03$& 98.80 & $-2.02$\\
\hline
\label{tab:sims}
\end{tabular}
\end{center}

\small {Notes. From left to right, the columns show: the simulation label; mass resolution [dark matter, gas, stars]; halo Virial mass ($M_{200}$); halo Virial radius ($r_{200}$); stellar mass ($M_\star$); absolute V-band magnitude ($M_V$); projected half-light radius ($r_{1/2}$); and iron abundance ([Fe/H]). The units for each column are marked in the second row. All properties are reported for the higher resolution version of these simulations, except for Halo 383. To study numerical convergence, we also compare our results with lower resolution simulations that have a mass resolution of $[945,161,300]$\,M$_\odot$. All simulations were first presented in \citet{orkney2021, orkney_edge_2023}, and \citet{gray_edge_2025} for Halo 383.} \par

\end{table*}

EDGE is a suite of high-resolution cosmological zoom simulations of isolated dwarf galaxies in a $\Lambda$CDM cosmology ($\Omega_{m}=0.309$, $\Omega_{\Lambda}=0.691$, $\Omega_{\rm{b}}=0.045$, and $\rm{H}_{0}=67.77 \rm{km}\,\rm{s}^{-1}\rm{Mpc}^{-1}$; \citealt{Planck20}). It uses the adaptive mesh refinement hydrodynamics code, \texttt{RAMSES} \citep{Teyssier02}, with a maximum spatial resolution of 3\,pc and a mass resolution of
$\rm{M}_{\rm{gas}}$=18\,$\rm{M}_{\odot}$, $\rm{M}_{\rm{\star}}$=300\,$\rm{M}_{\odot}$ and $\rm{M}_{\rm{DM}}$=117\,$\rm{M}_{\odot}$ for the higher resolution simulations we present here \citep{orkney2021} (We compare these
also with lower resolution simulations that have $\rm{M}_{\rm{gas}}$=161\,$\rm{M}_{\odot}$, $\rm{M}_{\rm{\star}}$=300\,$\rm{M}_{\odot}$ and $\rm{M}_{\rm{DM}}$=945\,$\rm{M}_{\odot}$, to test numerical convergence; \citealt{Rey19, rey2020edge}.). For Halo 383, there is no higher-resolution version available, so we use only the lower resolution \citep{gray_edge_2025}. The galaxy formation model includes subgrid prescriptions for star formation, stellar feedback, cosmic reionisation and gas cooling. Star formation occurs when the gas temperature falls below $T_{\rm SF} = 100$\,K and the gas density rises above $\rho_{\rm SF} = 300$\,atoms/cc. Stars are then formed stochastically at a rate $\dot{\rho_{\star}} = \varepsilon_{\rm{ff}} \frac{\rho_{\rm{g}}}{t_{\rm{ff}}}$, where $\varepsilon_{\rm{ff}} = 0.1$ is the star formation efficiency and $t_{\rm{ff}}$ is the local gas free-fall time. Each star particle has a mass $\sim300$\,M${_{\odot}}$ and represents a single stellar population that has a \citet{Kroupa01} initial stellar mass function. Energy from supernovae (Type Ia and Type II) is injected purely thermally, if the cooling radius is resolved (which is the case $>90\%$ of the time). If not, some additional momentum is also injected to compensate for gas over-cooling (see \citealt{agertz_edge_2020} for details). Stellar winds from massive stars ($>5$\,M$_{\odot}$) continuously inject energy, mass, and metals into the interstellar medium, as in \citet{Agertz13}. These EDGE simulations do not include radiative transfer (photo-ionisation or photo-heating) from young stars (such physics is included in EDGE2, without significant changes to stellar masses and sizes; \citealt{rey_edge_2025}). Finally, reionisation is modelled as a uniform, time-dependent, heat source \citep{haardt1995radiative,rey2020edge}.

%-------------------------------------------------------------------

\section{Methods}\label{sec:methods}

In this section, we describe our methodology. In Sect.~\ref{sec:rar}, we explain how we estimate the baryonic and total, `observed', acceleration for our sample of galaxies. In Sect.~\ref{sec:gravsphere}, we describe the mass modelling tool, \textsc{GravSphere} (\citealt{read2017}, \citealt{draco}, \citealt{genina}, \citealt{read2019}, \citealt{collins2021}) that we use to determine the cumulative mass profiles of our sample of galaxies from their photometric and kinematic data.

\subsection{The radial relation acceleration}\label{sec:rar}

Throughout this paper, we assume that our sample of dwarf galaxies is in dynamical pseudo-equilibrium (meaning that while they may deviate from strict equilibrium due to tidal or internal processes, their overall kinematics are sufficiently stable for equilibrium methods to apply), which is a good assumption given that they all orbit in a weak tidal field\footnote{The Small Magellanic Cloud is undergoing heavy tidal disruption \citep{deleo20}, much more extreme than any of the galaxies we study here. Yet, it can still be successfully mass modelled \citep{deleo23}.} (\citealt{Pace22} and see Sect.~\ref{sec:efe}). Furthermore, we assume spherical symmetry. This is not likely to be valid for our sample of galaxies \citep[e.g.][]{Orkney23,Goater24}. However, bias due to triaxiality, for the number of kinematic tracers we use in this paper, is expected to be small. \cite{read2017} apply \textsc{GravSphere} to a triaxial mock with axis ratios 1:0.8:0.6, consistent with expectations in the Standard Cosmology (e.g. \citealt{Orkney23}). They show that the assumption of spherical symmetry in \textsc{GravSphere} leads to bias when applied to triaxial mock data, but, for $\sim 1000$ tracer stars, this is typically smaller than their quoted 95\% confidence intervals. This conclusion is further supported by \cite{genina} and \cite{nguyen_trial_2025}. They apply \textsc{GravSphere} to mock satellites drawn from cosmological simulations \textsc{apostle} and \textsc{fire}, respectively. In both cases, \textsc{GravSphere} provides an excellent recovery of the truth within its quoted uncertainties. This demonstrates that our results will not be significantly biased by realistic departures from spherical symmetry or tidal forces from the Milky Way.
 
Assuming spherical symmetry, the observed acceleration then follows as:
\begin{equation}
    g_\mathrm{obs}(r) = -\nabla\Phi_{\mathrm{tot}}(r) = \frac{GM_\mathrm{tot}(<r)}{r^2},
\label{eq:gobs}
\end{equation}
where $G$ is the gravitational constant and $M_\mathrm{tot}(<r)$ is the total cumulative mass at radius $r$. The baryonic acceleration follows similarly from the baryonic mass:
\begin{equation}
    g_\mathrm{bar}(r) = \frac{GM_{\rm bar}(<r)}{r^2}.
\label{eq:gbar}
\end{equation}
where $M_{\rm bar}(<r) = M_\star(<r) + M_{\rm gas}(<r)$ is the cumulative baryonic mass (i.e. the sum of the cumulative stellar and gas masses).

\citet{lelli_one_2017} argue that one cannot estimate the gravitational field as a function of radius for dSph galaxies because of a well-known degeneracy between the projected stellar velocity dispersion, $\sigma_{\rm LOS}(R)$, and the velocity dispersion anisotropy, $\beta(r)$ -- the `mass-anisotropy degeneracy'. However, we now have line-of-sight velocities for many individual stars in each dwarf we model in this paper, and we have a mass modelling method, \textsc{GravSphere}, that can break the mass-anisotropy degeneracy by using information about the shape of the velocity distribution function (\citealt{read2017}, \citealt{draco}, \citealt{genina}, \citealt{read2019}, \citealt{collins2021}). We describe how we use \textsc{GravSphere} to estimate both $M_\mathrm{tot}(<r)$ and $M_\star(<r)$ at each radius, $r$, next.

\subsection{\textsc{GravSphere}} \label{sec:gravsphere}
To measure the mass profiles of our sample, we used the updated version of \textsc{GravSphere} Jeans modelling code\footnote{The newest version of this code is available for download at \url{https://github.com/justinread/gravsphere} \citep{collins2021}.}. A detailed explanation of the code can be found in \citet{read2017}, \citet{draco}, \citet{genina} and \citet{collins2021}. We briefly summarise this here for the reader's convenience. \par 
\textsc{GravSphere} solves the Jeans equation \citep{jeans} for our member stars while assuming that the stellar system is spherical, non-rotating, and in a steady state, given by:
\begin{equation}
\frac{1}{\nu_\star}\pdv{}{r} \left(\nu_\star\sigma_r^2\right)+\frac{2\beta(r)\sigma_r^2}{r} = -\frac{GM(<r)}{r^2},
\label{eq:jeans}
\end{equation}
where the velocity anisotropy profile $\beta(r)$ is defined as:
\begin{equation}
    \beta = 1 - \frac{\sigma_t^2}{\sigma_r^2},
\end{equation}
where $\sigma_t$ and $\sigma_r$ are the tangential and radial velocity dispersions respectively, and $\sigma_r$ is given by \citep{van_der_marel_velocity_1994, mamon_dark_2005}:
\begin{equation}
\sigma_r^2 = \frac{1}{\nu_\star(r)g(r)}\int^{\infty}_{r}\frac{GM(\tilde{r})\nu_\star(\tilde{r})}{\tilde{r^2}}g(\tilde{r})\, \mathrm{d}\tilde{r},
\label{eq:sigmar2}
\end{equation}
with
\begin{equation}
g(r) = \exp\left(2\int\frac{\beta(r)}{r}\, \mathrm{d}r\right)
\label{eq:g}
\end{equation}
and $M(<r)$ the total cumulative mass as a function of the radius, $r$. The tracer number density $\nu_\star$ characterises the radial density profile of a population of massless tracers (in our case, stars moving in a galaxy) that move in the gravitational potential of its cumulative mass distribution $M(<r)$ modelled either\footnote{The estimate of the baryonic acceleration is highly sensitive to the accuracy of the fitted $\nu_\star(r)$, so we fit both profiles for each galaxy and adopt the one that best reproduces the observed distribution.  The impact of adopting these different profiles on the inferred RAR is illustrated in Appendix~\ref{app:tracerdensity} for the case of Grus I, the least well-sampled system, showing that our results are robust against the choice of tracer density profile.} with a sum of $N_P$ Plummer spheres \citep{plummer}:
\begin{equation}
\nu_\star(r) = \sum_{j}^{N_P} \frac{3M_j}{4\pi a^3_j}\left(1+\frac{r^2}{a_j^2}\right)^{-5/2},
\label{eq:tracerdensity_plummer}
\end{equation}
where $M_j$ and $a_j$ are the mass and scale length of each individual component; or with the generic $\alpha\beta\gamma$ profile initially proposed by \cite{zhao_analytical_1996} and parametrized by \cite{banares-hernandez_new_2025} given by a double power-law model:
\begin{equation}
\nu_\star(r) = \frac{\rho_c}{(r/r_c)^\gamma(1+(r/r_c)^\alpha)^{(\gamma-\beta)/\alpha}}
\label{eq:tracerdensity_abg},
\end{equation}
where three exponent variables $\alpha, \beta, \gamma$ and the scale radius $r_c$ and density $\rho_c$ were introduced.

This allows us to recover the density profile $\rho(r)$ and the velocity anisotropy profile $\beta(r)$ of the studied stellar systems \citep{read2017}, modelled as:
\begin{equation}
    \beta(r) = \beta_0+(\beta_\infty-\beta_0)\frac{1}{1+(r_0/r)^\nu},
\label{eq:beta}
\end{equation}
where $\beta_0$ and $\beta_\infty$ are the values of the anisotropy profile at $r=0$ and $r=\infty$, $r_0$ is a transition radius and $\nu$ is the steepness of the transition. To avoid infinities, \textsc{GravSphere} uses a symmetrised version of $\beta(r)$
\begin{equation}
\Tilde{\beta}(r) = \frac{\beta}{2-\beta},
  \label{eq:betasim}  
\end{equation}
where $\Tilde{\beta} = -1$ and $\Tilde{\beta} = 1$ correspond to a fully tangential and a fully radial distribution, respectively. \par 
\textsc{GravSphere} harnesses higher order moments of the velocity distribution via the fourth order Virial Shape Parameters (VSPs) (\citealt{vsps}, \citealt{richardson_dark_2014}, \citealt{read2017}) given by:
\begin{equation}
    v_\mathrm{s1} = \frac{2}{5} \int_{\infty}^{0} GM(5-2\beta)\nu_\star\sigma_r^2r \, \mathrm{d}r \\ 
    = \int_{\infty}^{0} \Sigma \langle v^4_\mathrm{los}\rangle  R\, \mathrm{d}R
\label{vsp1}
\end{equation}
and:
\begin{equation}
 v_\mathrm{s2} = \frac{4}{35} \int_{\infty}^{0} GM(7-6\beta)\nu_\star\sigma_r^2r^3 \, \mathrm{d}r \\ 
    = \int_{\infty}^{0} \Sigma \langle v^4_\mathrm{los}\rangle R^3\, \mathrm{d}R.
\label{vsp2}
\end{equation}
to break the mass-anisotropy degeneracy. \par
The cumulative mass distribution $M(<r)$ is modelled with a cNFWt profile \citep{draco}. This is based on the Navarro-Frenk White profile (NFW; \citealt{nfw}) fitted to pure dark matter simulations:
\begin{equation}
\rho_{\text{NFW}}(r) = \frac{\rho_0}{(r/r_s)(1+r/r_s)^2},
\label{eq:nfw}
\end{equation}
To account for the possibility of cored profiles, as is observed in many dwarf galaxies (see e.g. \citealt{mcgaugh_2001_rotation}, \citealt{marchesini_2002_soft}), and the possibility of mass loss beyond the tidal radius for satellite dwarfs \citep[e.g.][]{read2006}, the cNFWt modifies the NFW profile at both small and large radii. The density profile of this model is given by
\begin{equation}
\rho_\mathrm{cNFWt}(r) =
\begin{cases}
    f^n \rho_{\mathrm{NFW}}+\frac{nf^n(1-f^2)}{4\pi r^2r_c}M_\mathrm{NFW}, r < r_t \\
    f^n \rho_{\mathrm{NFW}}+\frac{nf^n(1-f^2)}{4\pi r^2r_c}M_\mathrm{NFW}\left(\frac{r}{r_t}\right)^{-\delta}, r > r_t
\end{cases}
\label{eq:corenfwt-profile}
\end{equation}
where $M_\mathrm{NFW}$ is the mass profile of the NFW model and the function $f^n$ determines how cuspy the halo is below a core-size parameter, $r_c$:
\begin{equation}
    f^n = \left[\tanh\left(\frac{r}{r_c}\right)\right]^n.
\label{eq:fn}
\end{equation}
The cumulative mass of the cNFWt model is then given by:
\begin{equation}
\begin{split}
M_\mathrm{cNFWt}(<r) =
\begin{cases}
    M_\mathrm{NFW}(<r)f^n, r < r_t \\
    M_\mathrm{NFW}(r_t)f^n + \\ 4\pi\rho_\mathrm{cNFWt}(r_t)\frac{r_t^3}{3-\delta} \left[\left(\frac{r}{r_t}\right)^{3-\delta}-1\right], r > r_t
\end{cases}
\label{eq:mass-profile}
\end{split}
\end{equation}
where $r_t$ and $\delta$ model the effect of tidal stripping beyond $r_t$, with $r_t$ being the tidal stripping radius and $\delta$ the density slope beyond $r_t$.\par 
 
To assign the data to bins of the projected radius, $R$, we use an algorithm called \textsc{Binulator} \citep{collins2021}. After running the \textsc{Binulator} routine, we run \textsc{GravSphere} to determine the cumulative mass profile. \textsc{GravSphere} solves the Jeans equation for the projected velocity dispersion. It also fits the two VSPs, as determined from the stellar kinematic data by the \textsc{Binulator}. \par
\textsc{GravSphere} uses the ensemble sampler \textsc{Emcee} \citep{emcee} to fit the model to the data. Each Markov chain (walker) communicates with the other walkers at each step. We used the default number of walkers $N_\mathrm{walkers} = 250$ and steps $N_\mathrm{steps} = 50000$, since these values were sufficient to achieve convergence. The first half of the steps generated are always discarded as a conservative burn-in criterion.

\subsection{Priors}\label{sub:priors}
We kept the \textsc{GravSphere} priors to their default values, which we summarise here for convenience. We used priors on the symmetrised velocity anisotropy of $ -1 < \Tilde{\beta}_0 < 1$, $-1 < \Tilde{\beta}_\infty < 1$, $-1 < \log_{10}(r_0/\mathrm{kpc}) < 0$, and $1 < \eta < 3$. We used a flat prior on the baryonic mass of $0.75 M_\star < M_{\mathrm{bar}} < 1.25 M_\star$, where $M_\star$ corresponds to the measured stellar mass described in Table~\ref{tab:properties}. In cases where gas is also present, we added it to the stellar mass and used the total value as the prior for the baryonic mass. \par
For the cNFWt model, we fit the following free parameters: the halo mass $M_{200}$ and concentration parameter before infall $c_{200}$; the dark matter core-size parameter $r_c$; the tidal stripping radius $r_t$ and the logarithmic density slope beyond $r_t$, $\delta$. We assume flat priors of $10^{7.5}$ M$_\odot < M_{200} < 10^{11.5}$M$_\odot$ and $1 < c_{200} < 50$ (see \citealt{julio_muse-faint_2023} for a discussion of these choices), $-2 < \log_{10}(r_c/$kpc$) < 0.5$; $0.3 < \log_{10}(r_t/R_{1/2}) < 1$; and $3.5 <\delta <5$ on these parameters. Our results are not sensitive to these choices (refer to \citealt{zoutendijk_muse-faint_2021b, julio_muse-faint_2023} for a comprehensive discussion on the robustness of the results obtained using \textsc{GravSphere} in MUSE-Faint data).

Finally, it is important to note that the Jeans equations, which \textsc{GravSphere} solves, assume only that weak-field gravity is described by a scalar field, and that the gravitational force follows from the gradients of this scalar field. The Poisson equation does not actually enter when deriving the observed accelerations. It only enters if we choose to interpret the observed acceleration as owing to some enclosed dark matter mass. This is an important point because it means that we can directly compare our observed accelerations, derived using \textsc{GravSphere}, to alternative gravity models like MOND.

\subsection{Recovering the RAR from mock data}\label{sec:recovery}

\begin{figure*}[ht]
 \begin{minipage}[c]{0.75\textwidth}
    \includegraphics[width=\textwidth]{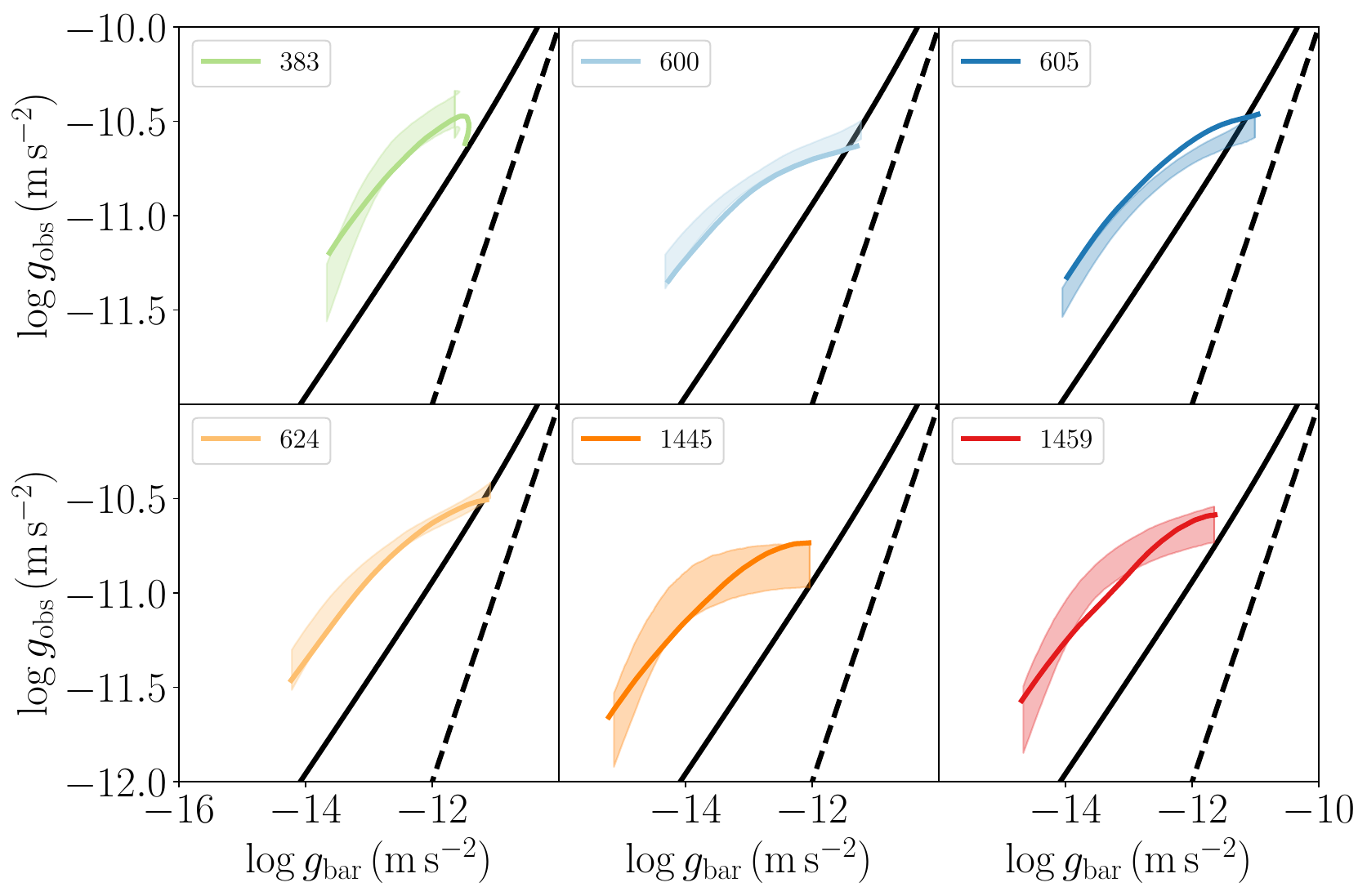}
  \end{minipage}\hfill
  \begin{minipage}[c]{0.23\textwidth}
\caption{Testing our recovery of the RAR by applying \textsc{GravSphere} to mock data drawn from a simulated EDGE dwarfs. For this test, we use all of the simulated star particles to fit the photometric light profile and a random 1000 stars to measure the line-of-sight velocity dispersion profile (similar to our more poorly sampled observed dwarfs; see Table \ref{tab:properties}). The shaded band marks the 68\% confidence intervals. The true answer (as determined directly from the simulations, using equations \ref{eq:gobs} and \ref{eq:gbar}, assuming spherical symmetry) is marked by the solid lines. The black solid line marks the RAR derived by \cite{mcgaugh_radial_2016} for SPARC local spiral galaxies. The blacked dashed line marks $g_\mathrm{obs} = g_\mathrm{bar}$.}
    \label{fig:RARrecovery}
  \end{minipage}
\end{figure*}

In this section, we test how well we can recover the RAR by applying \textsc{GravSphere} to mock data taken from the simulated EDGE dwarfs. For this test, we use all of the simulated star particles to fit the photometric light profile and a random 1000 stars to measure the line-of-sight velocity dispersion profile up to 1\,kpc (broadly representative of our observed dwarfs; see Table~\ref{tab:properties}). In addition, to reproduce the observational uncertainties in our mock data, we impose velocity errors of 2 km\,s$^{-1}$ on the individual stars. These data are passed to the \textsc{Binulator} and then to \textsc{GravSphere}, as described in Sect.~\ref{sec:gravsphere}. The results are shown in Figure \ref{fig:RARrecovery}. Based on an empirical rotation estimator applied to all galaxies, we find that the observed dwarfs are consistent with being non-rotating, while the EDGE dwarfs display significant rotation. To account for this behaviour in \textsc{GravSphere}, we introduce a new rotation parameter described in Appendix~\ref{app:rotation}. Notice that the true acceleration, directly determined from the simulations, matches within the $68\%$ confidence bands determined by \textsc{GravSphere}. 

\section{Results}\label{sec:results}
By using \textsc{GravSphere}, described in Sect.~\ref{sec:gravsphere}, and the equations described in Sect.~\ref{sec:rar}, in this section, we estimate the observed and baryonic accelerations for both our observed (Sect.~\ref{sec:obsdata}) and simulated (Sect.~\ref{sec:simulations}) data. In Sect.~\ref{sec:rar-observations}, we present the resulting RAR for our observations. In Sect.~\ref{sec:rar-simulations}, we present the same for the simulated EDGE dwarf galaxies.

\subsection{RAR of the observations}\label{sec:rar-observations}

\begin{figure*}[ht]
\begin{minipage}[c]{0.75\textwidth}
\includegraphics[width=\textwidth]{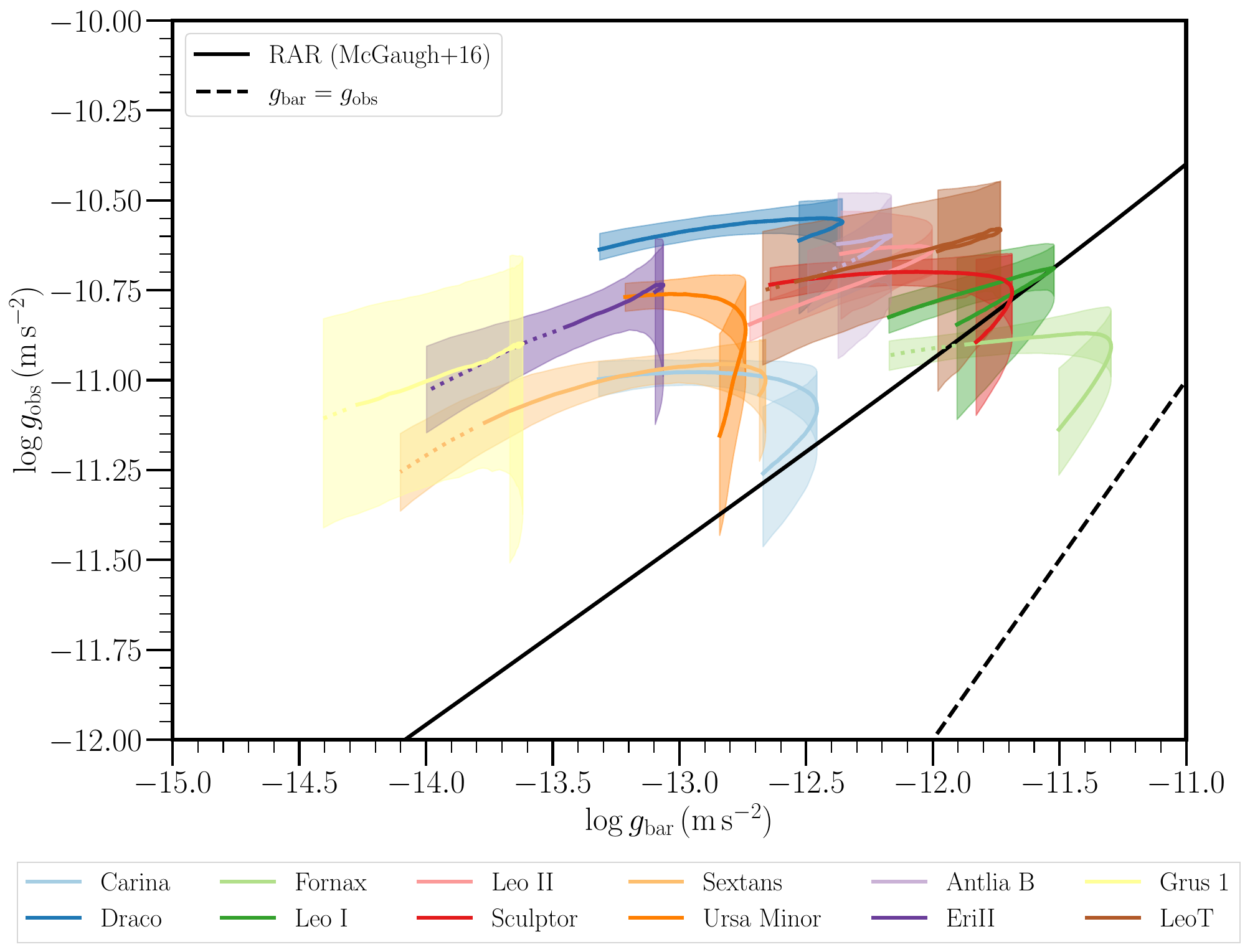}
  \end{minipage}\hfill
  \begin{minipage}[c]{0.23\textwidth}
\caption{The radial acceleration relation (RAR) for our sample of dwarf galaxies. Each colour corresponds to a different galaxy, where the shaded area corresponds to the $68\%$ confidence interval and the solid line marks the median value. The dotted lines correspond to extrapolations beyond the kinematic data range. The black solid line marks the RAR derived by \cite{mcgaugh_radial_2016} for SPARC local spiral galaxies. The blacked dashed line marks $g_\mathrm{obs} = g_\mathrm{bar}$.}
    \label{fig:rar-main}
      \end{minipage}
\end{figure*}

Figure~\ref{fig:rar-main} shows the RAR we obtain for our sample of 12 dwarf galaxies. We plot the $g_\mathrm{obs}-g_\mathrm{bar}$ relation as a function of radius within each galaxy, where each colour corresponds to a different galaxy, the shaded bands mark the $68\%$ confidence intervals, the lines show the median values, and the dotted lines show extrapolations outside of the data range to larger radii. We compare this with the RAR derived by \cite{mcgaugh_radial_2016} (solid black line). Notice that, with the exception of Fornax (light green), all of the dwarf galaxies lie systematically above the low mass extrapolation of the RAR from \cite{mcgaugh_radial_2016}.  This deviation becomes stronger for the galaxies with lower baryonic masses (i.e. lower $g_{\rm bar}$). Fornax stands out as an outlier in our sample, lying below the classical RAR, in contrast to the other dwarfs. Its substantially higher stellar mass and extended star formation history  (e.g. \citealt{boer_star_2012}) distinguish it from both the other observed dwarfs and the EDGE galaxies. These properties make Fornax particularly susceptible to repeated gas cooling and stellar feedback, which is expected to produce a dark matter core (e.g. \citealt{read_mass_2005,pontzen_how_2012,di_cintio_dependence_2014}). Indeed, there is already dynamical evidence for such a core reported in the literature (e.g. \citealt{goerdt_does_2006,cole_mass_2012,pascale_action-based_2018,read_dark_2019}, but see \citealt{genina22}). Cored dwarfs occupy a distinct locus in the RAR compared to those with a central cusp, behaving more similarly to MOND predictions \citep{eriksen_cusp-core-like_2021}. If Fornax is uniquely cored amongst our sample, then this provides a natural explanation for its more MOND-like locus in the RAR. However, more detailed work, left for future studies, will be needed to fully understand this difference.

Furthermore, notice that, in some galaxies, the radially resolved data exhibit upward or downward `hooks' in the RAR plane. These features are similar to those described in \citet{li_effect_2022} and \citet{Mercardo24}: low-mass galaxies with cuspy dark matter halo profiles should have upward-bending `hook' features that deviate from the observed RAR; while, conversely, downward hooks arise due to the non-monotonic behaviour of the total acceleration profile in the presence of a feedback-induced core, leading to multiple observed accelerations $g_\mathrm{obs}$ for a given baryonic acceleration $g_\mathrm{bar}$. Given the uncertainties, several of our systems allow for both scenarios, or even for a more monotonic relation. However, for most of them, there is no one-to-one map from $g_\mathrm{bar}$ to $g_\mathrm{obs}$, neither within each dwarf, nor in-between them as a population. 

\subsection{RAR of the EDGE simulations}\label{sec:rar-simulations}

\begin{figure*}[ht]
  \begin{minipage}[c]{0.75\textwidth}
    \includegraphics[width=\textwidth]{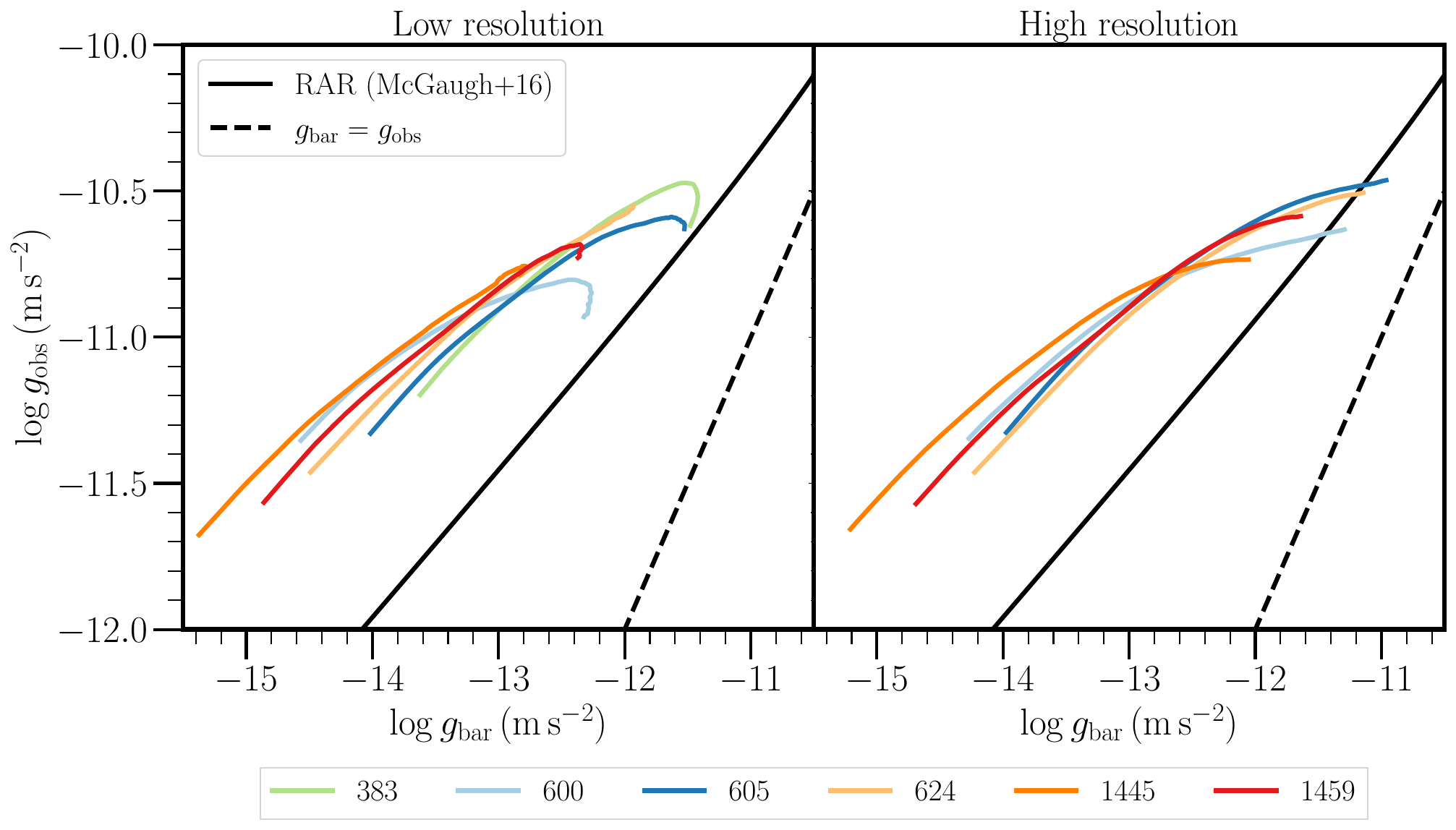}
  \end{minipage}\hfill
  \begin{minipage}[c]{0.23\textwidth}
        \caption{The radial acceleration relation for the EDGE simulated dwarf galaxies. The colours show each simulation as marked in the legend. The black solid line corresponds to the RAR derived by \cite{mcgaugh_radial_2016} for SPARC local spiral galaxies. The black dashed line marks $g_\mathrm{obs} = g_\mathrm{bar}$. Left: RAR of the lower resolution simulations. Right: RAR of the higher resolution simulations.}
    \label{fig:rar-simulations}
  \end{minipage}
\end{figure*}

Figure~\ref{fig:rar-simulations} shows the for the EDGE simulated galaxies. Notice that the low and high resolution results are in good agreement for all of the simulated dwarfs, though the higher resolution simulations (right panel) slightly deviate from the lower resolution ones (left panel) at low and high $g_\mathrm{bar}$\footnote{When extending the radial range of the high-resolution haloes down to $r=0.01$\,kpc, the hooks in the RAR become visible. Their apparent absence in our fiducial plots is thus not physical, but a consequence of the adopted radial cut at $r>0.1$\,kpc, (see Figure~\ref{fig:gm-1459} of Appendix~\ref{app:scatter}).}. Notice further that, like most observed dwarfs, the EDGE dwarfs lie systematically above the RAR from \cite{mcgaugh_radial_2016}. 

None of the EDGE dwarfs scatter below the \cite{mcgaugh_radial_2016} RAR line, and so none look similar to the data for Fornax. However, notice that the EDGE dwarfs are $\sim 10$ times less massive than Fornax (see Figure~\ref{fig:rh-stellarmass}) and so this does not (at least not yet) constitute a mismatch between theory and data. 

\begin{figure*}[ht]
  \begin{minipage}[c]{0.75\textwidth}
    \includegraphics[width=\textwidth]{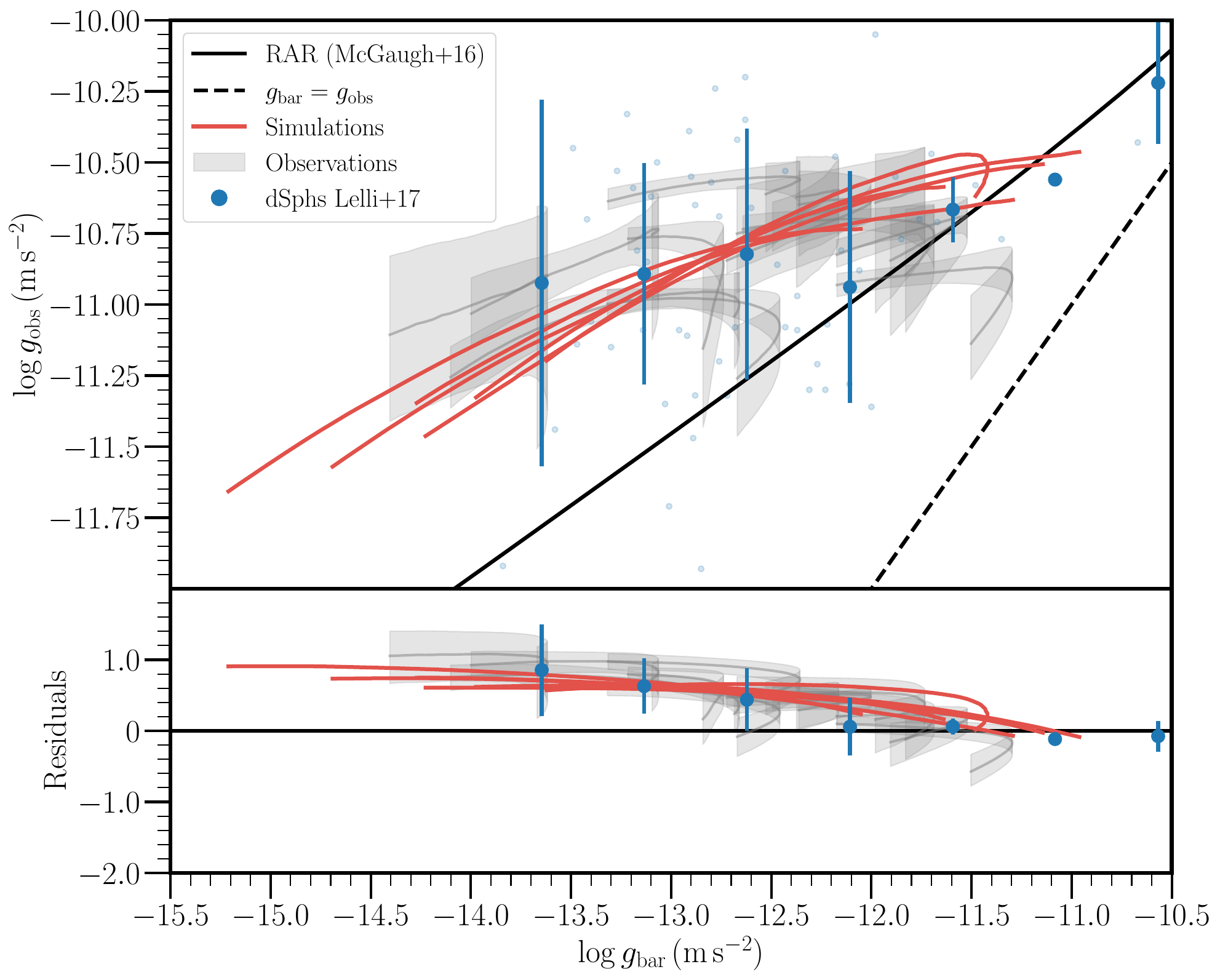}
  \end{minipage}\hfill
  \begin{minipage}[c]{0.23\textwidth}
    \caption{Quantitative comparison of the EDGE simulated RAR with our observed dwarf galaxy sample. The black solid line corresponds to the RAR derived by \cite{mcgaugh_radial_2016} for SPARC local spiral galaxies. The median and 68\% scatter in the observed $g_\mathrm{obs}-g_\mathrm{bar}$ relations are marked by the grey solid lines and shaded regions, respectively. The $g_\mathrm{obs}-g_\mathrm{bar}$ relations estimated directly from the simulations are represented by solid red lines. The blacked dashed line marks $g_\mathrm{obs} = g_\mathrm{bar}$. 
    The small blue points mark the positions of the dSphs on this relation \cite{lelli_one_2017}, while the blue circles indicate the mean and $1\sigma$ scatter of the binned data. At the bottom, the residuals relative to the \cite{mcgaugh_radial_2016} relation (black solid line) are represented for the observed, the simulated, and the \cite{lelli_one_2017} data.}
    \label{fig:rar-comparison-simulations}
        \end{minipage}
\end{figure*}

In Figure~\ref{fig:rar-comparison-simulations}, we quantitatively compare the EDGE simulated RAR with our observed dwarf galaxy sample. For this comparison, we plot just the higher resolution simulations (except for Halo 383, for which we have only the lower resolution version). The median and 68\% scatter in the observed $g_\mathrm{obs}-g_\mathrm{bar}$ relations are marked by the grey solid lines and shaded regions, respectively. The relations estimated directly from the simulations are represented by solid red lines. Notice that the agreement between the two is remarkably good, both in terms of the median relation and its scatter. Also shown, in blue, is the RAR for classical dSphs derived at their half-light radii from \citet{lelli_one_2017}. These measurements, although noisier and more sparsely sampled, are broadly consistent with both our results and the simulations, further reinforcing the conclusion that the standard $\Lambda$CDM cosmology reproduces the observed behaviour of galaxies in the RAR plane at the dwarf galaxy scale.

\section{Discussion} \label{sec:discussion}

\subsection{The origin of scatter in the RAR of low mass dwarf galaxies}\label{sec:scatter}

We may reasonably ask what the origin of the scatter in the EDGE simulated RAR is. As shown in \citet{Rey19} and \citet{Kim24}, the EDGE dwarfs show significant scatter in $M_*$ at a given $M_{200}$ ($\sim$0.5\,dex ($1\sigma$) at a halo mass of $\sim$10$^9$ M$_\odot$). This owes to reionisation combined with scatter in dwarf galaxy assembly histories. At fixed $M_{200}$, dwarfs that form earlier form more stars before reionisation quenching, leading to a higher $M_*$ today. As such, variations in the $M_*-M_{200}$ relation of low-mass dwarfs in $\Lambda$CDM naturally generate scatter in the $g_\mathrm{bar}-g_\mathrm{obs}$ plane. 

To illustrate this explicitly, we show in Figure~\ref{fig:gm-1459} of the Appendix~\ref{app:scatter} the `genetically modified' versions of Halo 1459, in which only reionisation and early assembly history were varied. These runs (MX02 and MX03) share the same halo mass but differ in their stellar mass, and exhibit $g_\mathrm{bar}-g_\mathrm{obs}$ tracks offset by an amount comparable to the scatter between distinct EDGE dwarfs. 

That said, the scatter in EDGE is smaller than in the observational data (see Figure~\ref{fig:rar-comparison-simulations}). Even after including observational uncertainties in the mock data (discussed in~\ref{sec:recovery}), the EDGE scatter does not reach the full amplitude seen in the Local Group dwarfs (see Figure~\ref{fig:rar-errors} in Appendix~\ref{app:scatter}). This does not necessarily indicate a problem for $\Lambda$CDM, since additional sources of scatter are expected in the observed sample. The Local Group dwarfs are satellites, whereas EDGE dwarfs are isolated galaxies; and quenching on infall \citep[e.g.][]{read_erkal19} and tidal effects \citep[e.g.][]{read2006} can both increase the scatter. Moreover, the observed dwarfs span a wider range of pre-infall halo masses than our EDGE comparison sample. This is particularly relevant for Fornax, which is among the most discrepant galaxies in the data and likely originates from a substantially more massive progenitor.

While the EDGE scatter alone falls short of matching the data, the combination of reionisation, assembly history, observational uncertainties, satellite-specific processes, and a broader halo mass range might explain the observed dispersion in the dwarf galaxy RAR within $\Lambda$CDM.

\subsection{Can we compare isolated EDGE dwarfs to observed satellite dwarfs?}

Our simulated EDGE dwarfs are {\it isolated}, whereas our sample of observed dwarfs all orbit in the tidal field of a larger host galaxy. As such, we may worry that the two populations may not be directly comparable. We have selected our observed dwarf galaxies such that they orbit in a weak tidal field today, with a tidal radius that lies far beyond their stellar light profile (\citealt{Pace22}; see Table~\ref{tab:properties} and Sect.~\ref{sec:efe}). As such, the tidal field of the host galaxy is not likely to significantly influence the stellar light profiles of our sample of dwarfs \citep{read2006,Goater24}. However, it is likely that some of our observed dwarfs -- particularly the more massive ones like Fornax -- had their star formation shut down by ram-pressure from their host galaxy's hot gaseous corona \citep[e.g.][]{Gatto13}. This will lower their total stellar mass as compared to a similar but isolated counterpart, lowering $g_\mathrm{bar}$ for a given $g_\mathrm{obs}$ and inducing more scatter in the RAR plane. As such, we can consider the scatter that we see in the RAR plane for our EDGE dwarfs as a {\it lower bound} on the scatter expected for satellite dwarfs in a $\Lambda$CDM cosmology.

\subsection{The external field effect}\label{sec:efe}

We have shown that our observed dwarf galaxies do not follow any simple RAR (see Figure~\ref{fig:rar-main}). However, as discussed in Sect.~\ref{introduction}, alternative weak-field gravity theories like MOND are typically non-linear, leading to an `external field effect'. Since our sample of dwarfs all orbit in the tidal field of a larger host galaxy, we may worry that the scatter in the RAR is actually expected in MOND -- a result of each dwarf experiencing a distinct external field effect (EFE).

To test this idea, we follow the approach in \cite{mcgaugh_andromeda_2013} to estimate the relative size of the internal, $a_\mathrm{in}$, and external, $a_\mathrm{ex}$, acceleration (the EFE) for each dwarf in our observational sample. To determine the internal field, that depends only on the baryonic properties of the dwarf, we follow \citet{mcgaugh_andromeda_2013}, who derive: 

\begin{equation}
a_\mathrm{in} \sim \frac{\left(\frac{4}{9} G M_{\rm bar} a_0\right)^{1/2}}{R_{1/2}}
\end{equation}
valid for MOND. We assume here that $a_0 = 1.2 \times 10^{-10}$m\,s$^{-2}$ \citep{mcgaugh_andromeda_2013}.

The EFE ($a_\mathrm{ex}$) depends on the total baryonic mass of the host galaxy and, similarly to \cite{pawlowski_new_2015}, we adopt the empirical Milky Way (MW) model of \cite{mcgaugh_milky_2008}, using $a_\mathrm{ex} = V^2_\mathrm{MW}/D$, where $V_\mathrm{MW}$ is the circular velocity of the MW and $D$ is the Galactocentric distance of the object (this is reported in Table~\ref{tab:properties}). Ant B is not a satellite galaxy of the MW, but is instead a satellite of NGC 3109. As such, for this specific case, we used a circular velocity of 52\,kms$^{-1}$ \citep{huchtmeier_distribution_1973} and a distance of $D = 72$ kpc \citep{sand2015}.

We report our values for $a_\mathrm{in}$ and $a_\mathrm{ex}$, derived as described above, in Table \ref{tab:efe}. A system is considered to be in the MOND regime if $a_\mathrm{ex} < a_\mathrm{in}$$\ll a_0$ and in the EFE regime if $a_\mathrm{ex} > a_\mathrm{in}$. As can be seen, five of our dwarfs are clearly in the MOND regime (Fornax, Leo I, Leo II, Ant B, Leo T), five are clearly in the EFE regime (Carina, Draco, Sextans, Ursa Minor, Grus I) and the remainder are at the boundary (Sculptor, Eri II). This means that we do indeed expect scatter in the RAR for these dwarf galaxies in MOND. However, galaxies in the EFE regime should scatter {\it below} the RAR, while the opposite is what we see. Furthermore, we see no systematic trend between galaxies in the MOND or EFE regimes. Draco and Carina are a particularly illustrative example, as was already pointed out by \citet{read_dark_2019}. Both galaxies are in the EFE regime on very similar orbits around the Milky Way. Both have very similar baryonic mass distribution and, therefore, similar $g_\mathrm{bar}$. Yet, as can be seen in Figure \ref{fig:rar-main}, they have very different $g_\mathrm{obs}$ at high statistical significance. This difference might be due to their distinct evolutionary paths: Draco is a quenched system with a predominantly old stellar population and very slow chemical enrichment, consistent with a system that ceased star formation early, likely around the epoch of reionisation \citep{cohen_chemical_2009, weisz_star_2014}. Its high dark matter content makes it strongly dark-matter-dominated at all radii \citep{read_dark_2019}. Carina, by contrast, experienced extended and bursty star formation episodes with multiple distinct episodes of enrichment \citep{de_boer_episodic_2014}, which likely drove repeated gas outflows and feedback events. Such processes can reduce the central dark matter density, producing a lower observed acceleration at fixed baryonic content \citep{read_dark_2019}. These contrasting star formation and feedback histories naturally produce the observed offset in their RARs within $\Lambda$CDM. However, this suggests that the scatter in the RAR observed in our sample of dwarf galaxies poses a challenge for a MOND-like explanation, since galaxies with the same baryonic mass distribution and external field should fall on the same RAR. It remains to be seen if other alternative gravity theories that show different non-linear weak-field behaviour could explain these data.

\begin{table}[ht]
\begin{center}
\caption{MOND internal and external accelerations of our observed sample of dwarf galaxies.}

\begin{tabular}{lcc}
\hline
         Galaxy  & $a_\mathrm{in}$ & $a_\mathrm{ex}$ \\
                 & \multicolumn{2}{c}{($10^{-12}$m\,s$^{-2}$)} \\ \hline
Carina        &  $6.72$   & $9.90$ \\
Fornax        &  $25.17$  & $7.61$ \\
Draco         &  $6.64$   & $13.81$ \\
Leo I         &  $25.47$  & $3.27$ \\
Leo II        &  $13.32$  & $3.56$ \\
Sculptor      &  $14.61$  & $12.21$ \\
Sextans       &  $2.60$   & $12.21$ \\
Ursa Minor    &  $8.11$   & $13.81$ \\
\hline
Antlia B      & $10.18^*$ & $1.22$ \\
Eridanus II   & $2.95$    & $2.27$ \\
Grus 1        & $1.81$    & $8.4$   \\
Leo T         & $17.58^*$ & $2.03$   \\
\hline
\label{tab:efe}
\end{tabular}
\end{center}
\small {$(^*$) The estimation of $a_\mathrm{in}$ requires the mass of the system, so for these galaxies we considered both the stellar and gas mass.} 
\end{table}

\subsection{Challenges to interpretation}\label{sec:caveats}
During our analysis with \textsc{Gravsphere}, we assume that the systems under study have spherical symmetry. As previously pointed out, this might not be valid for our sample. Even though it was shown that the expected departures from spherical symmetry bias \textsc{GravSphere} mass models by less than their quoted 95\% confidence intervals for stellar kinematic samples of up to $\sim 1000$ stars, for some systems -- namely for the MUSE-Faint dwarfs -- we have less than 100 stars. Furthermore, for these particular dwarfs, the data is mostly concentrated inside their half-light radius, so the baryonic distribution is extrapolated from their surface brightness profile in the outskirts. As a result of the undersampling, both in the number of stars and the radius range in which they are concentrated, the RAR of these galaxies should be interpreted cautiously. However, within their error bars, they mostly agree with the dSphs for which we have a significant amount of data across a large range of radii. Moreover, our overall results remain unchanged even when these systems are excluded from the analysis — their inclusion simply amplifies the observed trends.

As shown in Table~\ref{tab:properties}, Antlia B and Leo T have gas. We assume, for simplicity, that the gas content of Antlia B and Leo T follows the same mass distribution as the stars, and this assumption underlies the results presented in Figure~\ref{fig:rar-main}. We show in Appendix~\ref{app:gas}, in Figure~\ref{fig:gascomparsion}, how the RAR of these two galaxies change when we exclude their gas content. As expected, since we assume that the gas and stars follow the same mass distribution, omitting the gas leads to a uniform decrease in the baryonic acceleration at all radii, effectively shifting the RAR to the left. While this approach provides a useful first-order approximation, it is important to note that, in reality, the gas content of a galaxy can be significantly more extended than its stellar distribution. This would increase the baryonic content of the galaxy at large radii, resulting in increased baryonic and total accelerations in the outer regions. Nonetheless, this would not change our conclusions.

Another potential caveat is the impact of binary stars, which can artificially increase the observed stellar velocity dispersions, leading to overestimated dynamical masses and elevated $g_{\rm obs}$ at fixed $g_{\rm bar}$ (e.g.~\citealt{pianta_impact_2022,gration_stellar_2025}). In our analysis, we do not explicitly correct for binaries. However, the MUSE-Faint data were obtained over multiple epochs and combined into period-averaged velocities, which reduces the binary contamination (discussed for Leo T in \citealt{vaz_muse-faint_2025}; and for the remaining MUSE-Faint UFDs in Vaz et al., in prep.). For the classical dSphs, binaries have only a minor impact on the velocity dispersion at the current levels of precision (e.g.~\citealt{mcconnachie_revisiting_2010,wang_unraveling_2023}). In these cases, we rely on the published membership catalogues, which in many instances make use of multi-epoch data to identify and remove likely binaries (e.g.~\citealt{walker_internal_2006})
Furthermore, for the more massive dwarfs, the higher intrinsic dispersions further suppress the contribution from binaries. We therefore expect that while binaries may introduce some additional scatter, it is unlikely that their contamination alters our main results.

Finally, tidal interactions may have influenced the past evolution of some systems. Yet, since all of our galaxies orbit in relatively weak present-day tidal fields, we do not expect tides to significantly affect our main conclusions. Moreover, our estimates of the tidal radius are based on the dynamical masses from \cite{pace_local_2025}, which were derived within the 3D half-light radius using the dynamical mass estimator in \cite{wolf_accurate_2010}. This likely underestimates the true masses of some dwarfs, and therefore also underestimates their tidal radii, making them appear more susceptible to tides than they actually are. In that sense, our assumption is conservative. Given that even heavily disrupted systems like the Small Magellanic Cloud can still be reliably mass modelled \citep{deleo20, deleo23}, we expect our main results to remain robust despite the possible influence of tides.

\section{Conclusions}\label{sec:conclusions}
We used literature line-of-sight velocities for the classical dwarf spheroidals of the Milky Way, combined with MUSE-Faint velocities of ultra-faint dwarfs, to measure the radially resolved radial acceleration relation of galaxies down to baryonic accelerations $\log g_\mathrm{bar} \sim -14$m\,s$^{-2}$, for the first time. 

We estimated the baryonic and observed accelerations, $g_\mathrm{bar}$ and $g_\mathrm{obs}$, using the \textsc{Gravsphere} Jeans mass modelling code, assuming that our galaxies are spherical, non-rotating, and in steady state. We tested these assumptions by applying \textsc{GravSphere} to mock data for simulated dwarf galaxies taken from the EDGE project, showing that we can successfully recover the RAR of these mock dwarfs within our quoted uncertainties.

We then applied our method to real data for 12 nearby dwarfs. Our dynamical analysis reinforced previous results that in the lowest mass galaxies, the $g_\mathrm{obs}-g_\mathrm{bar}$ relation does not follow the RAR calibrated on higher-mass galaxies \citep{mcgaugh_radial_2016}, but is systematically higher. Furthermore, we found significant scatter from galaxy to galaxy, while, when considering distinct radii within individual galaxies, we found that some galaxies can have two different values of $g_\mathrm{obs}$ at the same $g_\mathrm{bar}$.

We considered whether these results could be explained by the `external field effect' (EFE) that can occur in non-linear weak-field gravity theories like MOND. However, we showed that the EFE has the wrong sign, causing galaxies to scatter {\it below} the RAR, opposite to what is observed. Furthermore, we found no correlation between galaxies in the EFE regime as compared to galaxies outside. And, we showed that there exist galaxy `twins' like Carina and Draco that have similar EFE, similar orbits around the Milky Way, and similar baryonic mass distribution ($g_\mathrm{bar}$), yet statistically significantly distinct $g_\mathrm{obs}$.

Our results demonstrate that the RAR for low mass dwarf galaxies is not universal, and that the baryonic acceleration does not contain enough information, on its own, to derive the observed acceleration for these objects.

Finally, we compared our results with the RAR predicted in high resolution simulations of isolated dwarf galaxies in a $\Lambda$CDM cosmology, taken from the EDGE project. We showed that the higher and lower resolution versions of these simulations agreed well with one another, demonstrating good numerical convergence. We then showed that the simulated dwarfs behave similarly to our observed sample in lying systematically above the low-mass extrapolation of the RAR from \citet{mcgaugh_radial_2016}. However, the scatter in EDGE, which is caused by a combination of reionisation and the scatter in dark matter halo assembly histories \citep{Kim24}, is smaller than that seen in the Local Group dwarfs. While part of this difference can be attributed to additional dispersion introduced by \textsc{GravSphere} and observational uncertainties, further contributions are expected from the fact that the Local Group dwarfs are satellites rather than isolated dwarfs, and from their broader range of pre-infall halo masses. This makes the low-mass end of the RAR a powerful probe of both cosmology and the complex evolutionary paths of dwarf galaxies.

\begin{acknowledgement}
MPJ and MSP acknowledge funding via a Leibniz-Junior Research Group (project number J94/2020). OA acknowledges support from the Knut and Alice Wallenberg Foundation, the Swedish Research Council (grant 2019-04659), the Swedish National Space Agency (SNSA Dnr 2023-00164), and the LMK Foundation. Based on observations made with ESO Telescopes at the La Silla Paranal Observatory under programme IDs 0100.D-0807, 0101.D-0300, 0102.D-0372 and 0103.D-0705. This research has made use of Astropy \citep{astropy}, corner.py \citep{corner}, matplotlib \citep{matplotlib}, NASA’s Astrophysics Data System Bibliographic Services, NumPy \citep{numpy}, SciPy \citep{scipy}. Based on observations made with the NASA/ESA Hubble Space Telescope, and obtained from the Hubble Legacy Archive, which is a collaboration between the Space Telescope Science Institute (STScI/NASA), the Space Telescope European Coordinating Facility (ST-ECF/ESA) and the Canadian Astronomy Data Centre (CADC/NRC/CSA). JIR would like to acknowledge support from STFC grants ST/Y002865/1 and ST/Y002857/1.
\end{acknowledgement}

\bibliographystyle{aa}
\bibliography{RAR}

\begin{appendix}
\section{Adding rotation to GravSphere}\label{app:rotation}
\begin{equation}
    g(r)' = \exp\left(-2 \int_{0}^{r}\frac{\beta}{r'}\,dr'\right),
\label{eqn:ffuncdash}
\end{equation}
where
\begin{eqnarray}
\beta' & = & 1-\frac{\sigma_\theta^2 + \sigma_\phi^2 + \overline{v}_\phi^2}{2\sigma_r^2} \\
& = & \beta + \frac{\overline{v}_\phi^2}{2\sigma_r^2}
\end{eqnarray}
accounts for the mean streaming motion in the plane, $\overline{v}_\phi$. This, therefore, accounts for any unmodelled rotation.

We roughly parameterise the effect of $\beta'$ using a dimensionless `rotation parameter',  $A_{\rm rot}$, such that the streaming motion rotation rises linearly with radius:

\begin{equation}
\frac{\overline{v}_\phi^2}{2\sigma_r^2} = A_{\rm rot}\left(\frac{r}{R_{1/2}}\right)
\end{equation}
where $R_{1/2}$ is the projected half radius. This way, $A_{\rm rot} = 0$ corresponds to no rotation while $A_{\rm rot} = 1/2$ corresponds to an equal balance of rotation and 1D pressure support at $R_{1/2}$.

Using the above simplified model for $\overline{v}_\phi$, Equation~\ref{eqn:ffuncdash} becomes:

\begin{equation}
g(r)' = g(r) \exp\left(-2 A_{\rm rot} \frac{r}{R_{1/2}}\right)
\end{equation} 
We note that this is just a lower limit of the rotation, since we do not account for the inclination of a galaxy.

\section{Additional sources of scatter}\label{app:scatter}
Here, we aim to clarify both the physical and observational contributions to scatter in the RAR. 

\begin{figure}
    \centering
    \includegraphics[width=\columnwidth]{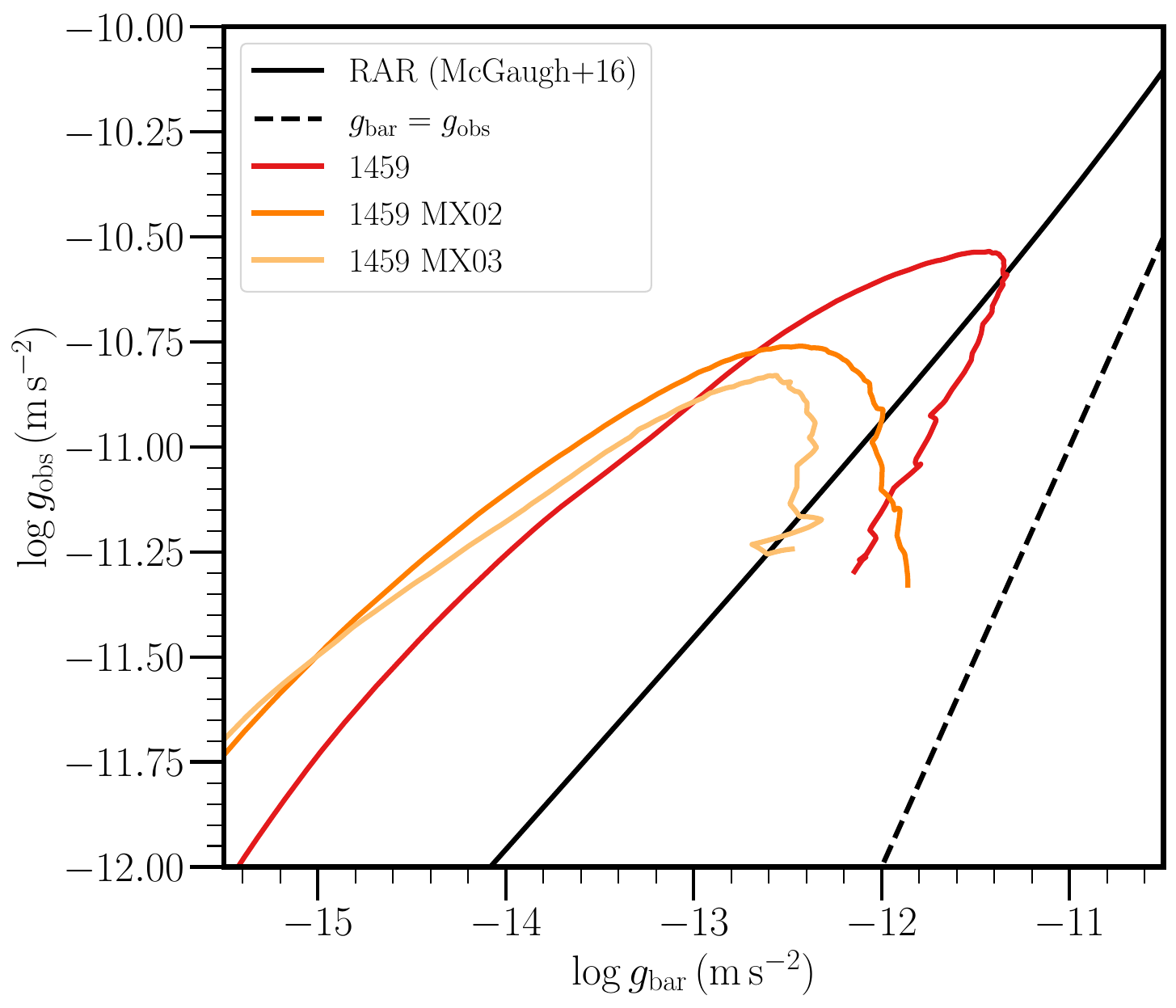}
    \caption{$g_\mathrm{obs}-g_\mathrm{bar}$ relations for genetically modified versions of Halo 1459 (MX02, MX03) extended down to $r=0.01$\,kpc. The black solid line marks the RAR derived by \cite{mcgaugh_radial_2016} for SPARC local spiral galaxies. The blacked dashed line marks $g_\mathrm{obs} = g_\mathrm{bar}$.}
    \label{fig:gm-1459}
\end{figure}

Figure~\ref{fig:gm-1459} shows results of the genetically modified (GM) versions of the EDGE Halo 1459 extended down to $r=0.01$\,kpc, which were designed to vary only the formation time relative to reionisation. This demonstrates that differences in assembly history at fixed halo mass can induce scatter in the $g_\mathrm{bar}$–$g_\mathrm{obs}$ relation comparable to that seen between different EDGE dwarfs. This supports the interpretation that the scatter in EDGE primarily arises from a combination of reionisation and assembly history effects.

\begin{figure}[!htbp]
    \centering
    \includegraphics[width=\columnwidth]{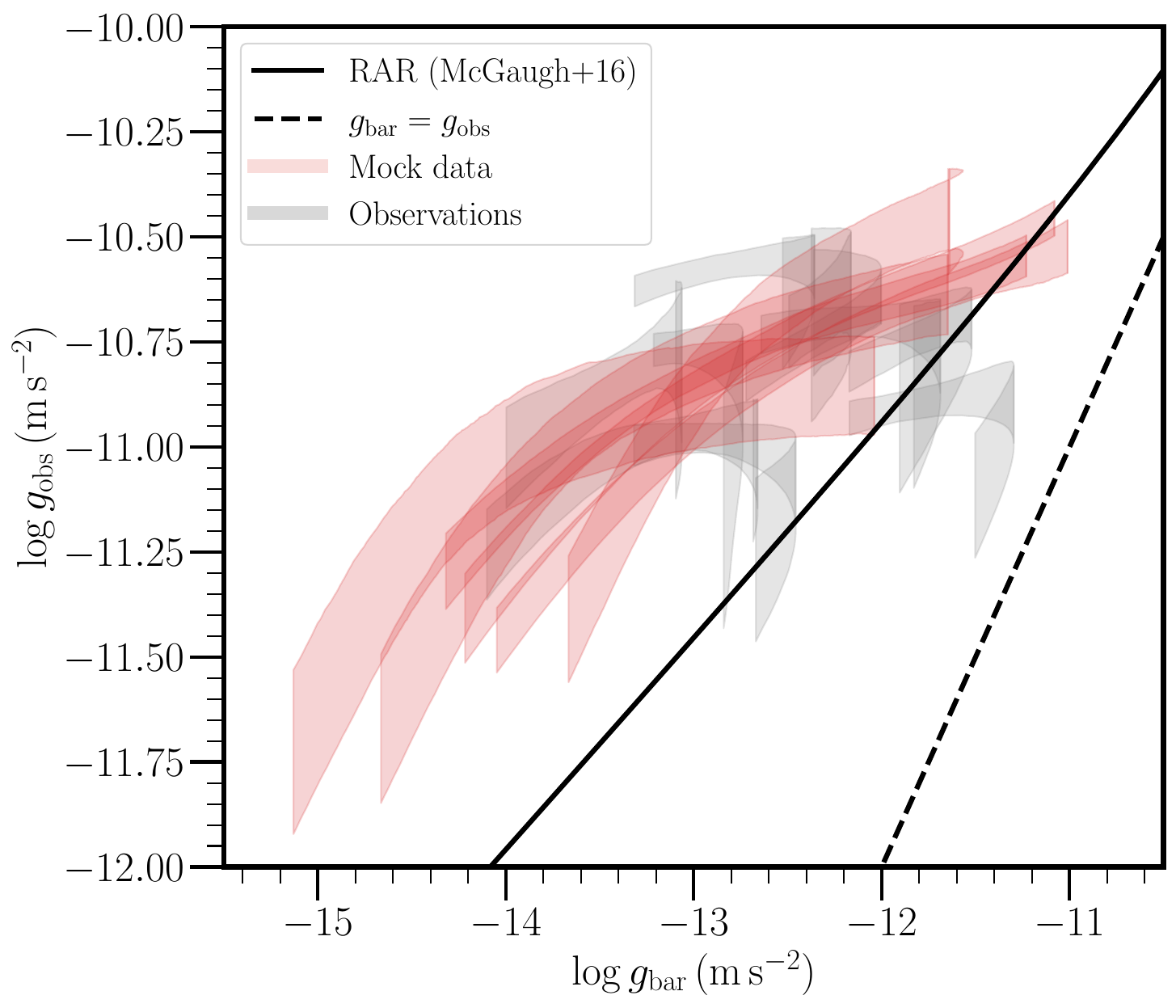}
    \caption{Effect of the mass-modelling pipeline and observational uncertainties on the RAR. The 68\% scatter obtained for the observed and for the mocks from the simulated $g_\mathrm{obs}-g_\mathrm{bar}$ relations are marked by the grey and red shaded regions, respectively. The black solid line marks the RAR derived by \cite{mcgaugh_radial_2016} for SPARC local spiral galaxies. The blacked dashed line marks $g_\mathrm{obs} = g_\mathrm{bar}$.}
    \label{fig:rar-errors}
\end{figure}

Figure~\ref{fig:rar-errors} illustrates the additional dispersion introduced by our mass-modelling pipeline \textsc{GravSphere} and observational uncertainties, quantified by re-running the modelling on mock data. This shows that observational effects can increase the apparent scatter in the RAR. Once accounting for observational uncertainties, the scatter in the EDGE simulations and the real data is comparable for lower baryonic accelerations ($g_{\rm bar} < -13.5$ m\,s$^{-2}$). At higher baryonic accelerations, however, the data show larger scatter than the simulations. This could be explained by the fact that the simulations are of isolated dwarfs while the data are for satellite dwarfs that span a wider range of pre-infall masses than EDGE (\citealt{read_erkal19}; \citealt{rey_edge_2025}). We will explore this in more detail in future work.

\section{GravSphere fits}\label{app:fits}
This appendix presents the \textsc{GravSphere} model fits to our observational data, allowing an assessment of how well the models reproduce the key dynamical and photometric properties of the galaxies. Figures~\ref{fig:fits}, \ref{fig:fits2}, and \ref{fig:fits3} show the results for each galaxy in our sample. For each object, we display the line-of-sight velocity dispersion profile (left column), the surface brightness profile (centre column), and the cumulative mass profile (right column). 

\begin{figure*}[h!]
    \centering
    \includegraphics[scale=0.265]{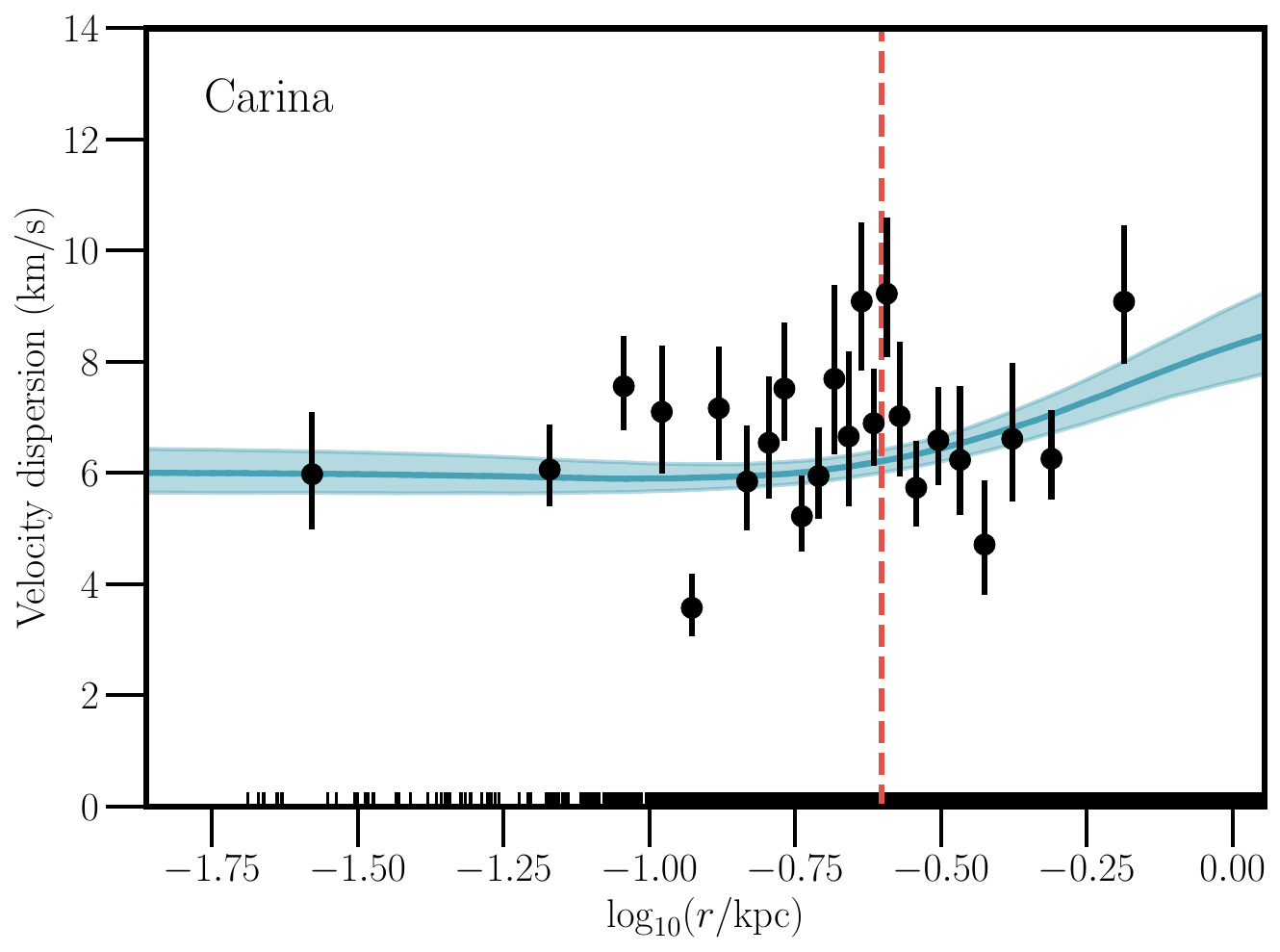}
    \includegraphics[scale=0.265]{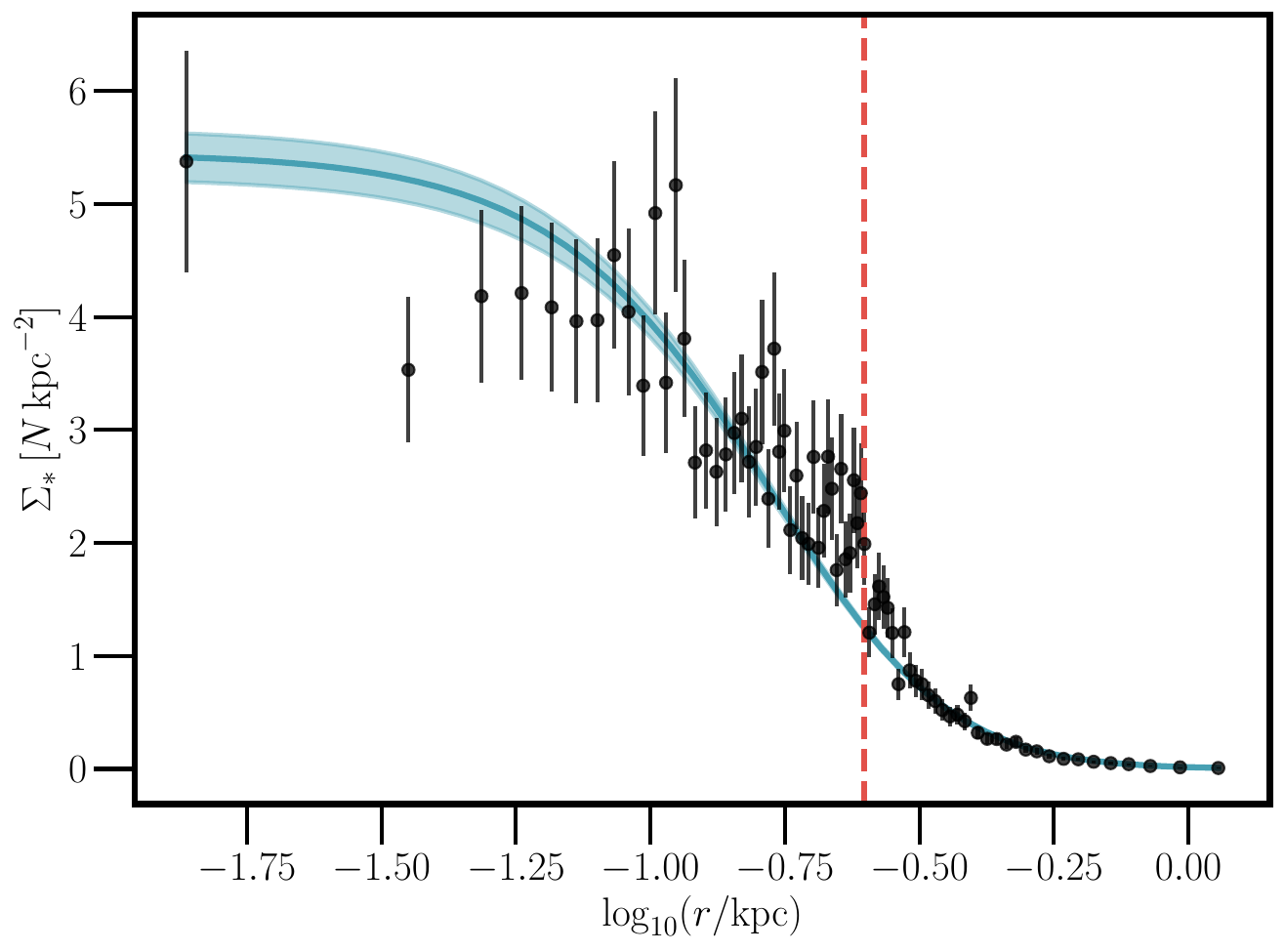}
    \includegraphics[scale=0.265]{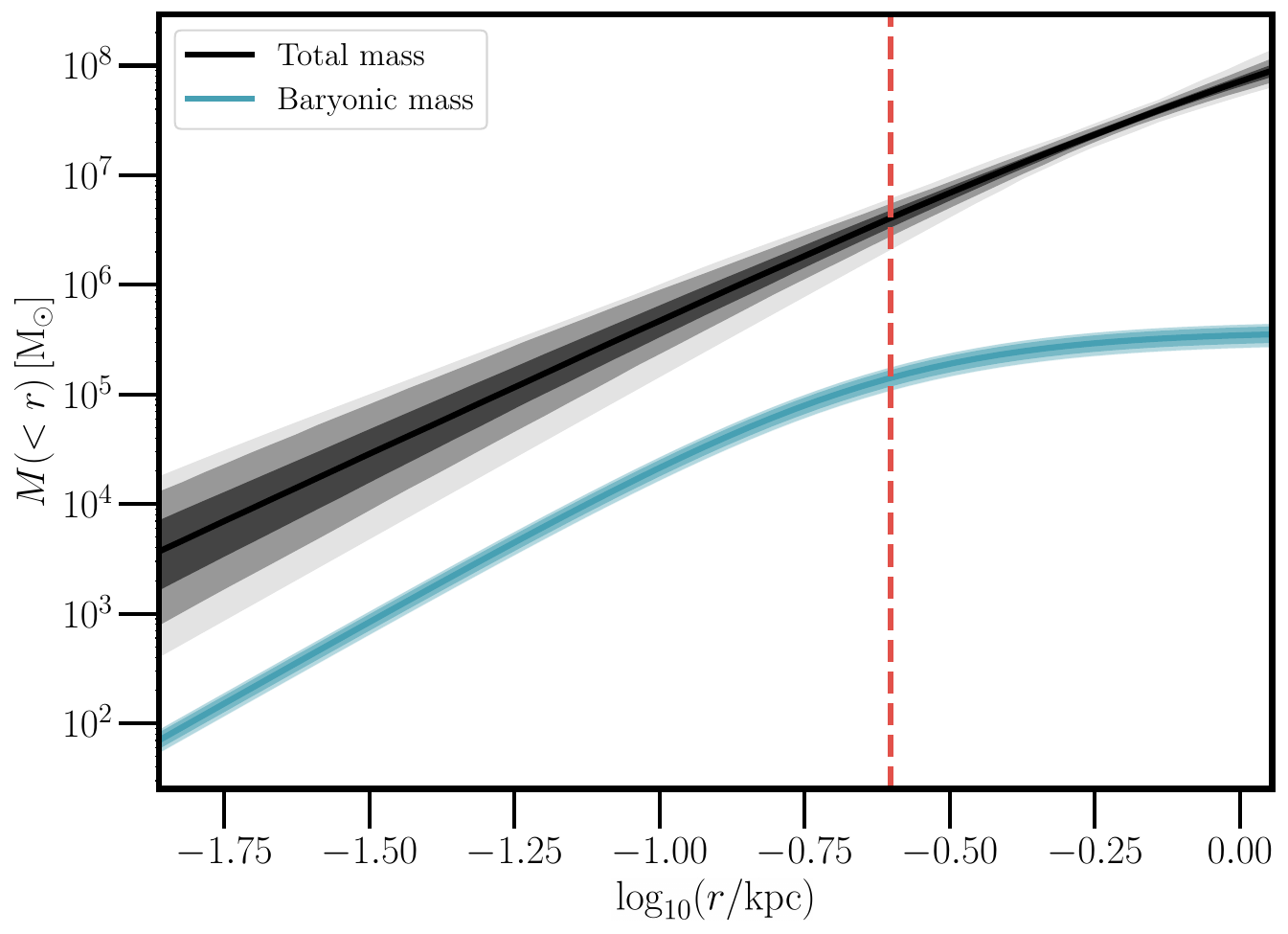}
    \includegraphics[scale=0.265]{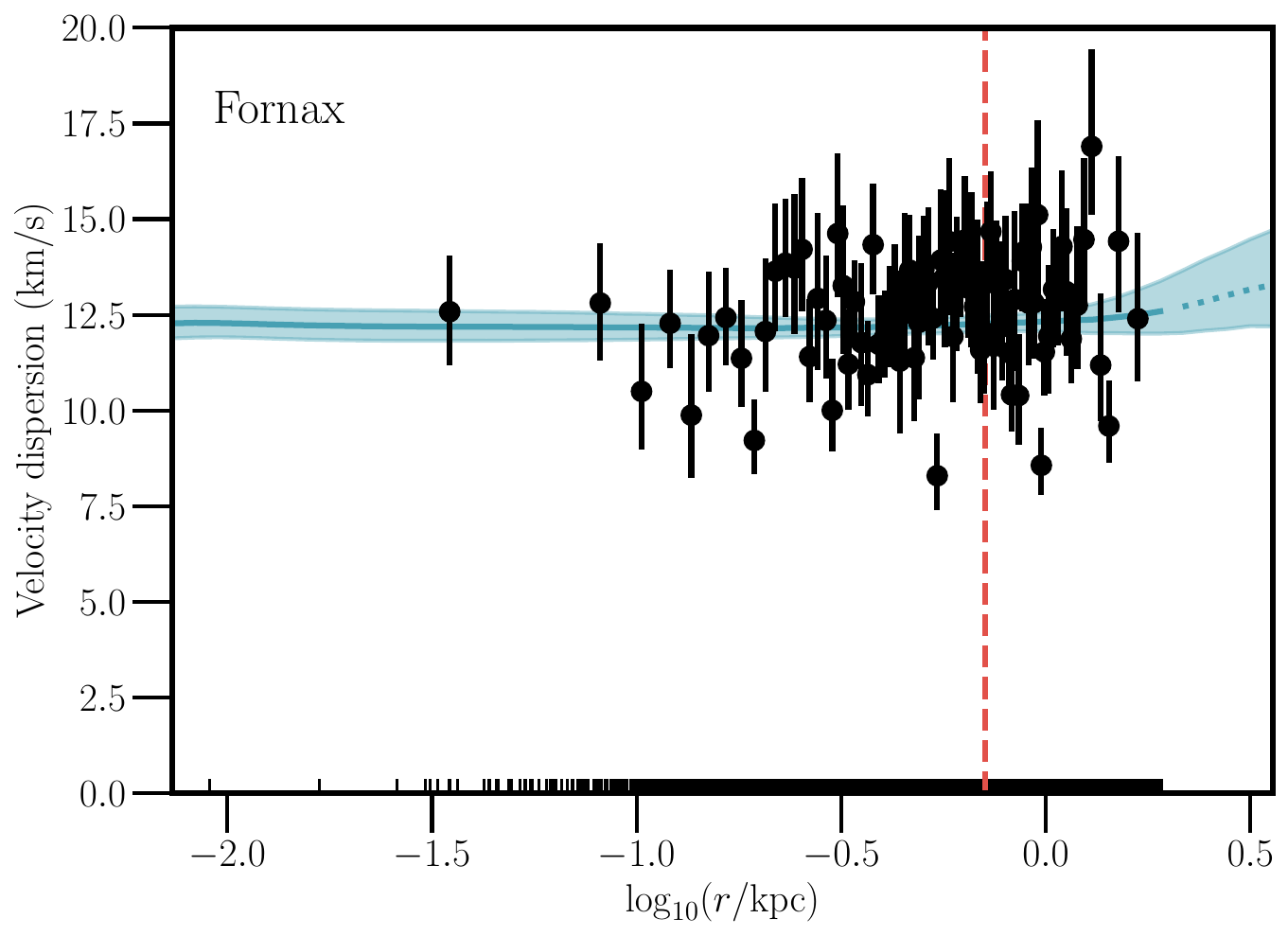}
    \includegraphics[scale=0.265]{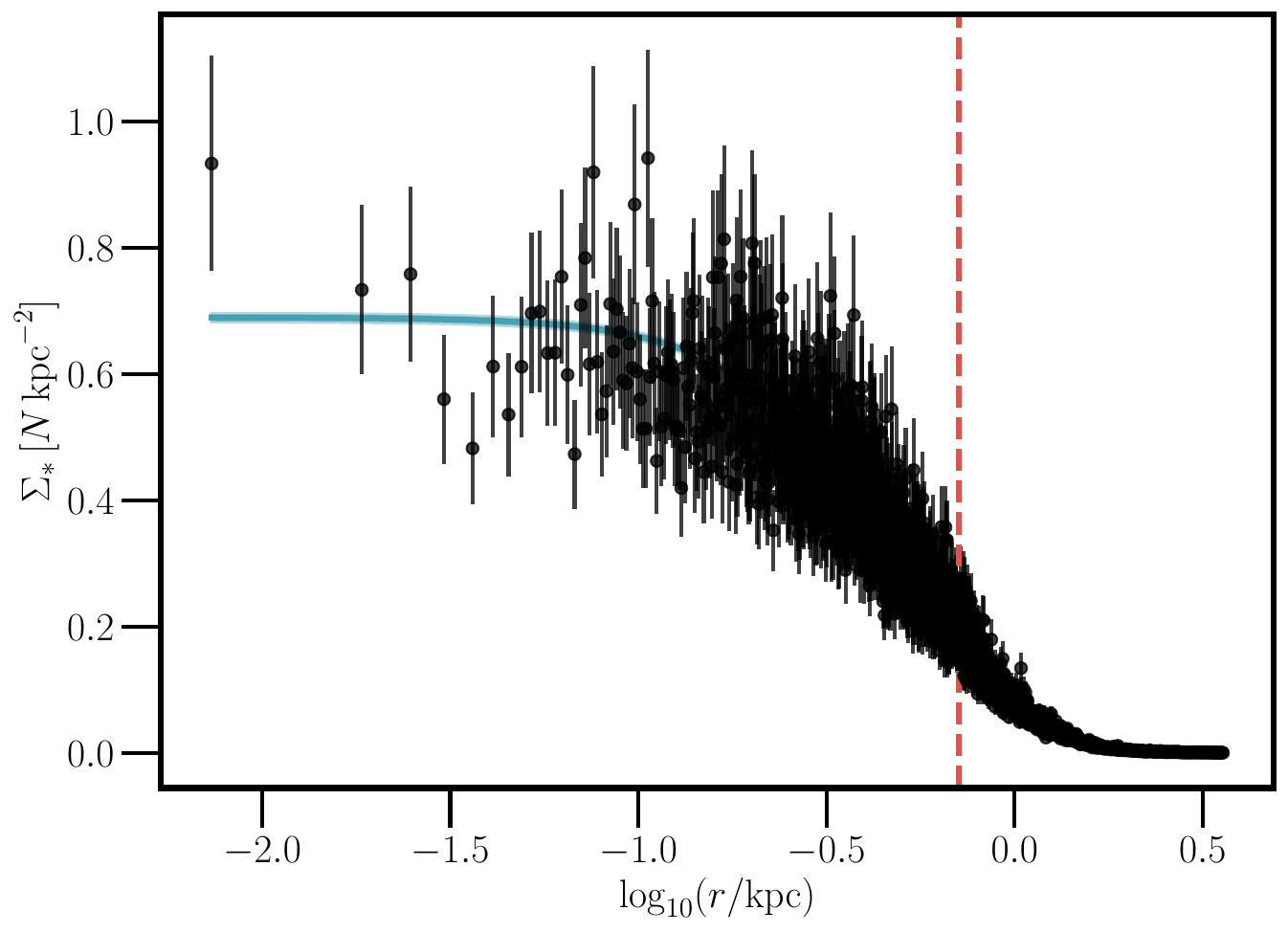}
    \includegraphics[scale=0.265]{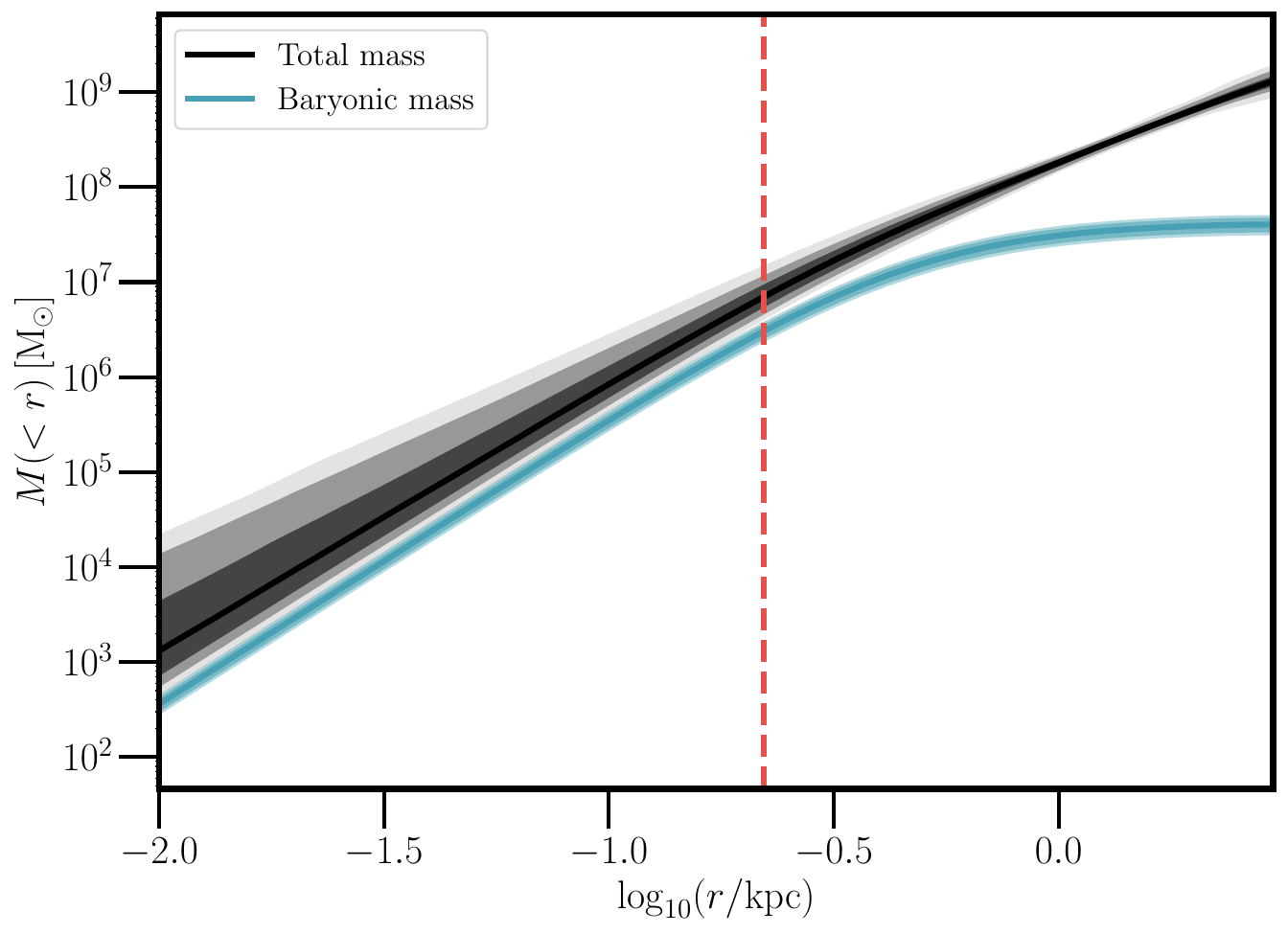}
    \includegraphics[scale=0.265]{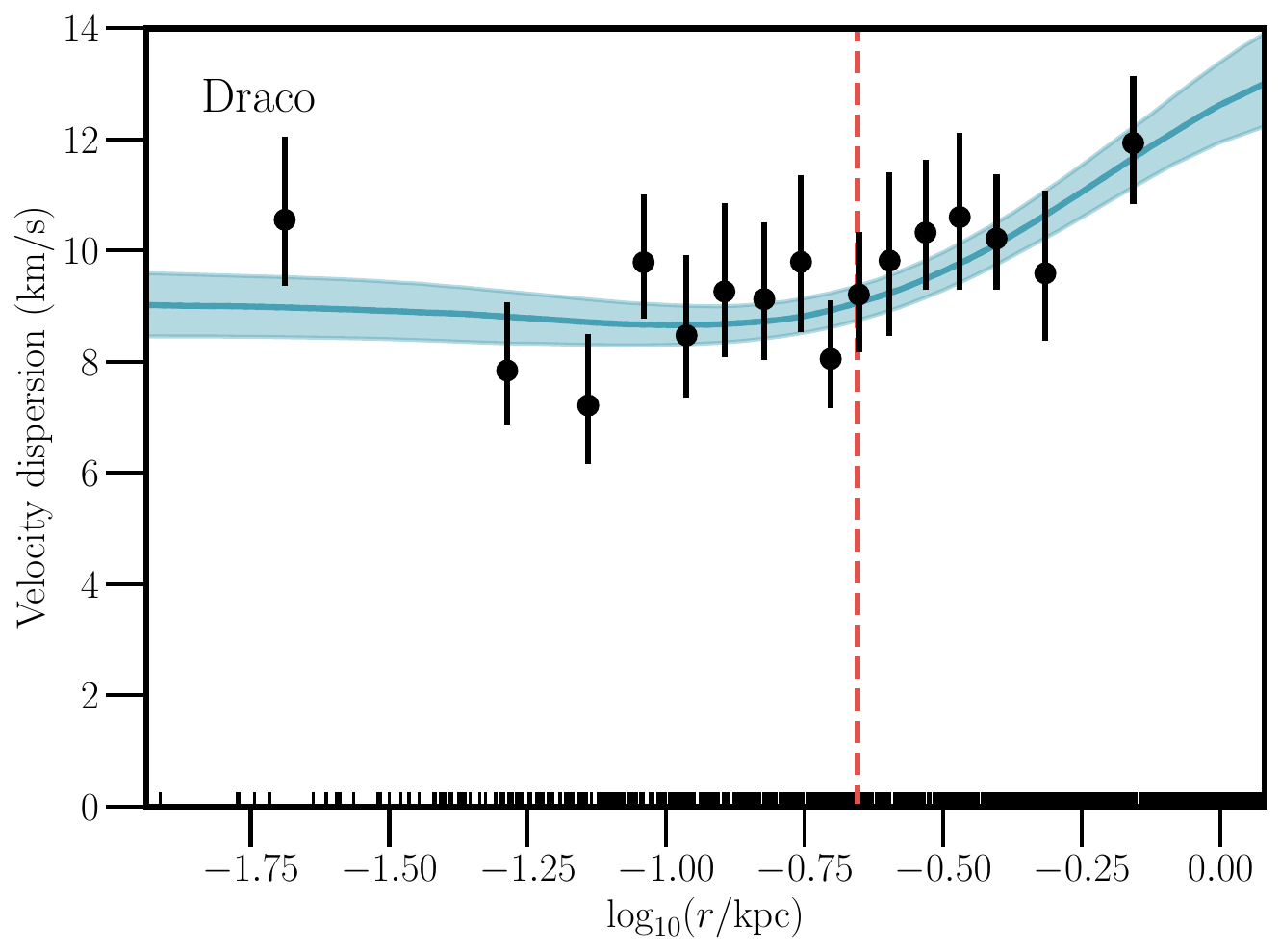}
    \includegraphics[scale=0.265]{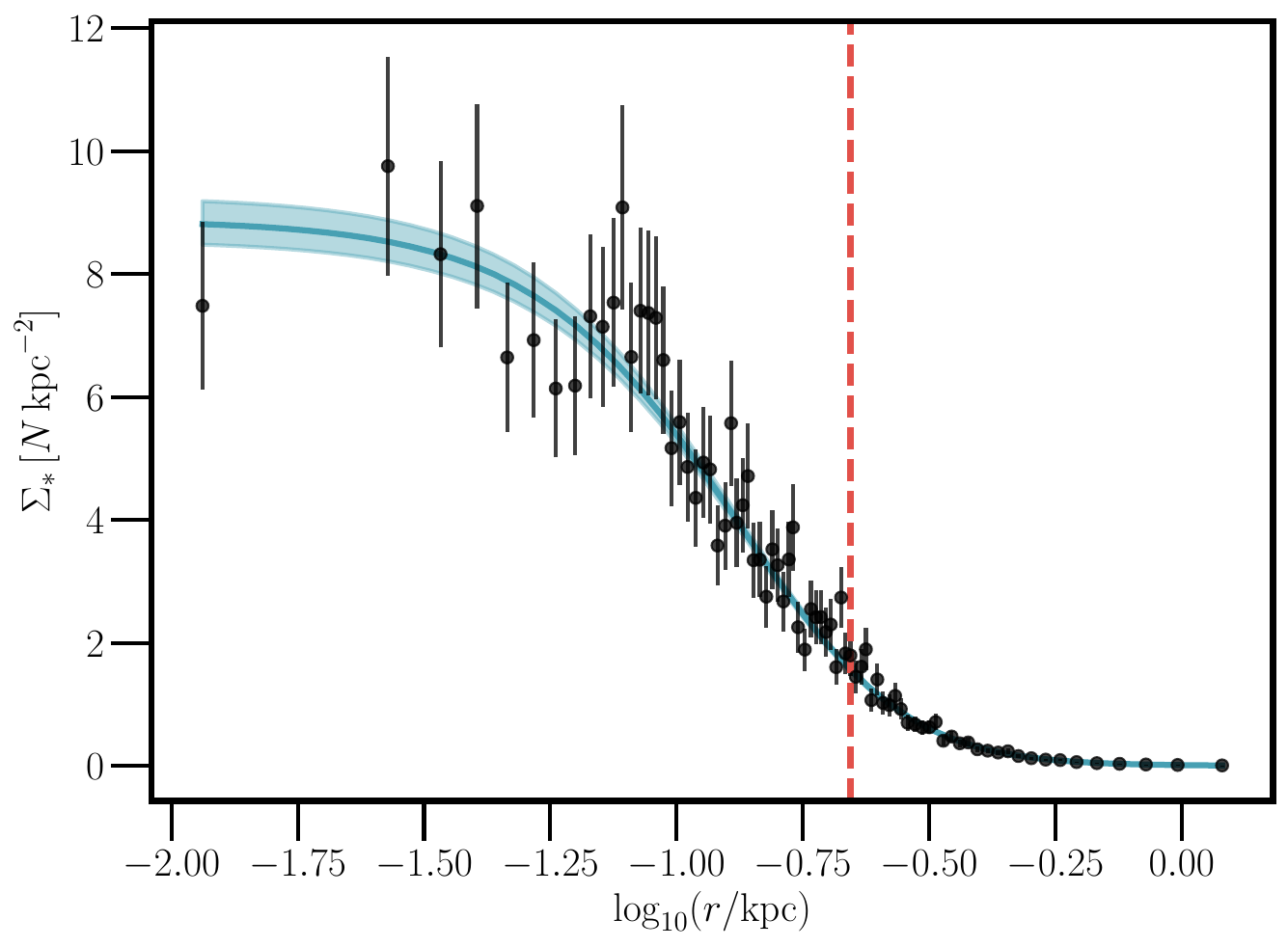}
    \includegraphics[scale=0.265]{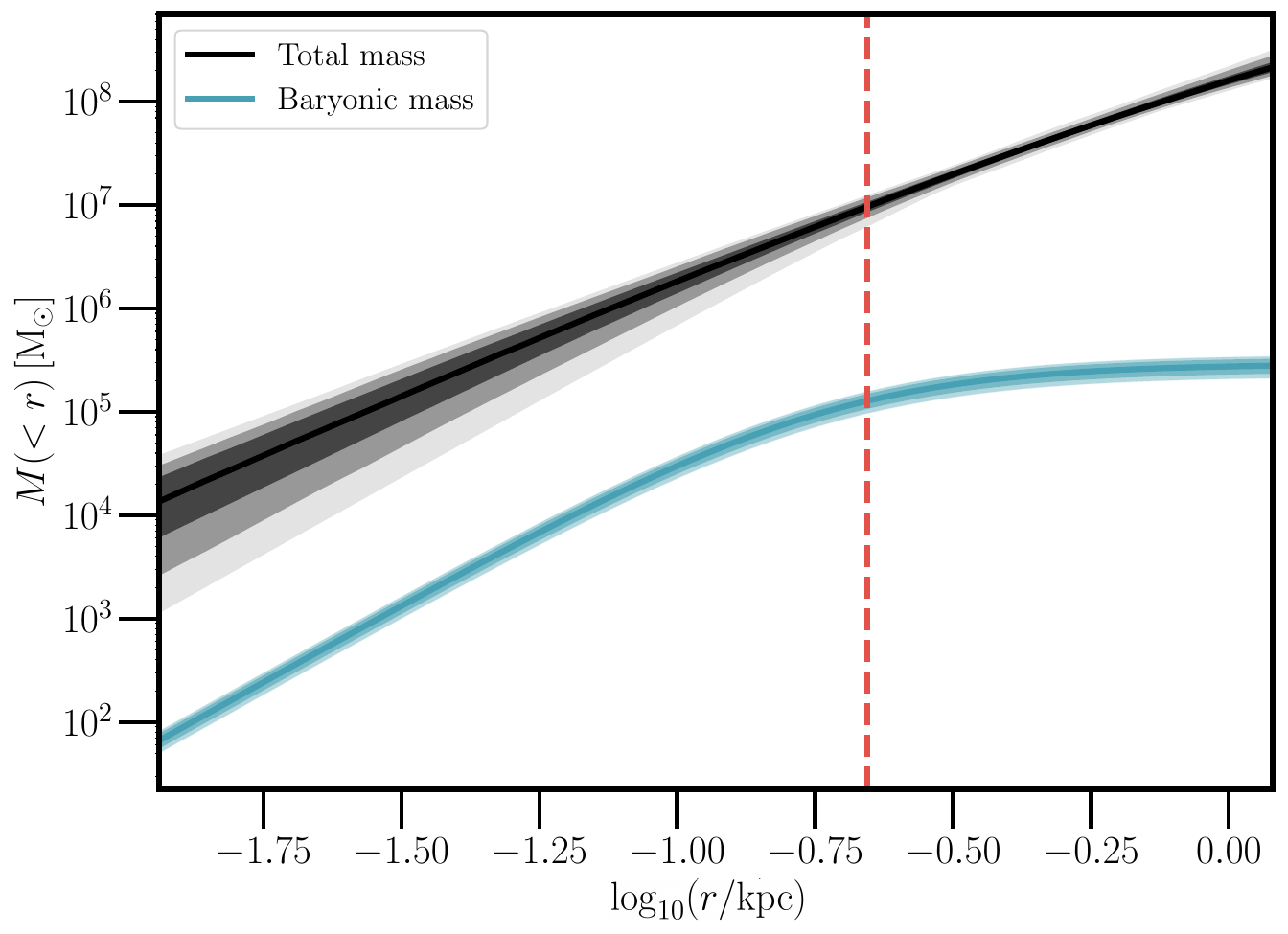}
    \includegraphics[scale=0.265]{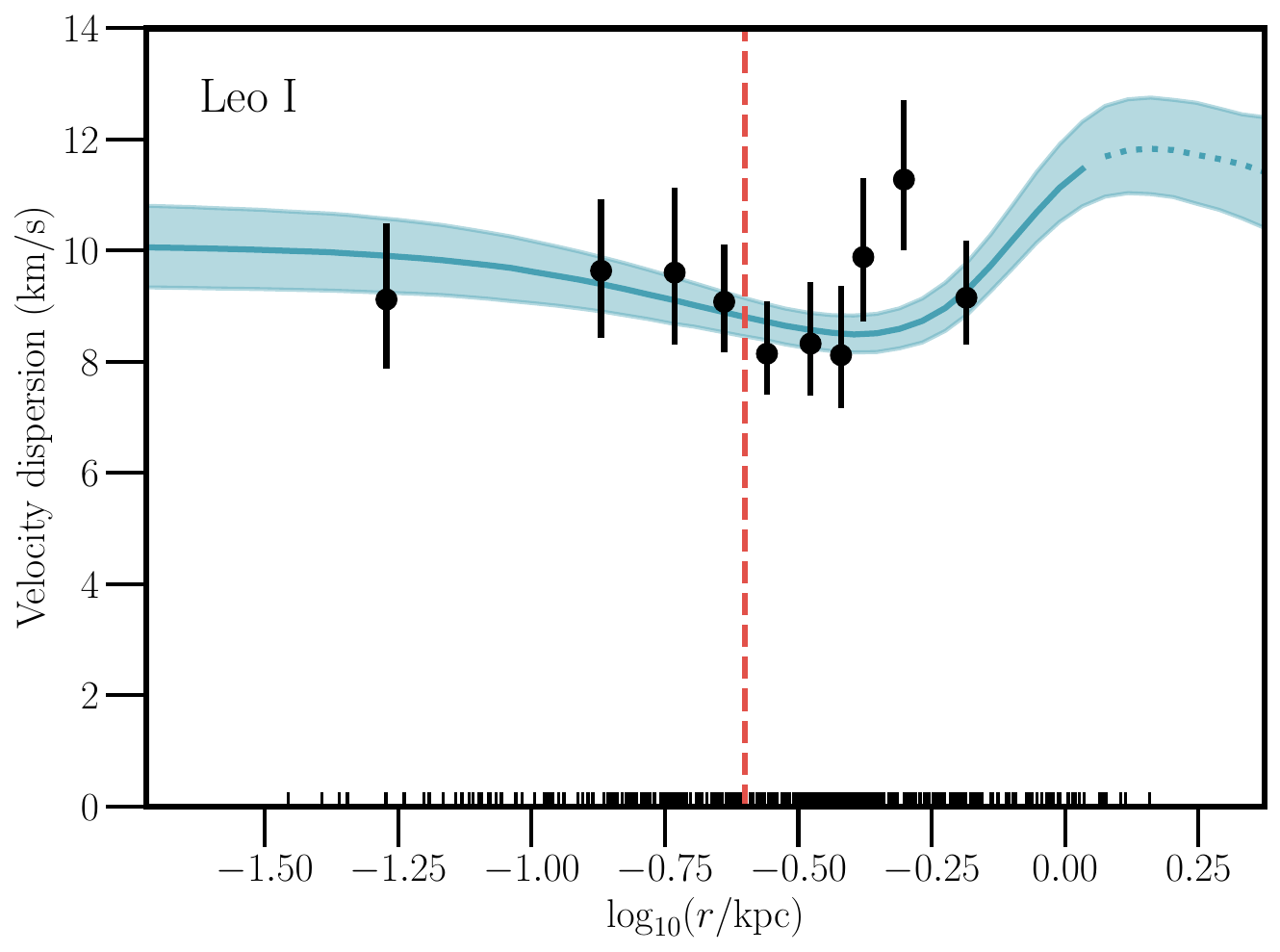}
    \includegraphics[scale=0.265]{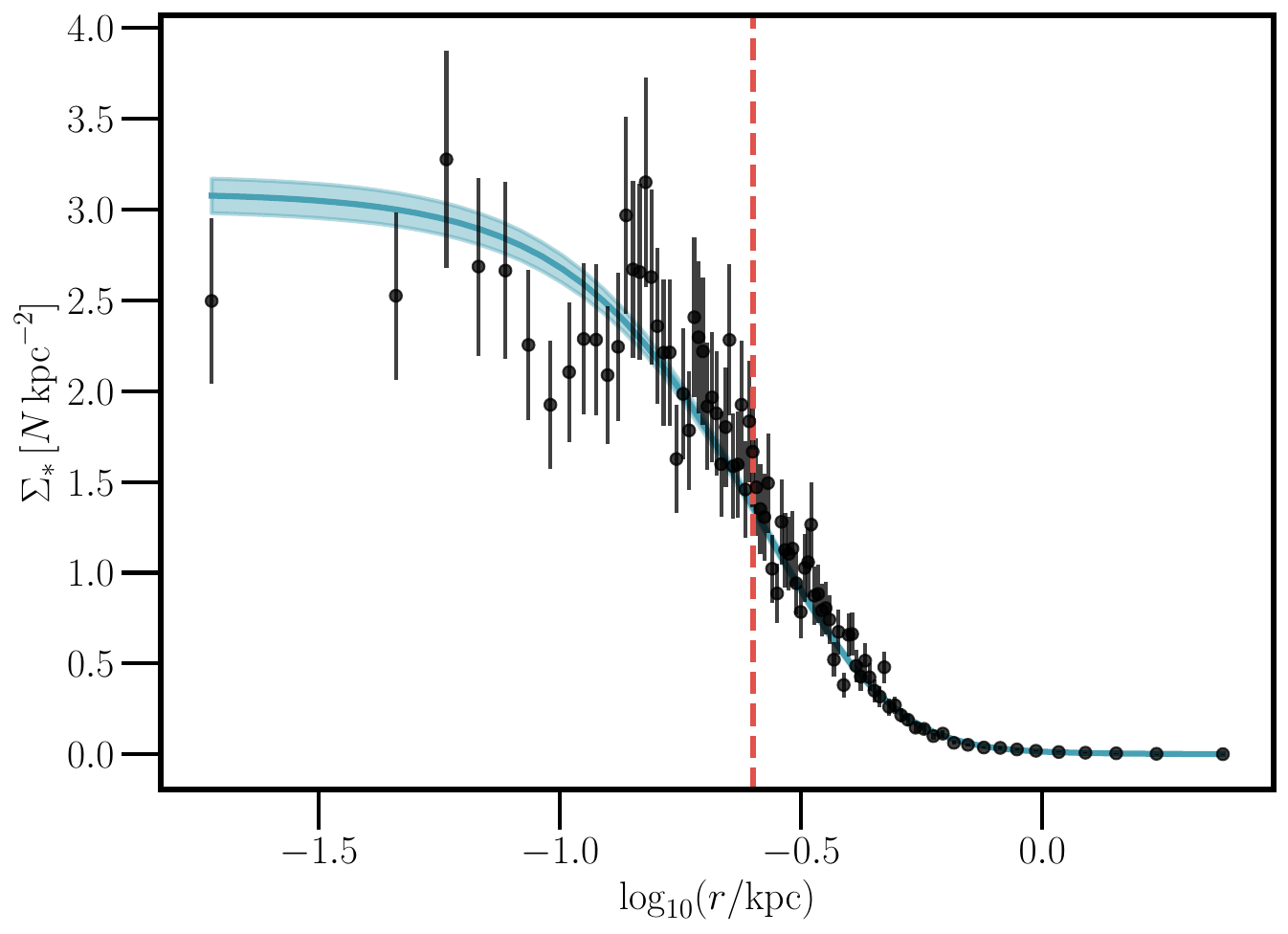}
    \includegraphics[scale=0.265]{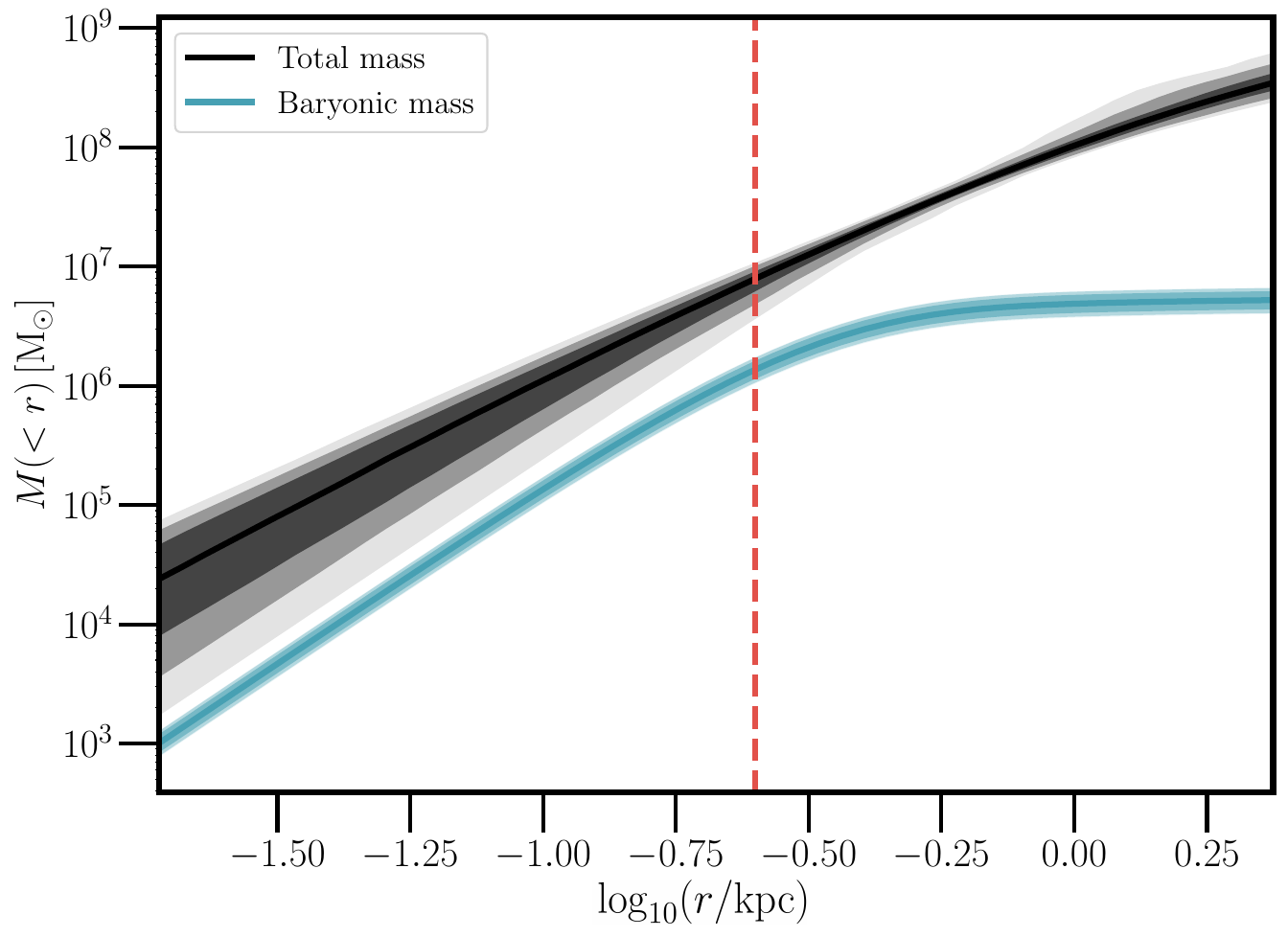}
    \includegraphics[scale=0.265]{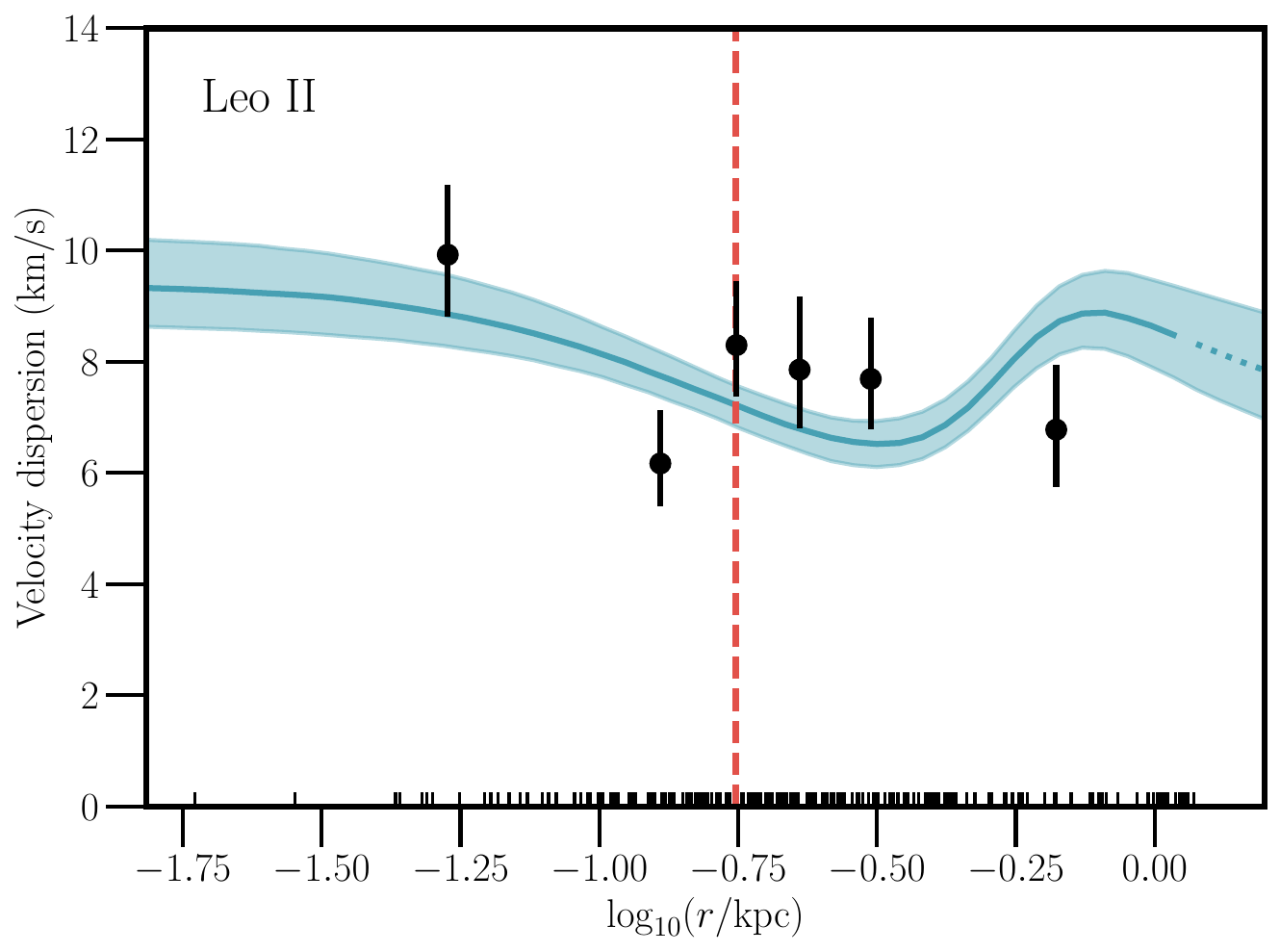}
    \includegraphics[scale=0.265]{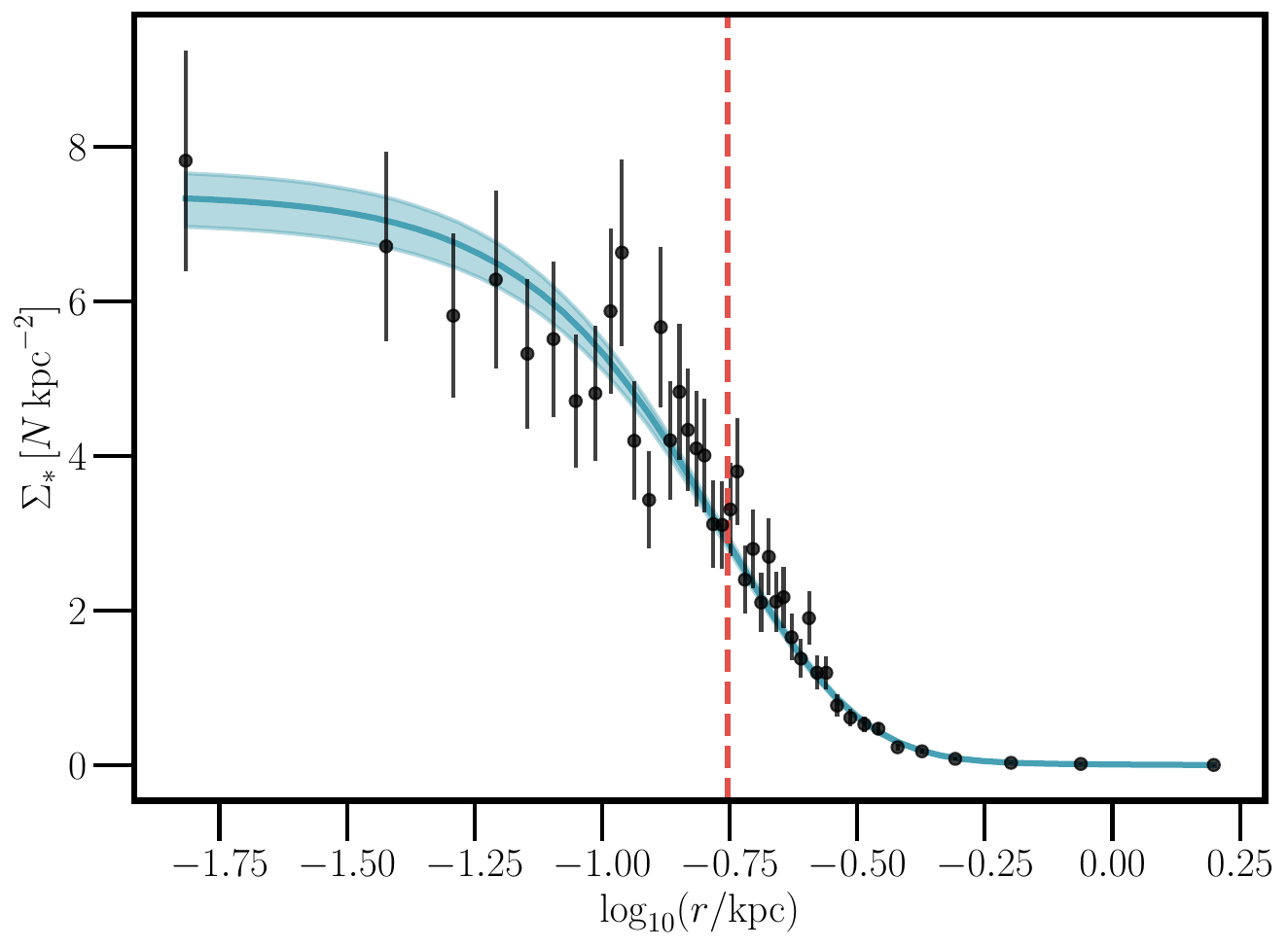}
    \includegraphics[scale=0.265]{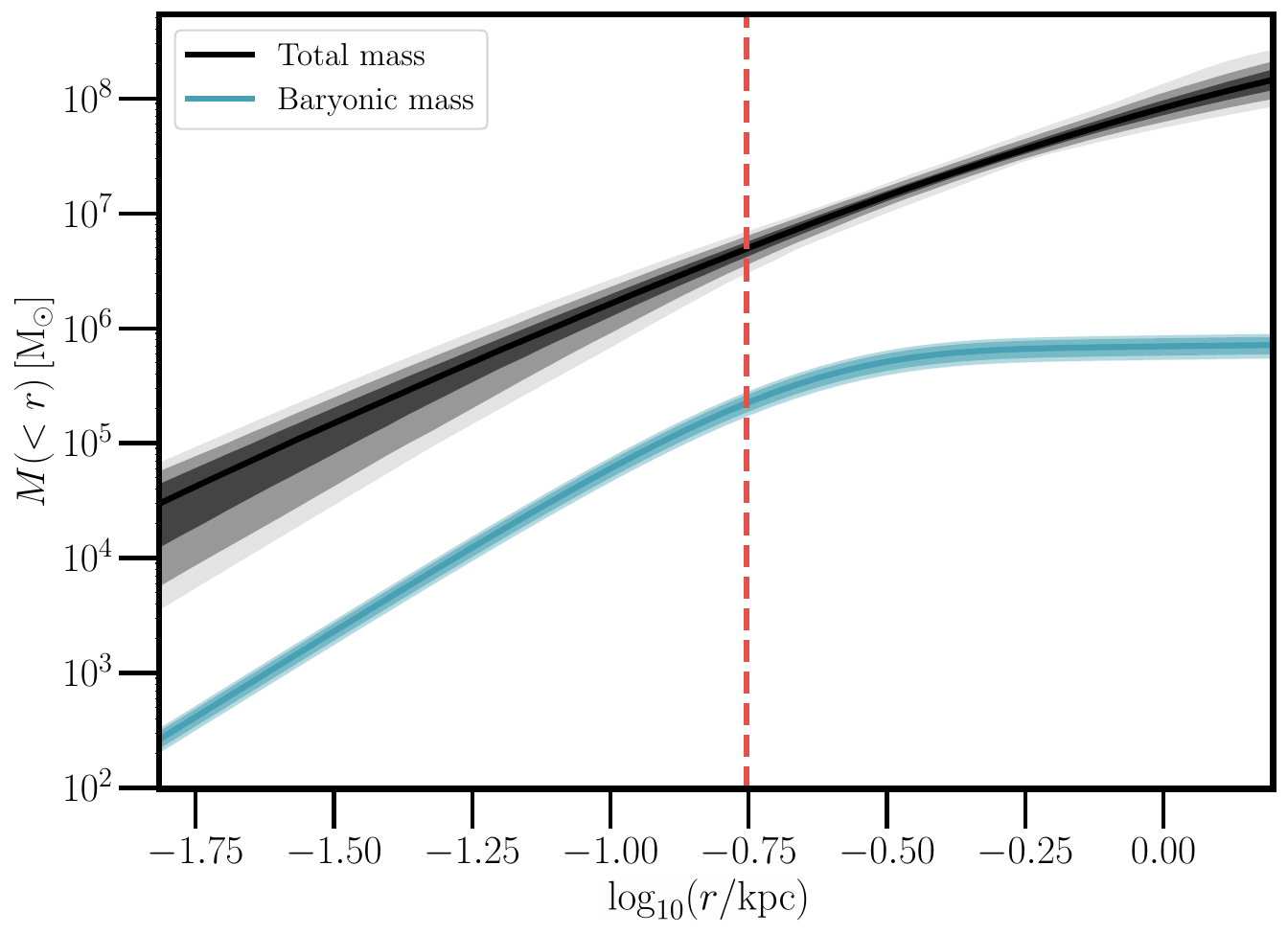}

    \caption{\textsc{GravSphere} fits for the velocity dispersion profile (left), surface brightness profile (centre), and mass profile (right) of the observed dwarf galaxies under study. The vertical red line in each plot represents the half-light radius.}
    \label{fig:fits}
\end{figure*}

\begin{figure*}[h!]
    \centering
    \includegraphics[scale=0.265]{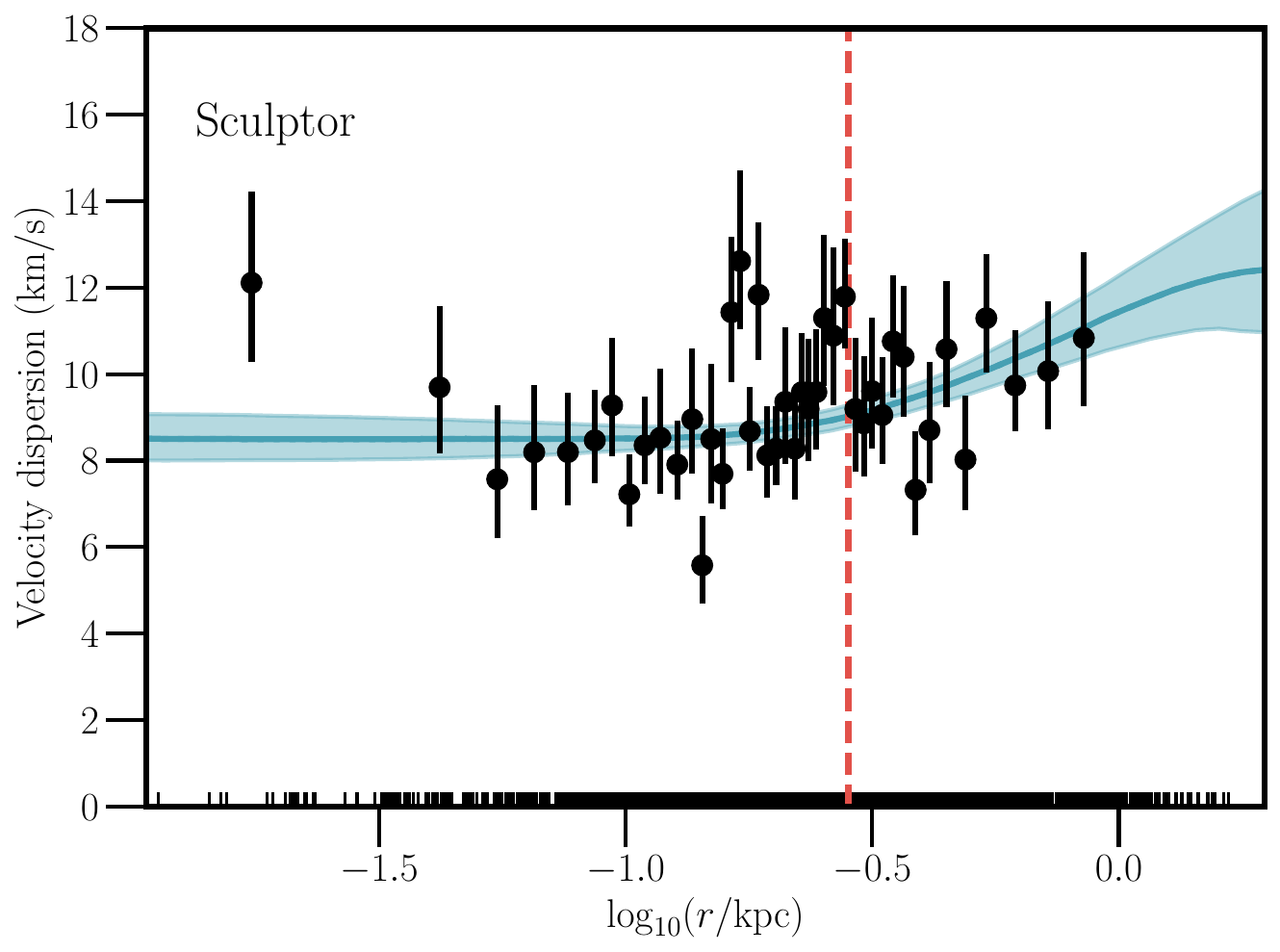}
    \includegraphics[scale=0.265]{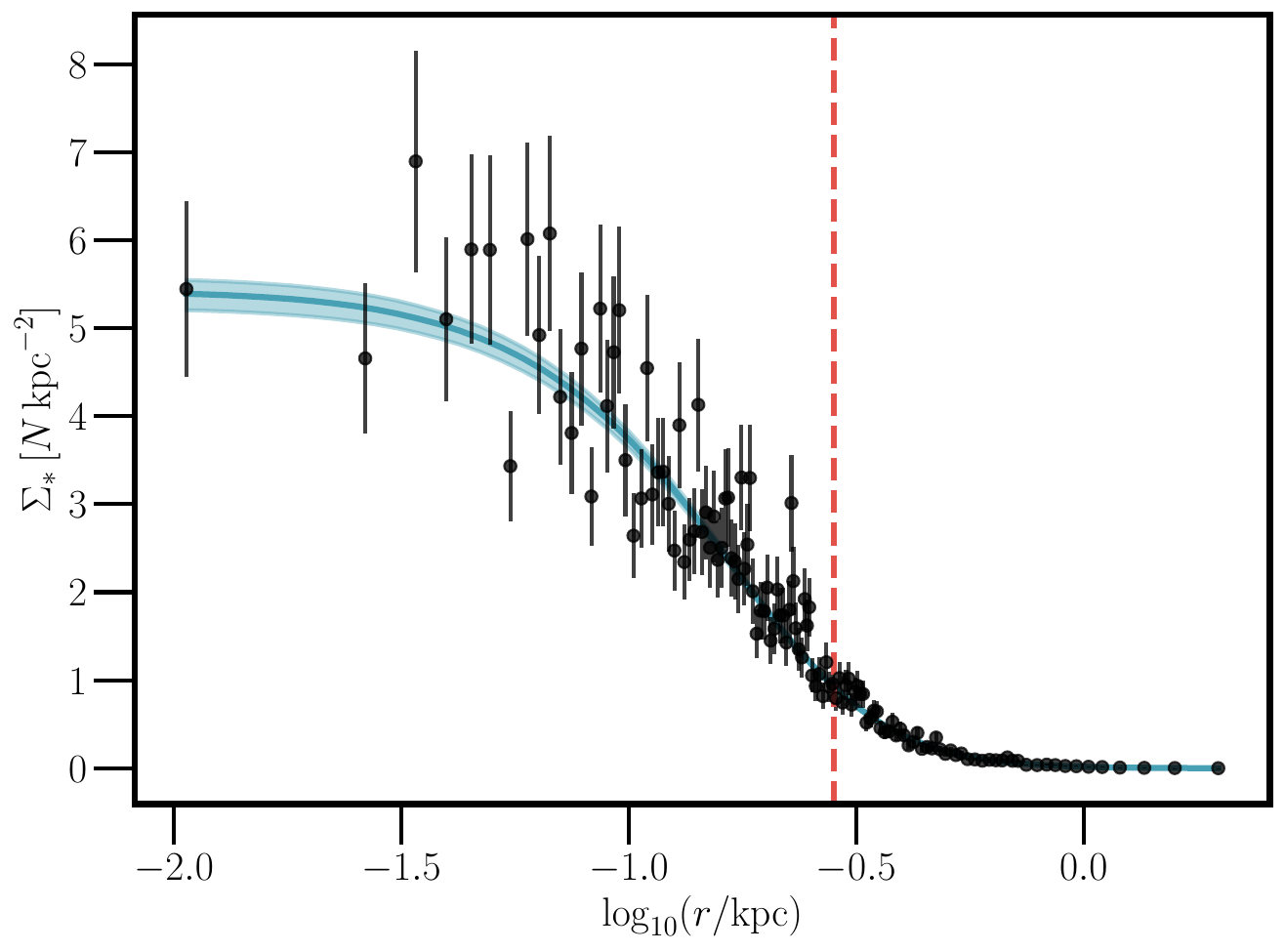}
    \includegraphics[scale=0.265]{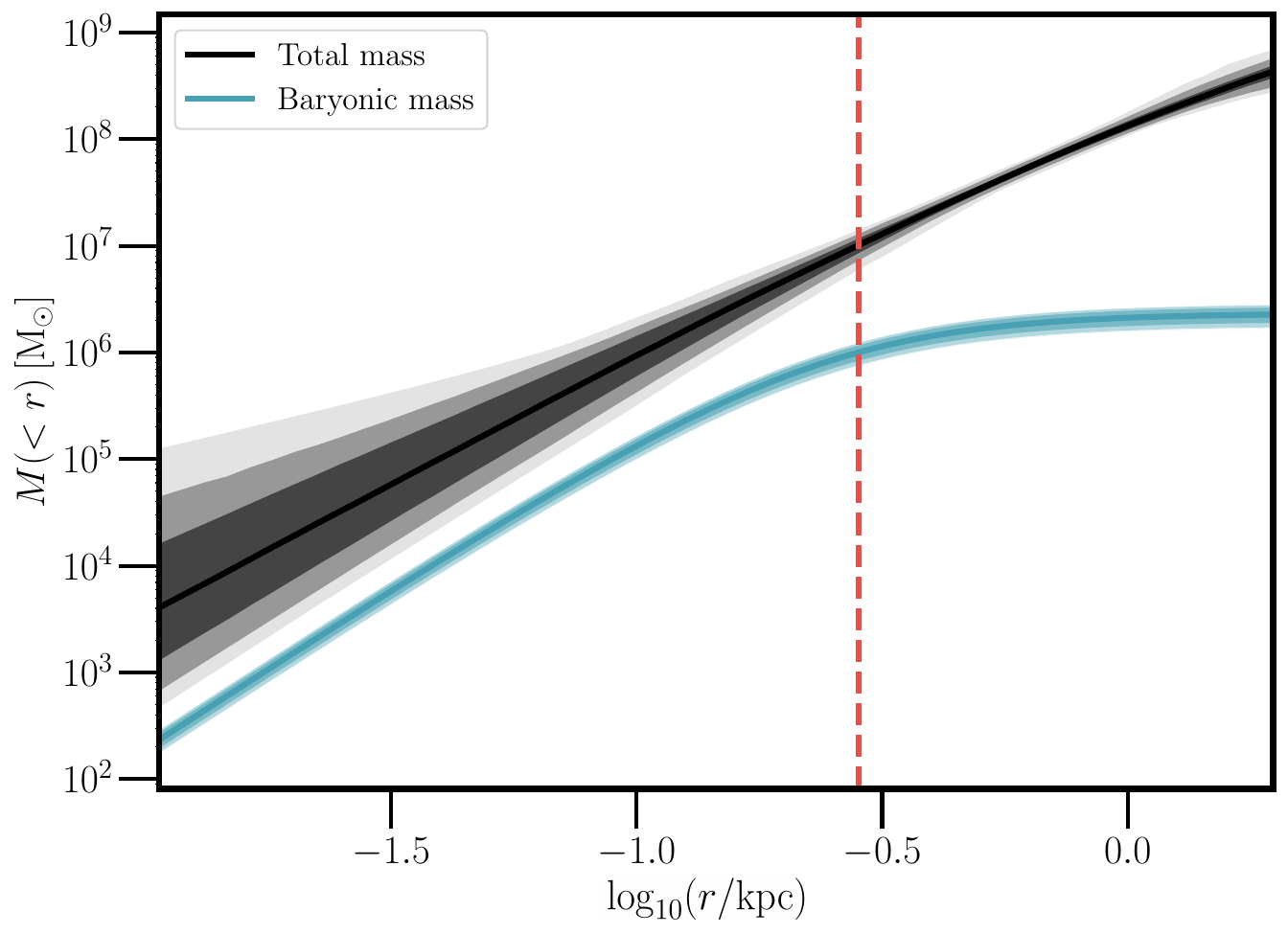}
    \includegraphics[scale=0.265]{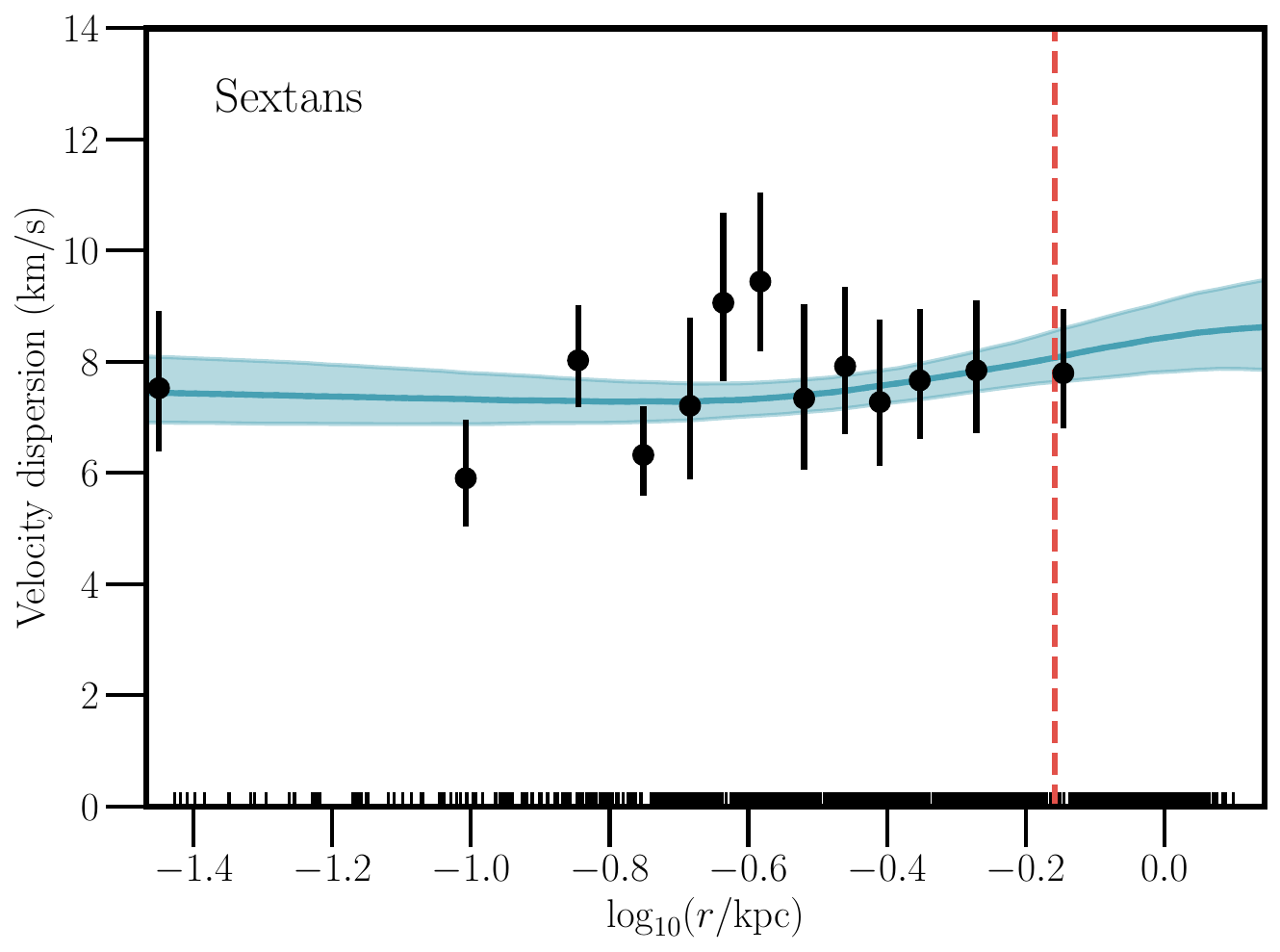}
    \includegraphics[scale=0.265]{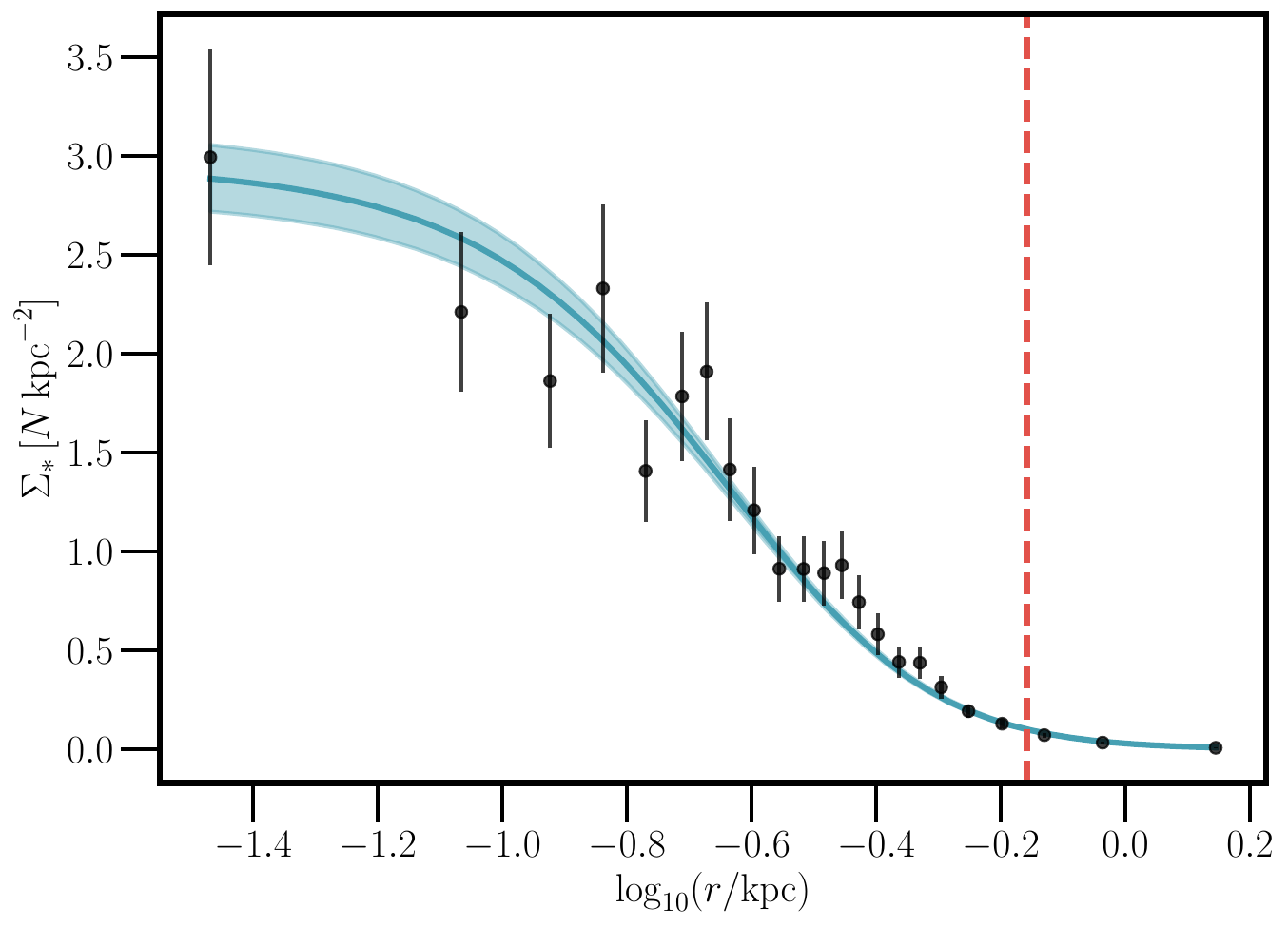}
    \includegraphics[scale=0.265]{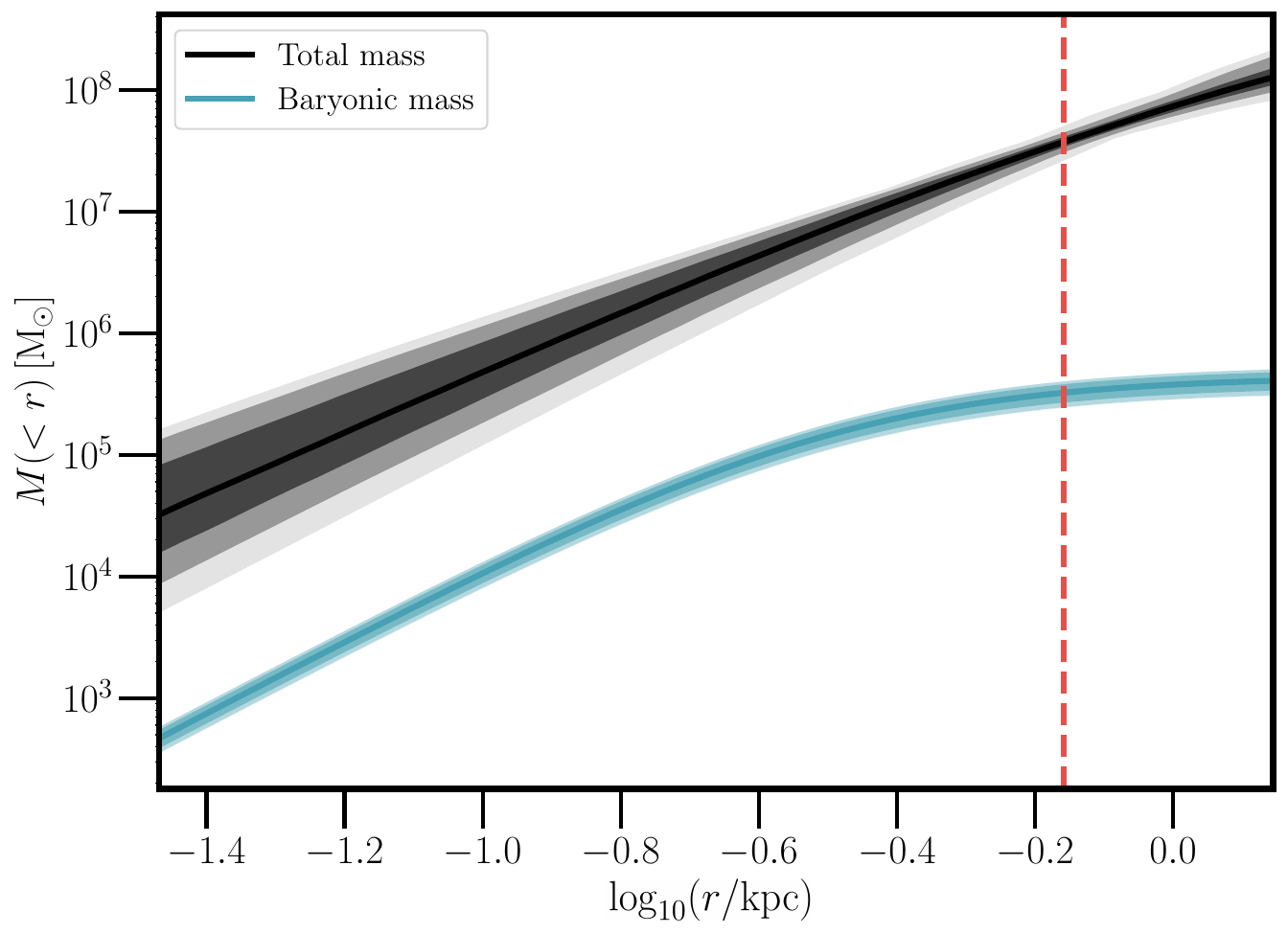}
    \includegraphics[scale=0.265]{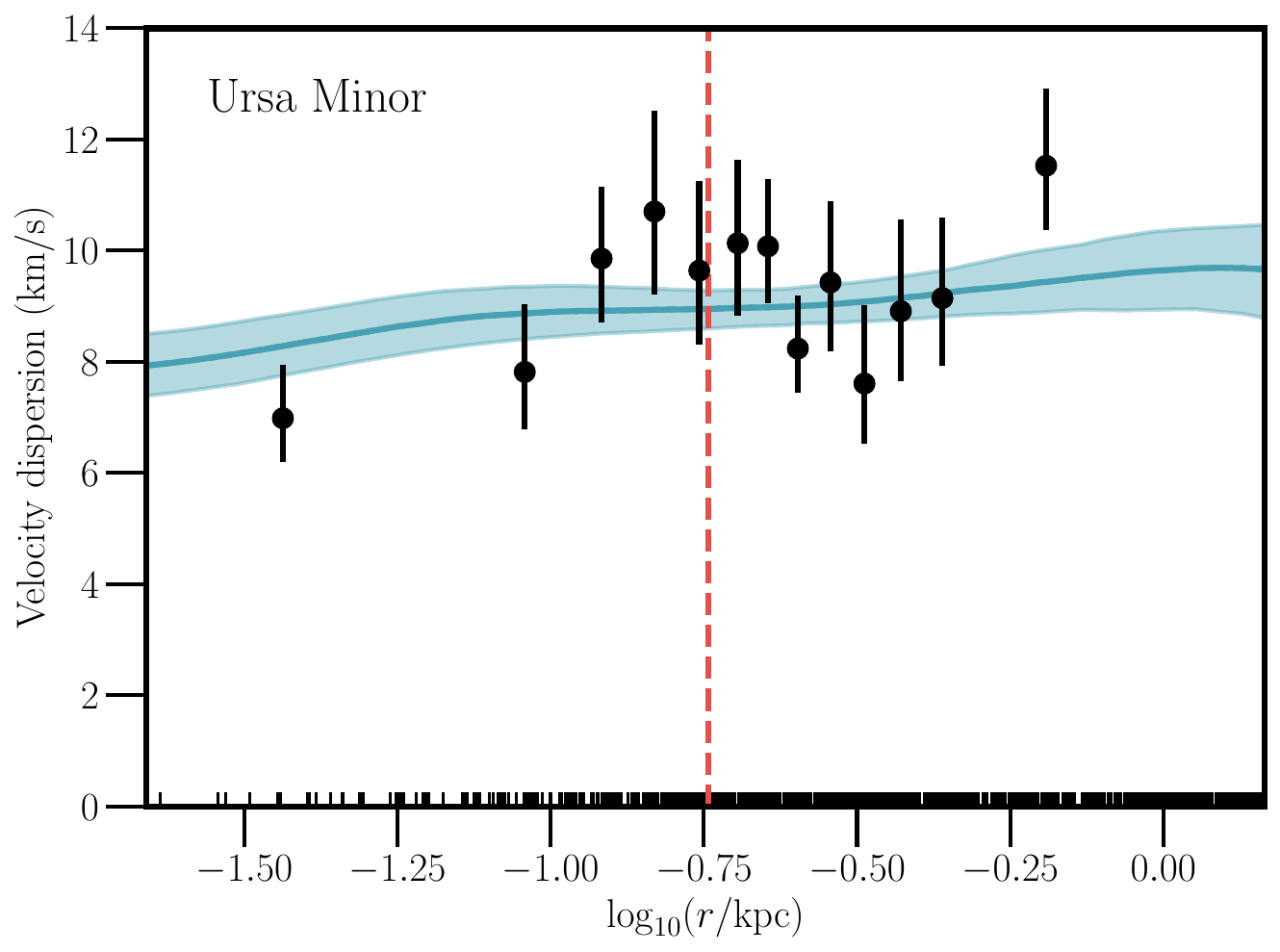}
    \includegraphics[scale=0.265]{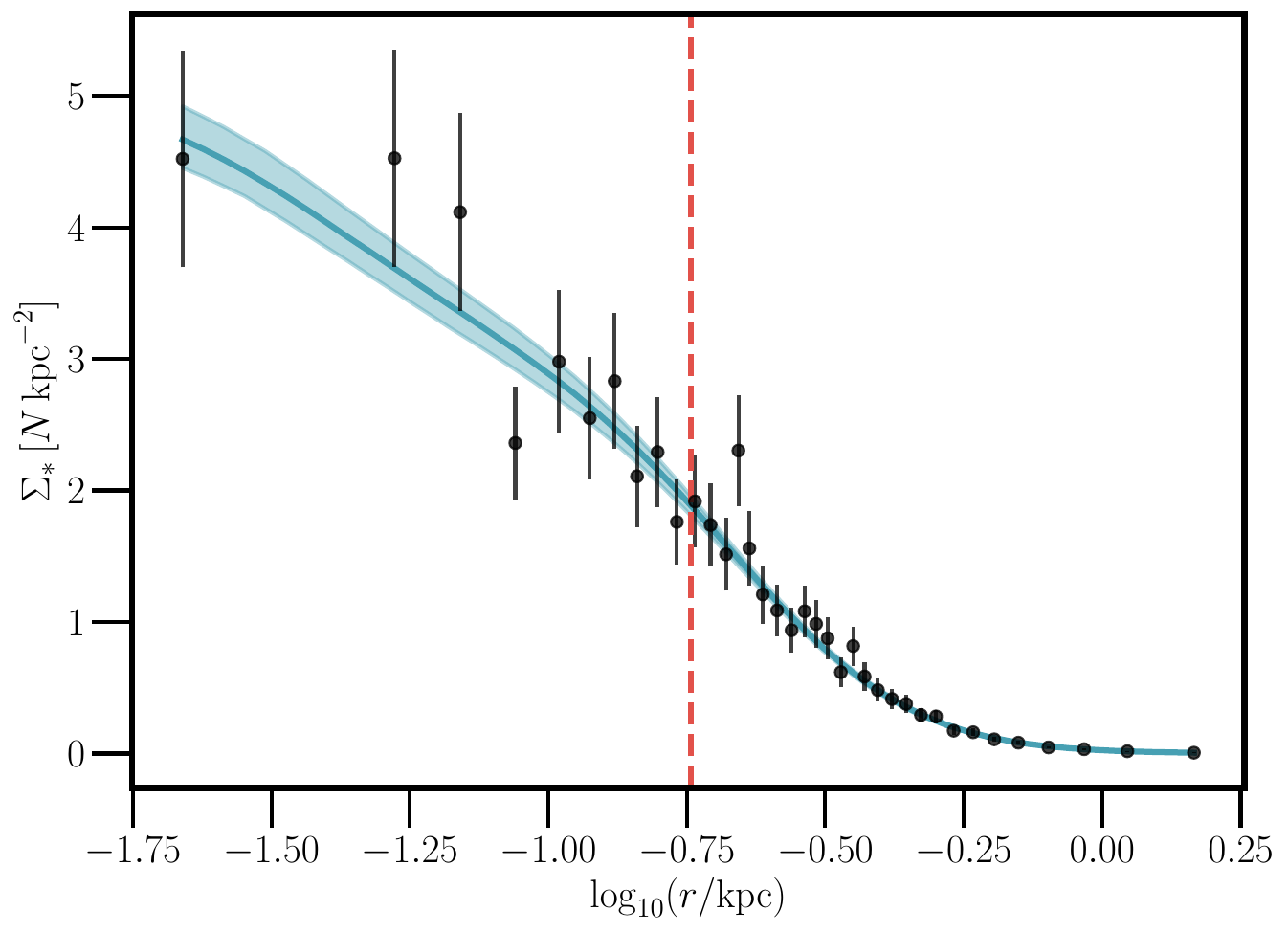}
    \includegraphics[scale=0.265]{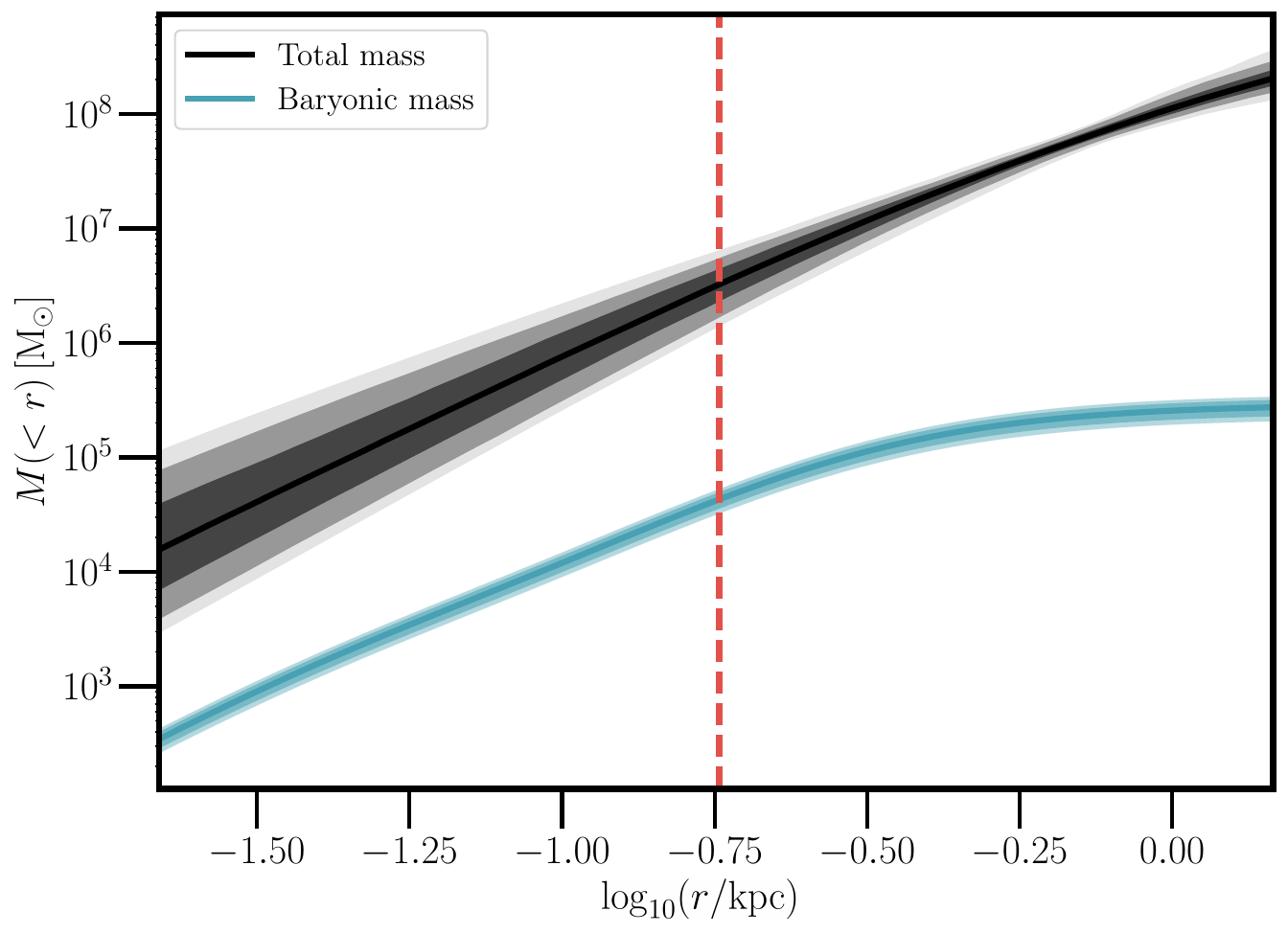}
    \includegraphics[scale=0.265]{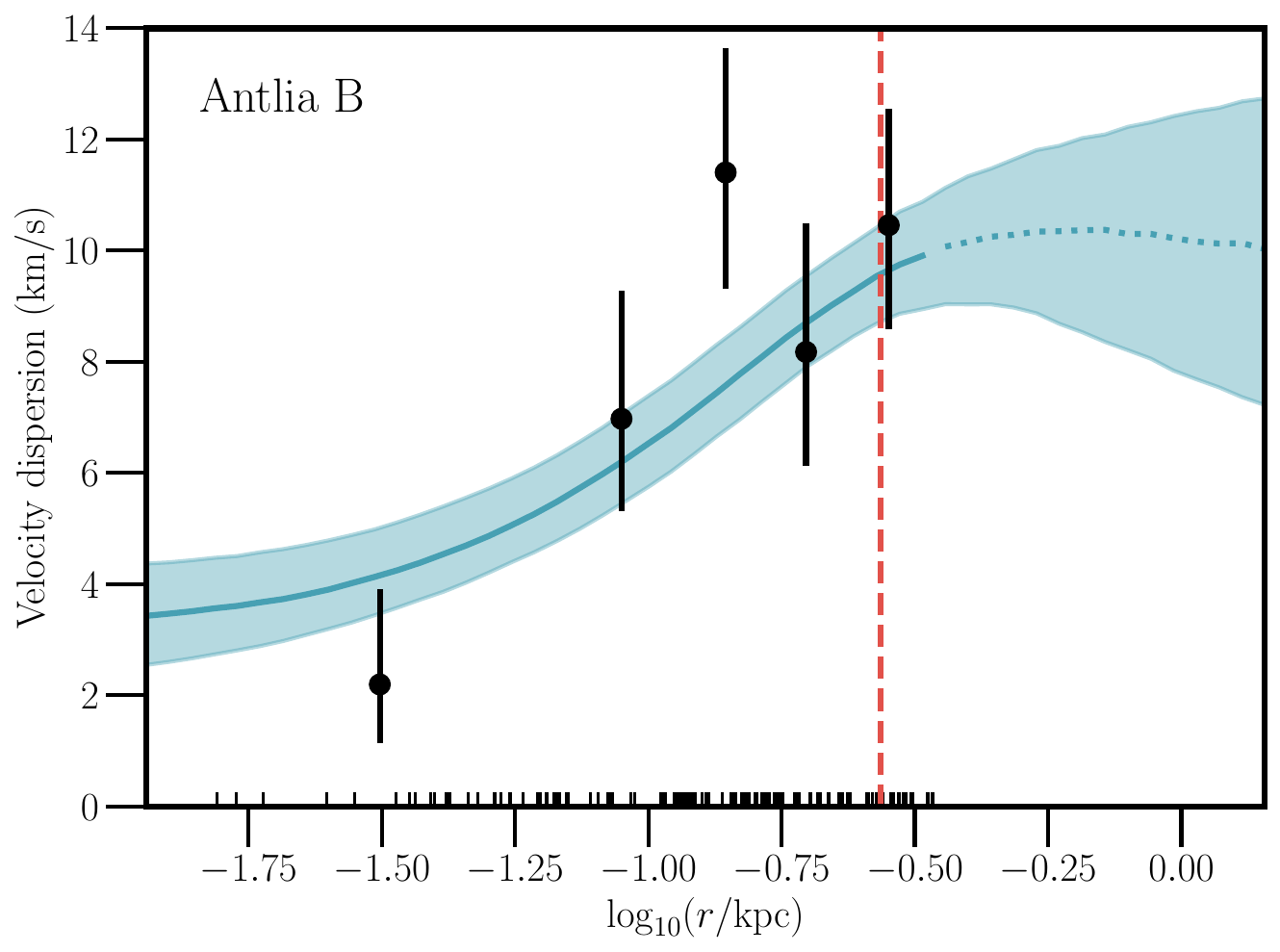}
    \includegraphics[scale=0.265]{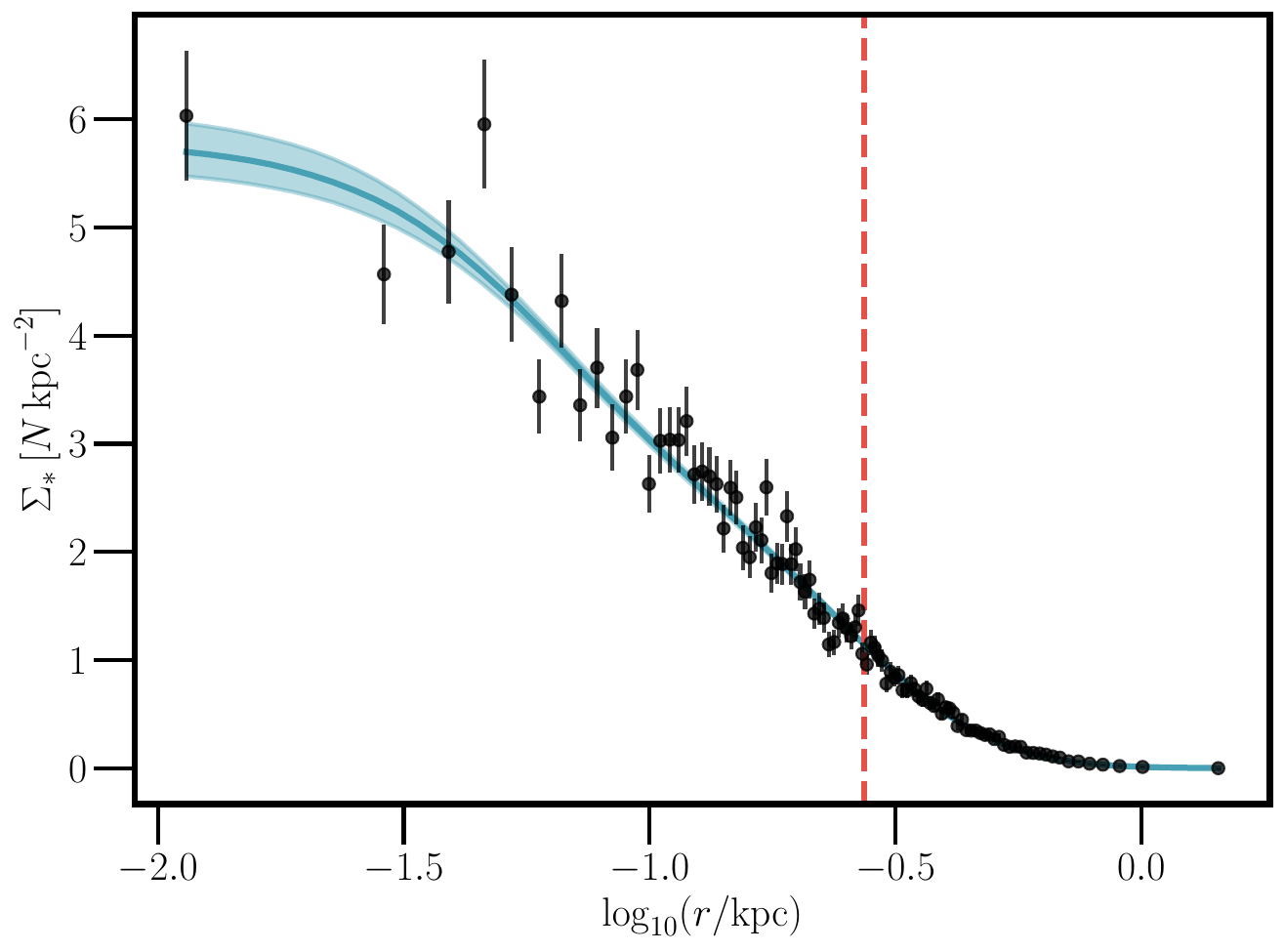}
    \includegraphics[scale=0.265]{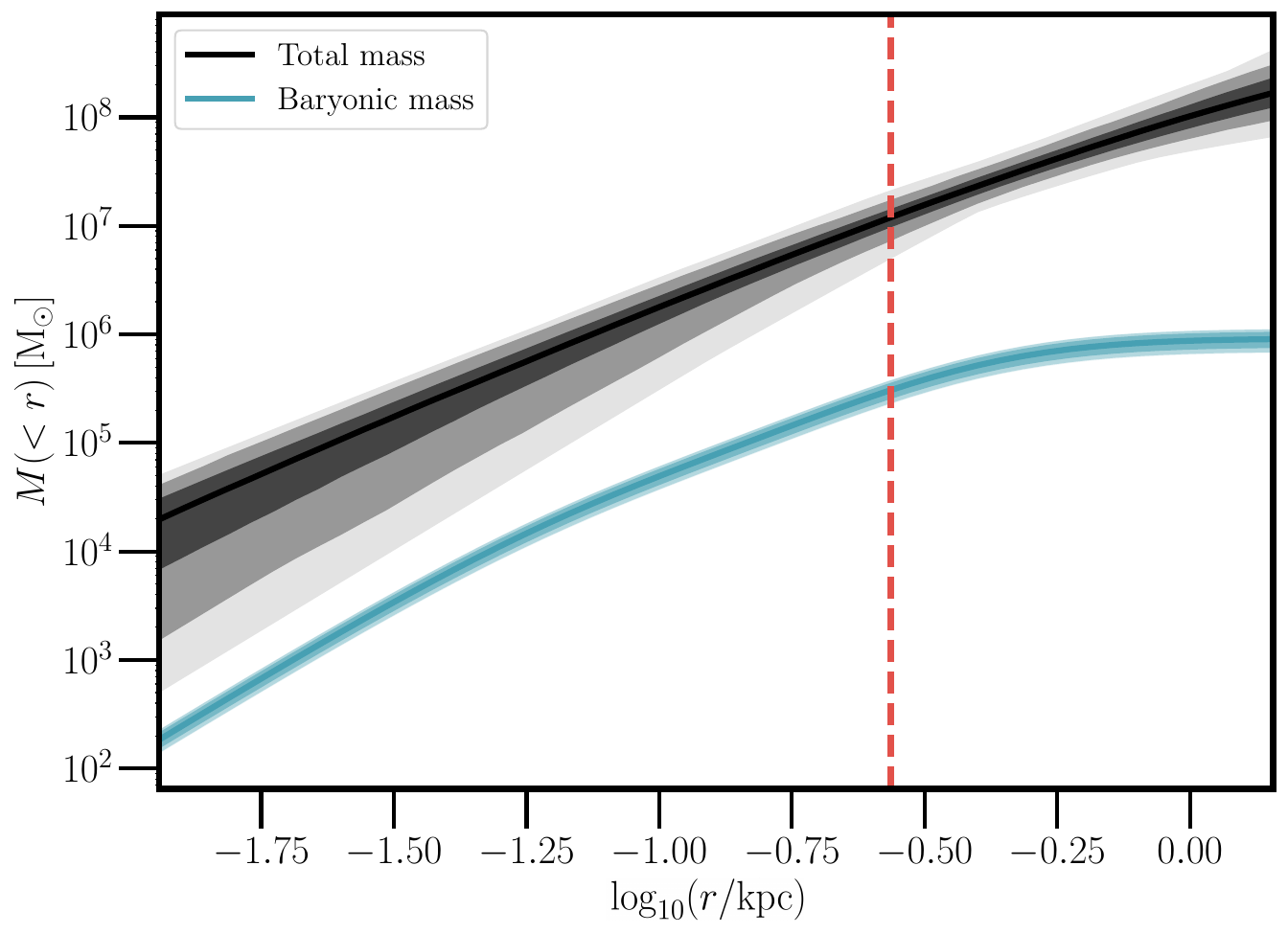}
    \includegraphics[scale=0.265]{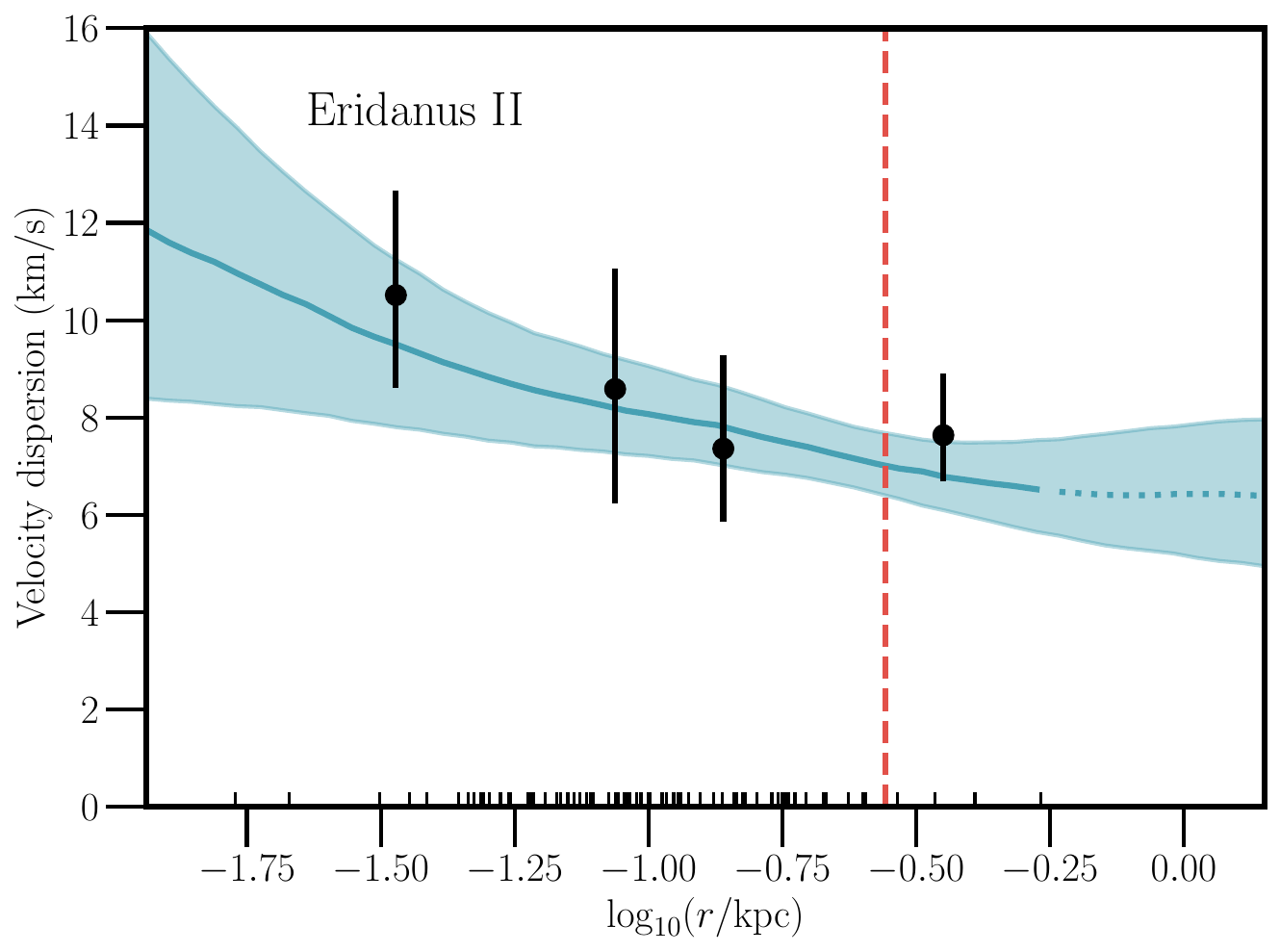}
    \includegraphics[scale=0.265]{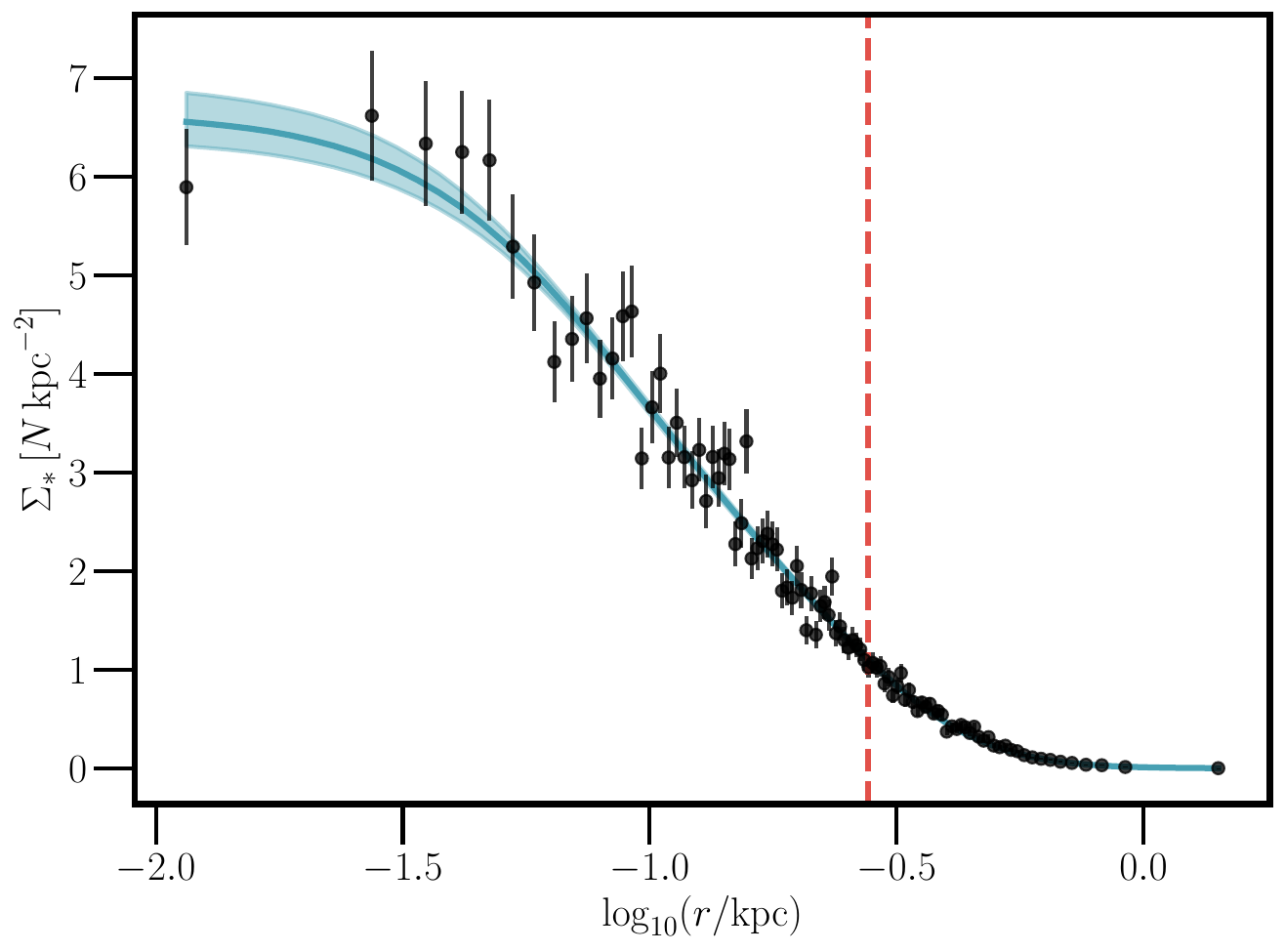}
    \includegraphics[scale=0.265]{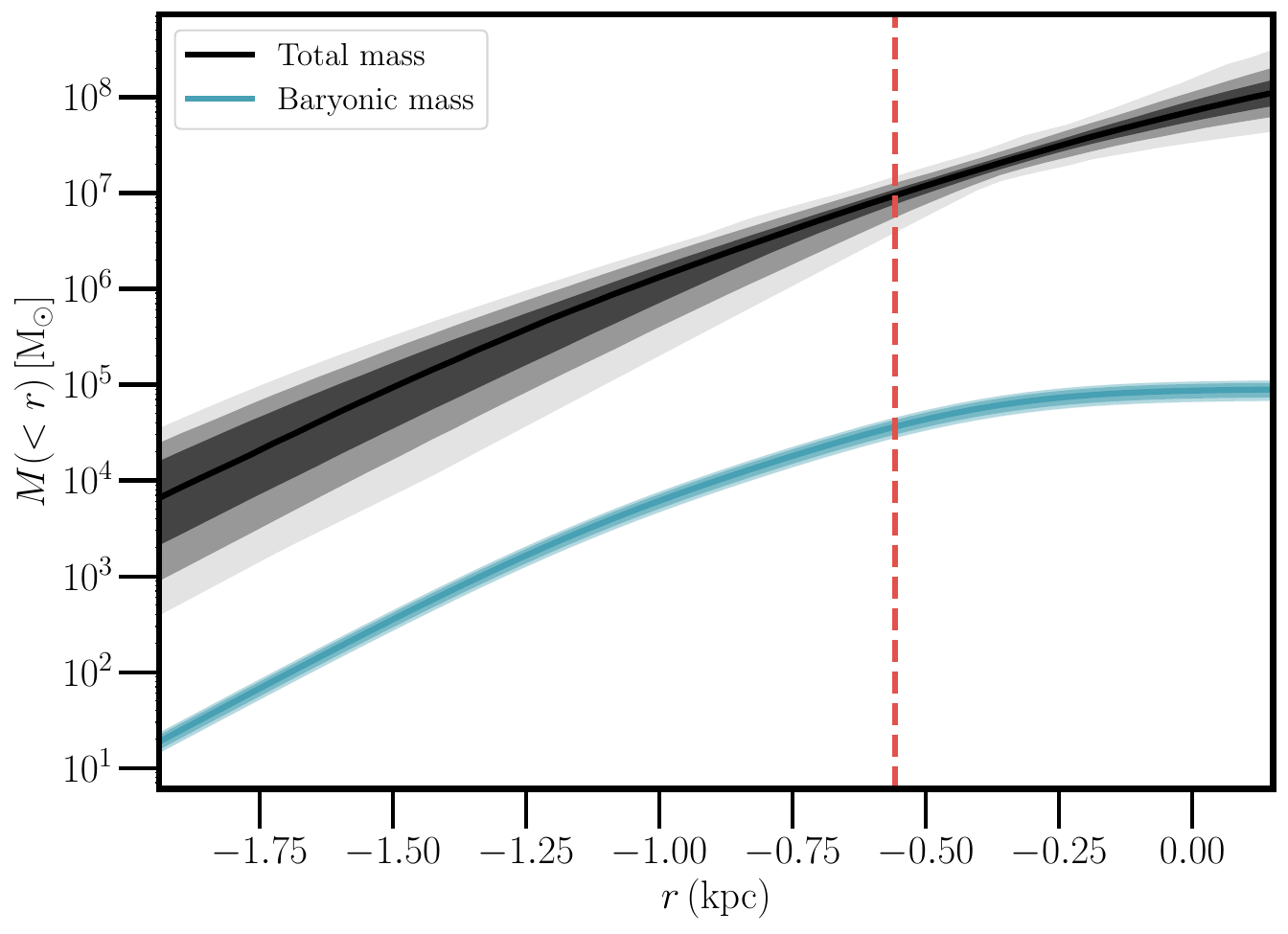}
    \caption{Continuation.}
    \label{fig:fits2}
\end{figure*}

\begin{figure*}[h!]
    \centering
    \includegraphics[scale=0.265]{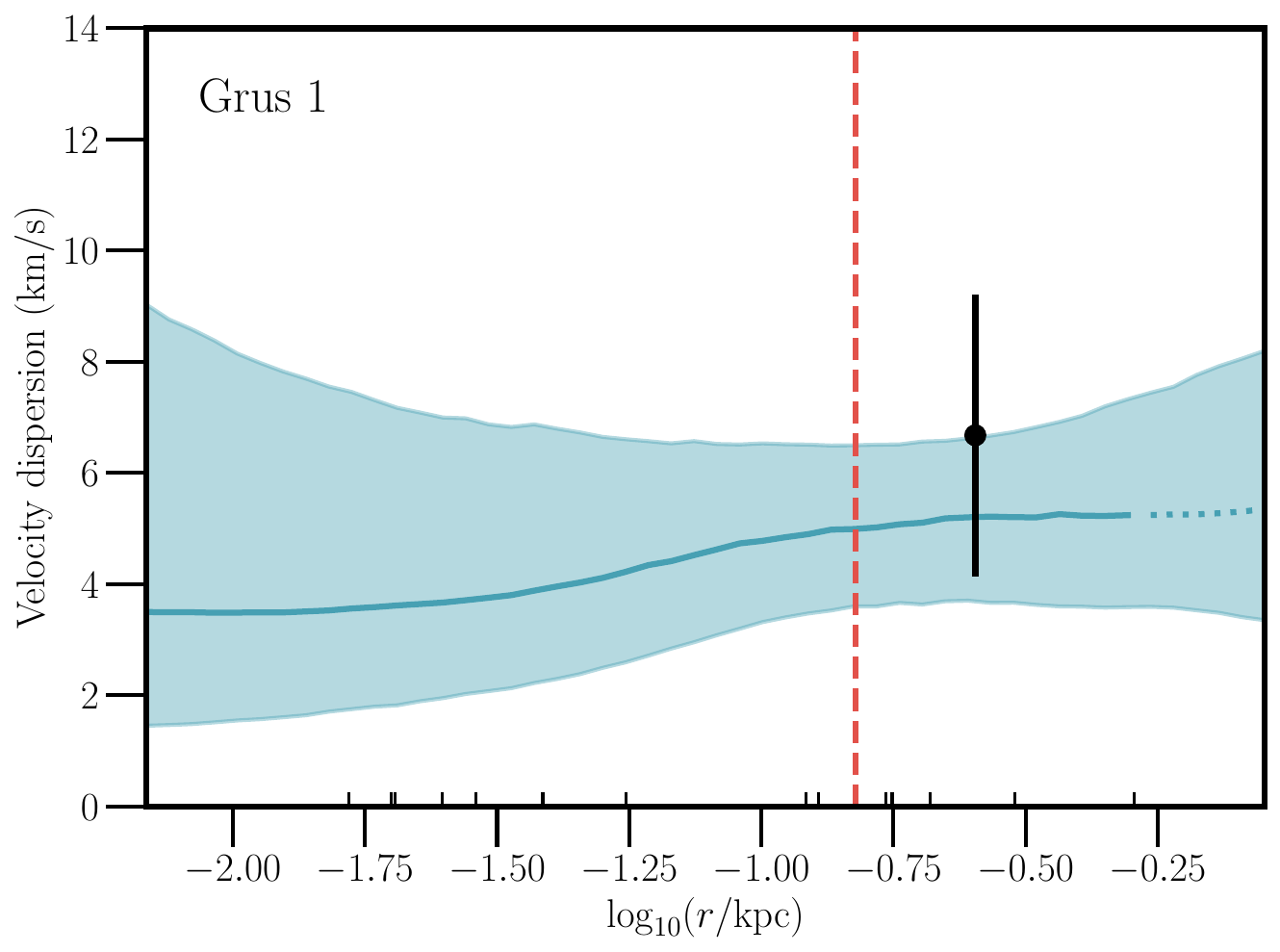}
    \includegraphics[scale=0.265]{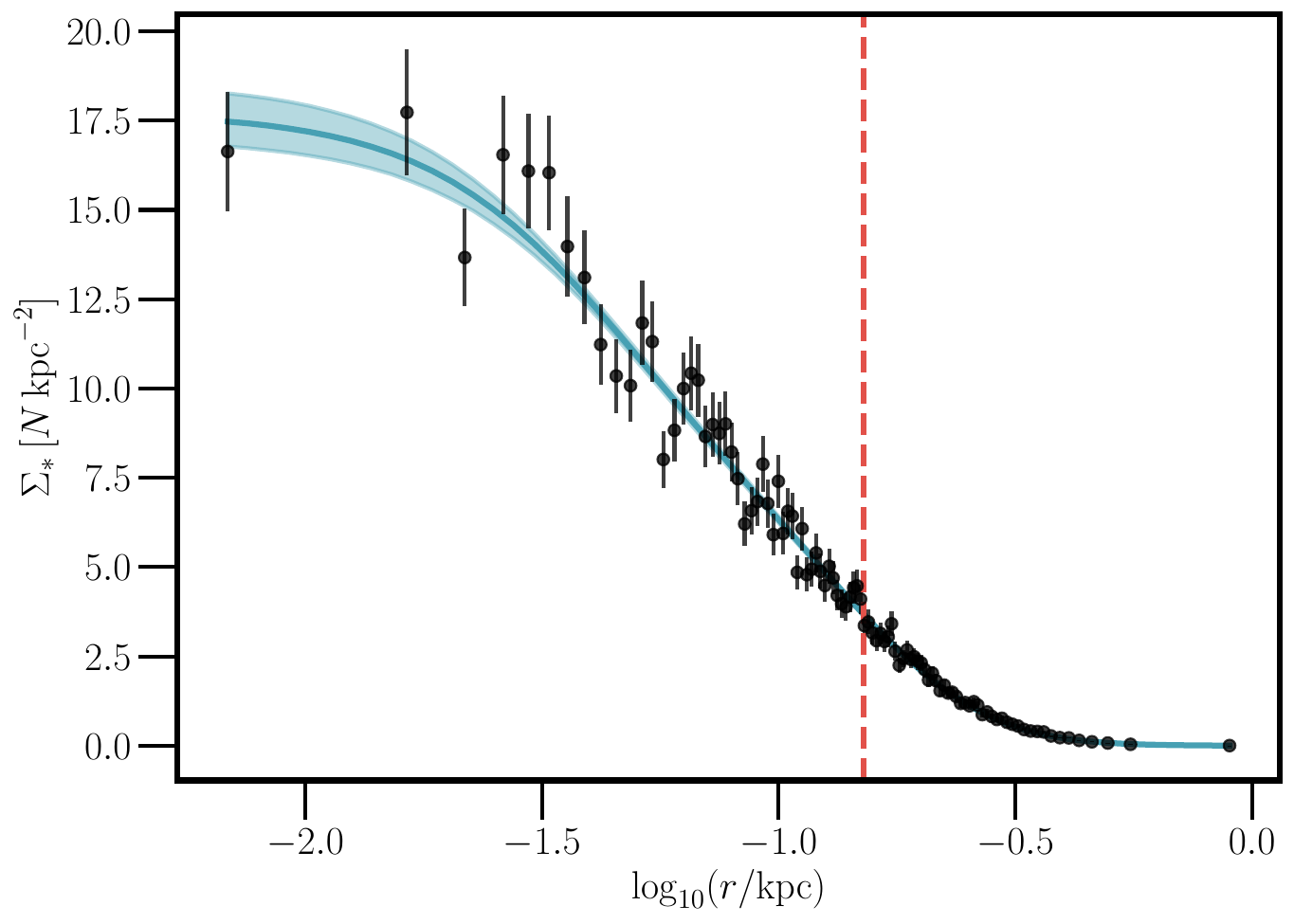}
    \includegraphics[scale=0.265]{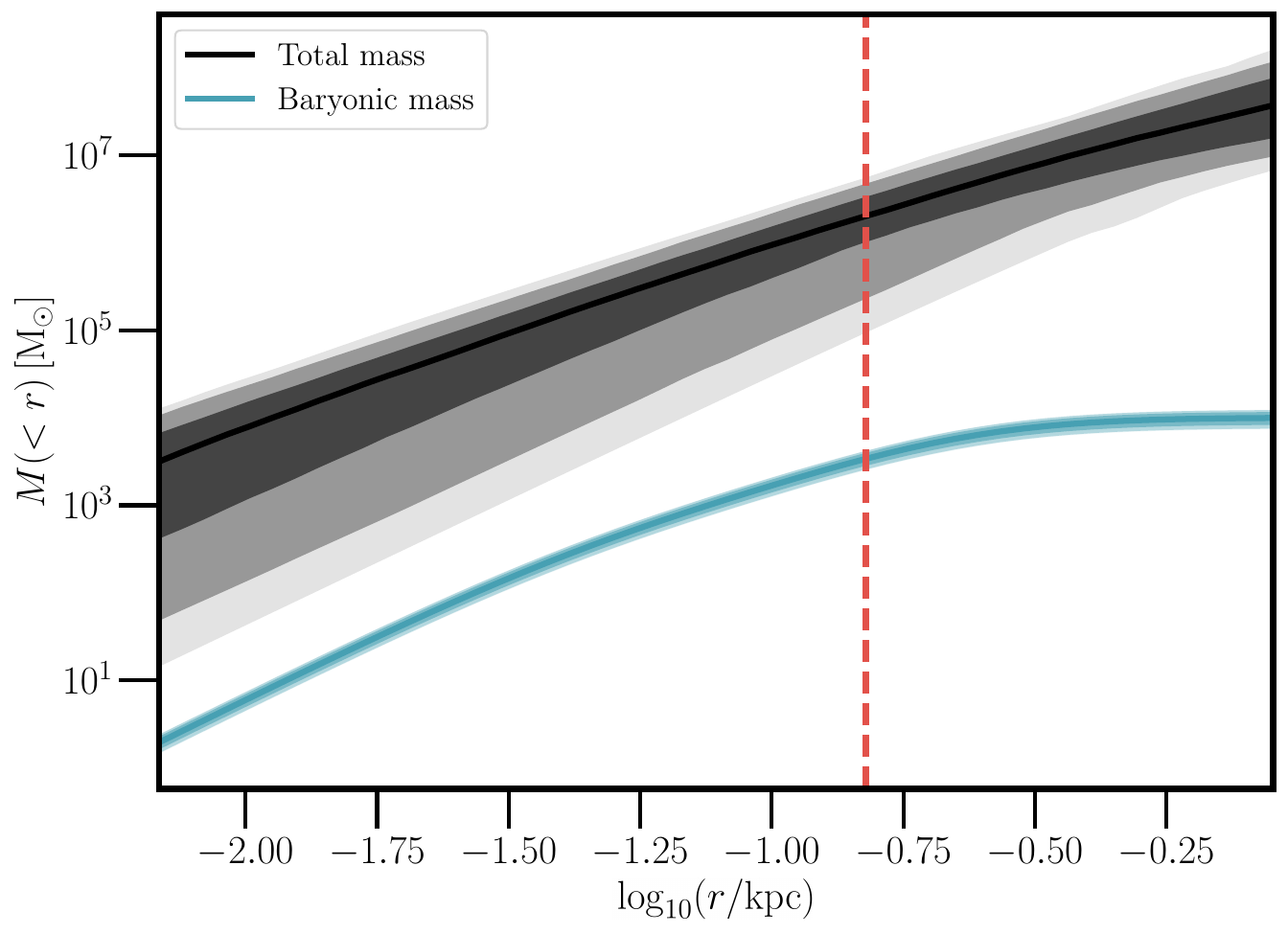}
    \includegraphics[scale=0.265]{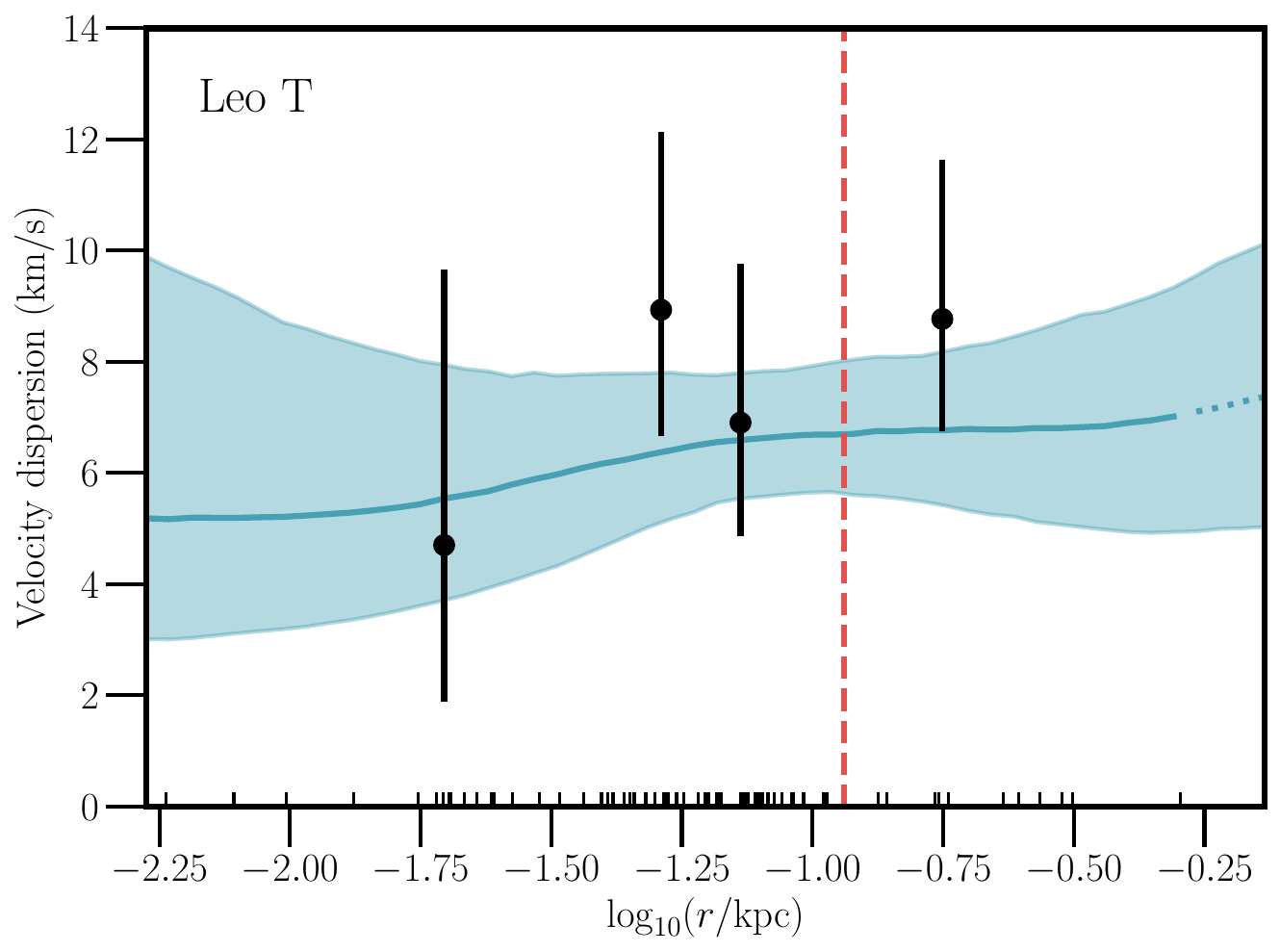}
    \includegraphics[scale=0.265]{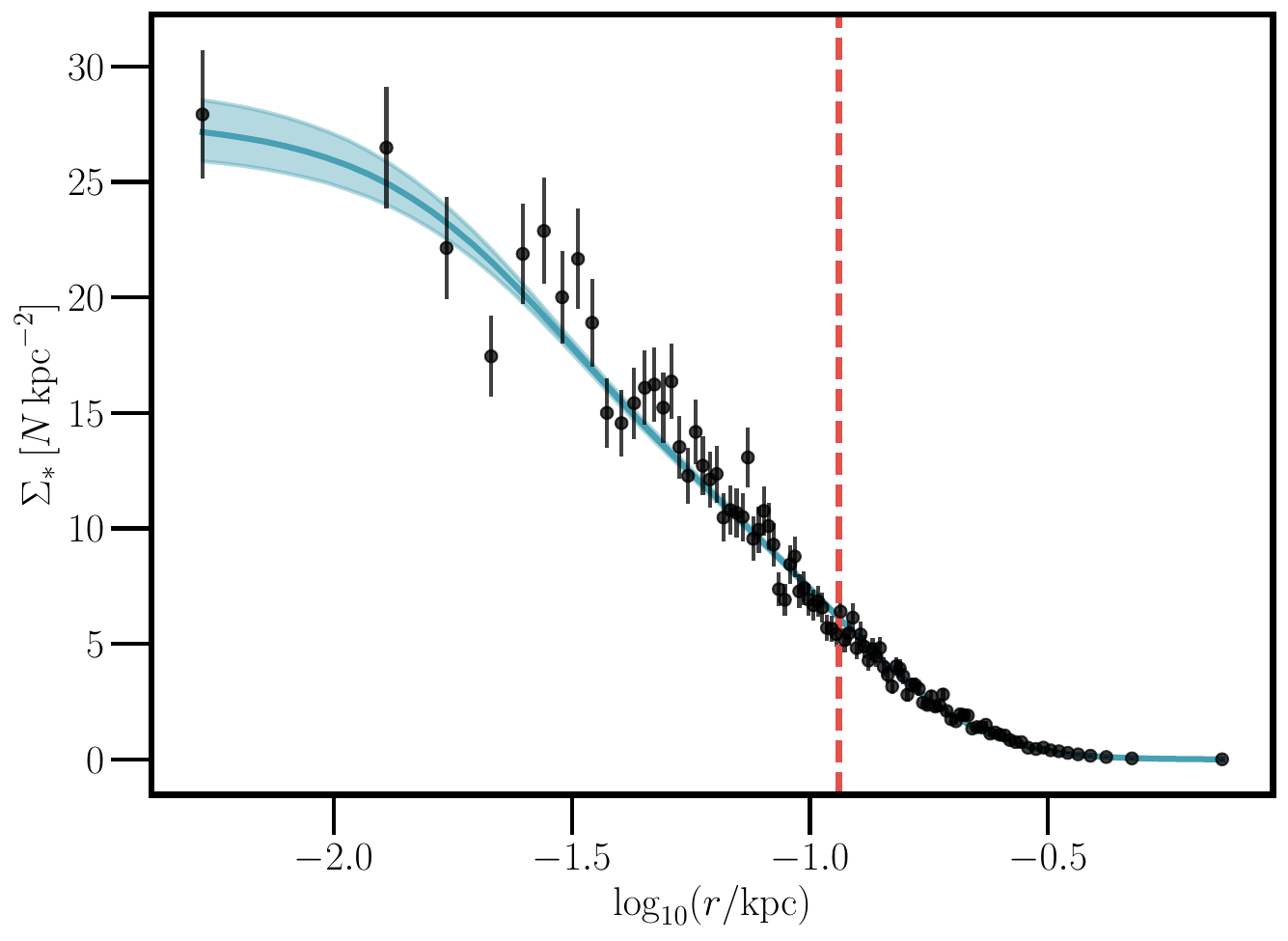}
    \includegraphics[scale=0.265]{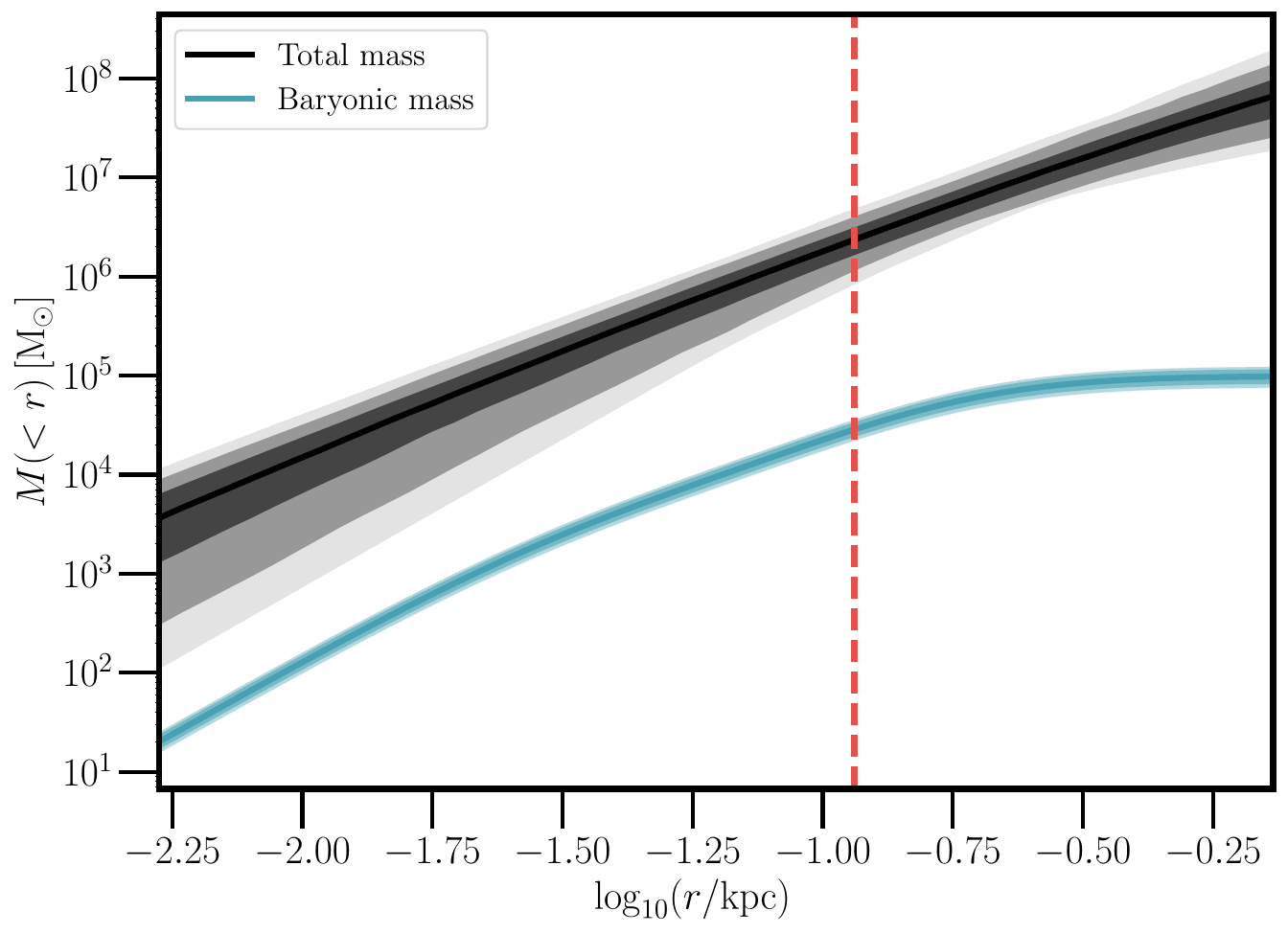}
    \caption{Continuation.}
    \label{fig:fits3}
\end{figure*}

\hfill \break \break \break \break

\section{Tracer density profile}\label{app:tracerdensity}
\begin{figure}[!htbp]
    \centering
    \includegraphics[width=7.8cm]{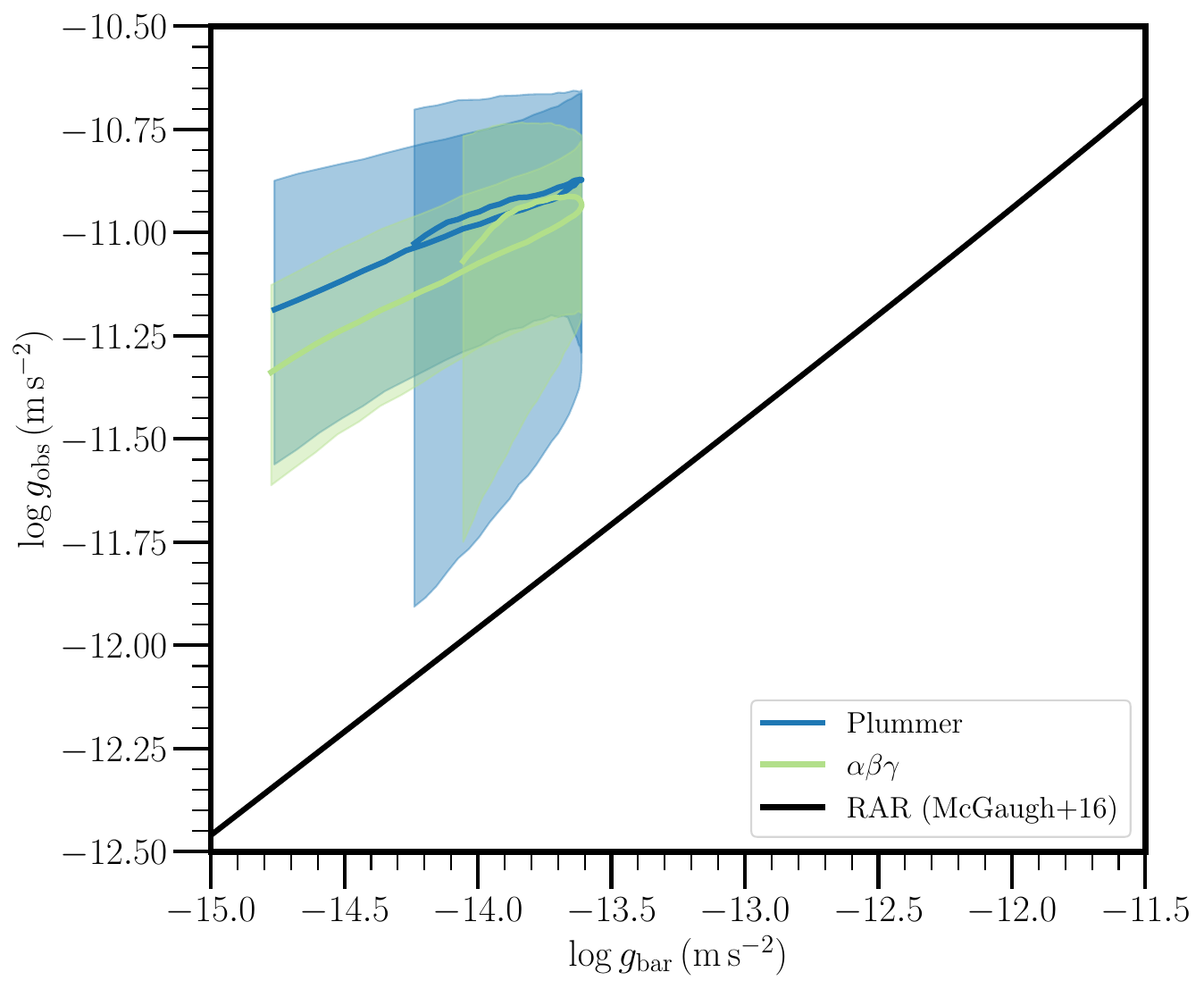}
    \caption{Comparison of the RAR of Grus I assuming two different tracer density profiles. The solid lines represent the estimated median RAR: the $\alpha\beta\gamma$ profile for the tracers in green, and a Plummer profile in blue. The shaded area corresponds to the $68\%$ confidence interval.} \label{fig:tracerdensityprofilecomparsion}
\end{figure}

Figure~\ref{fig:tracerdensityprofilecomparsion} illustrates how the RAR of Grus I, the UFD with the fewest tracers (14 stars), is affected by changing the assumed tracer density profile $\nu_\star(r)$. The blue represents the case where a Plummer profile (Equation~\ref{eq:tracerdensity_plummer}) is adopted, while the green correspond to an alternative $\alpha\beta\gamma$ profile (Equation~\ref{eq:tracerdensity_abg}). Despite this change in parametrisation, the qualitative behaviour of the RAR remains consistent, and the results agree within the $68\%$ confidence intervals. This shows that even for the least well-sampled systems, our conclusions are robust against the choice of tracer profile. For dwarfs with larger tracer samples, the sensitivity to the assumed functional form is expected to be even lower. 

\section{Galaxies with gas}\label{app:gas}
\begin{figure}[!htbp]
    \centering
    \includegraphics[width=7.8cm]{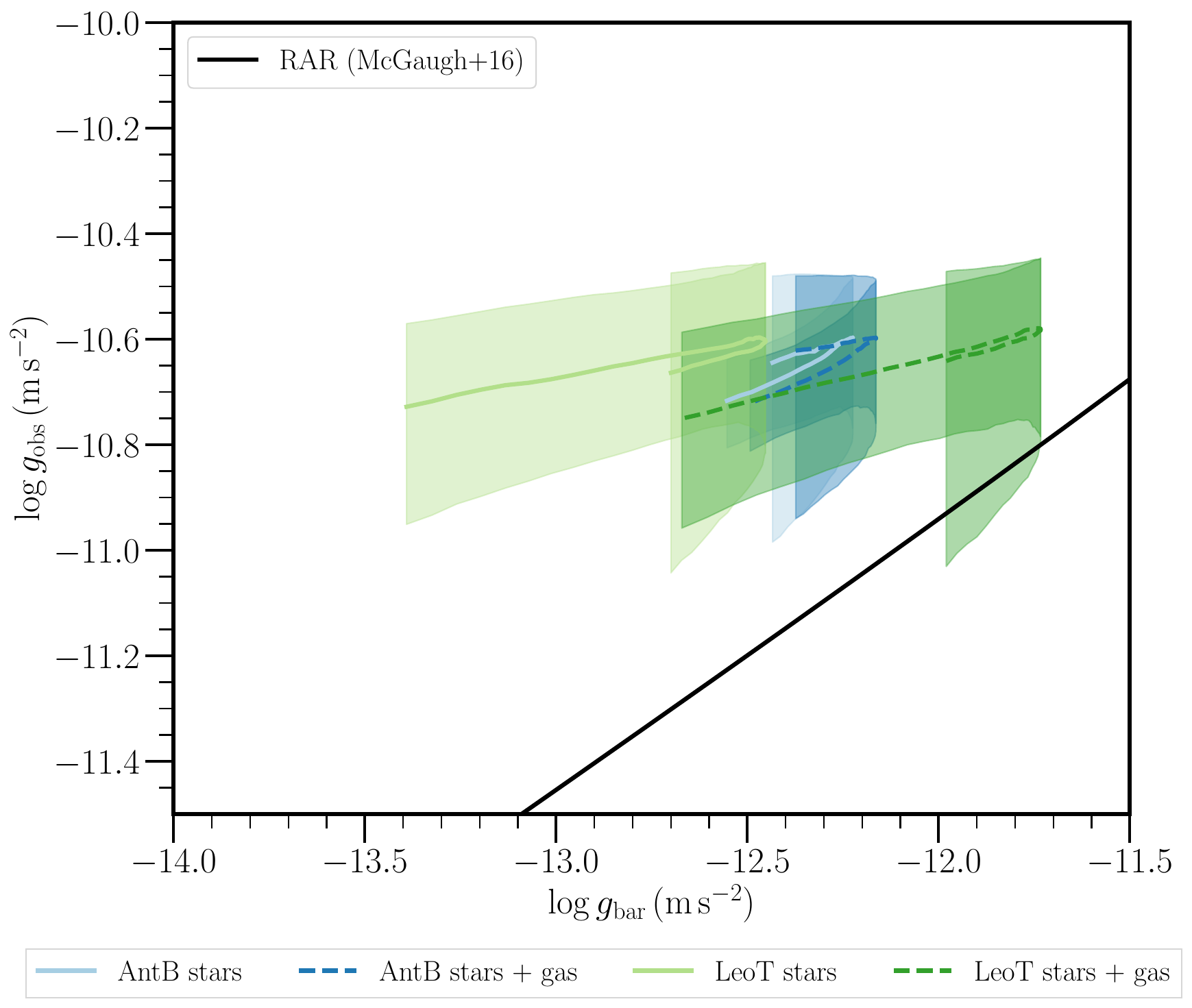}
    \caption{Comparison of the RAR of AntB and LeoT with and without their gas content. The solid lines represent their estimated median RAR considering only the mass of the stars and the dashed lines represent their estimated median RAR considering also their gas content. The shaded area corresponds to the $68\%$ confidence interval. For this comparison, we used the gas and stellar mass described in Table~\ref{tab:properties}.} \label{fig:gascomparsion}
\end{figure}

Figure~\ref{fig:gascomparsion} illustrates how the RARs of Antlia B and Leo T are affected by the inclusion or exclusion of their gas content. The dashed lines represent the first case, in which the gas is assumed to follow the same mass distribution as the stars, while the solid lines correspond to the scenario where only the stellar component is considered. As expected under our simplifying assumption, removing the gas shifts the RAR horizontally toward lower baryonic accelerations, without altering the overall shape of the relation. This shift reflects the reduced baryonic content at all radii when the gas is omitted. Despite this change, the qualitative behaviour of the RAR remains consistent, and our conclusions are unaffected.

\end{appendix}
\end{document}